\documentclass[conference]{IEEEtran}

\usepackage{graphicx}
\usepackage{subcaption}
\usepackage{epstopdf}
\usepackage{multirow}
\newcommand\blfootnote[1]{%
 \begingroup
 \renewcommand\thefootnote{}\footnote{#1}%
 \addtocounter{footnote}{-1}%
 \endgroup
}
\usepackage[]{footmisc}
\ifCLASSOPTIONcompsoc
  \usepackage[nocompress]{cite}
\else
  \usepackage{cite}
\fi
\makeatletter
\def\footnoterule{\kern-3\p@
  \hrule \@width 2in \kern 2.6\p@} 
\makeatother

\begin{document}
%
%
%
%
%

\title{Exploring Fault-Tolerant Erasure Codes for Scalable All-Flash Array Clusters}

\author{\IEEEauthorblockN{Sungjoon Koh\IEEEauthorrefmark{1},
Jie Zhang\IEEEauthorrefmark{1}\IEEEauthorrefmark{2}, Miryeong Kwon\IEEEauthorrefmark{1}, Jungyeon Yoon\IEEEauthorrefmark{4}, David Donofrio\IEEEauthorrefmark{3},\\
 Nam Sung Kim\IEEEauthorrefmark{5} and Myoungsoo Jung\IEEEauthorrefmark{1}}
\IEEEauthorblockA{\IEEEauthorrefmark{1}Computer Architecture and Memory Systems Laboratory, Korea Advanced Institute of Science and Technology,\\\IEEEauthorrefmark{2}School of Integrated Technology, Yonsei Institute Convergence Technology, Yonsei University
\\\IEEEauthorrefmark{3}Lawrence Berkeley National Laboratory,\IEEEauthorrefmark{4}Georgia Institute of Technology,\IEEEauthorrefmark{5}University of Illinois Urbana-Champaign\\
\small\IEEEauthorrefmark{1}skoh@camelab.org,\IEEEauthorrefmark{1,2}jie@camelab.org,\IEEEauthorrefmark{1}mkwon@camelab.org,\IEEEauthorrefmark{4}jungyeon@gatech.edu,\IEEEauthorrefmark{3}ddonofrio@lbl.gov,
\\\IEEEauthorrefmark{5}nskim@illinois.edu,\IEEEauthorrefmark{1}mj@camelab.org
}}

\IEEEtitleabstractindextext{%
\begin{abstract}
Large-scale systems with all-flash arrays have become increasingly common in many computing segments. 
To make such systems resilient, we can adopt erasure coding such as Reed-Solomon (RS) code as an alternative to replication 
because erasure coding incurs a significantly lower storage overhead than replication.
To understand the impact of using erasure coding on the system performance and other system aspects such as CPU utilization and network traffic, 
we build a storage cluster that consists of approximately 100 processor cores with more than 50 high-performance solid-state drives (SSDs), 
and evaluate the cluster with a popular open-source distributed parallel file system, called Ceph. Specifically, we analyze the behaviors of a system adopting erasure coding from the following five viewpoints, and compare with those of another system using replication: 
(1) storage system I/O performance;
(2) computing and software overheads; 
(3) I/O amplification;
(4) network traffic among storage nodes, and
(5) impact of physical data layout on performance of RS-coded SSD arrays. 
For all these analyses, we examine two representative RS configurations, used by Google file systems, and compare them with triple replication employed by a typical parallel file system as a default fault tolerance mechanism. 
Lastly, we collect 96 block-level traces from the cluster and 
release them to the public domain for the use of other researchers.
\end{abstract}

}

\maketitle
\makeatletter

\IEEEdisplaynontitleabstractindextext

%
\IEEEpeerreviewmaketitle

\section{Introduction}
\label{sec:introduction}

\blfootnote{This paper is published at 2018 IEEE Transactions on Parallel and Distributed Systems. This document is presented to ensure timely dissemination of scholarly and technical work.}

\IEEEPARstart{T}{he} explosive increase of data across all market segments, including internet of things (IoT), edge computing, endpoints, and data-centers, has created a demand for scalable and high-performance distributed storage systems. The amount of data generated by various sources is expected to continue to grow and reach 163 zetabytes in 2025 \cite{reinsel2017data}. 
It has been reported that 50\% of such data are generated by datacenters and enterprise edges to manage all other data \cite{rob2017edge}, which makes the storage bandwidth more important than ever before. Thus, high-performance computing (HPC) systems and datacenters have begun to adopt distributed storage systems comprising powerful computing resources with arrays of many solid-state drives (SSDs) instead of traditional spinning hard disk drives (HDDs). 
The latency and bandwidth of SSDs in such distributed storage systems are approximately 2$\times$ shorter and larger than those of enterprise HDDs, while SSDs consume significantly less power than HDDs. 
The low-power consumption and high-performance of SSDs are desirable for scalable and high-performance distributed storage systems.

\begin{figure*}
	\centering
	\includegraphics[width=\linewidth]{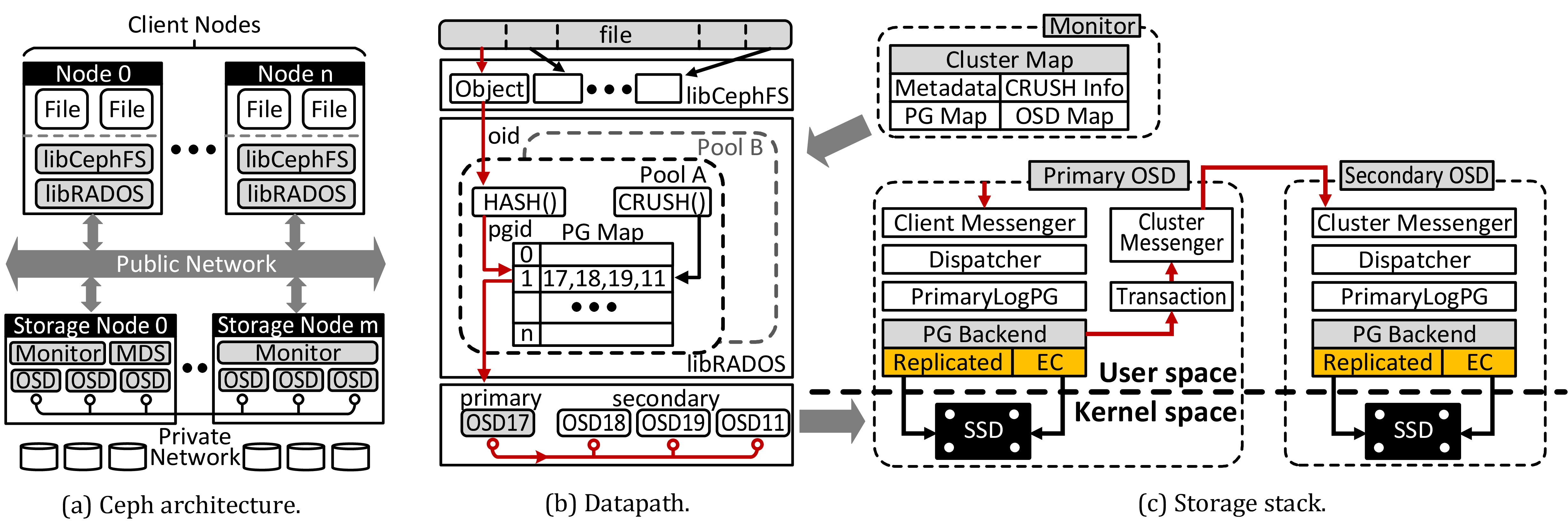}
	\vspace{-20pt}
	\caption{Overview of distributed storage systems with SSD arrays and a Ceph parallel file system.}
	\vspace{-15pt}
	\label{fig:ceph_bg}
\end{figure*}

SSD arrays (e.g., all-flash arrays) in distributed storage systems can accelerate block I/O services and increase the speed at which latency-sensitive applications process a large amount of data.  
Unfortunately, the existing configurations of distributed storage systems are often poorly tuned and unaware of the properties of the underlying SSDs. 
In addition, the distributed storage systems require strong fault tolerance mechanisms since their underlying storage devices suffer from failures; such fault tolerance mechanisms not only can significantly reduce the availability of SSD arrays but also seriously degrade their overall performance. For example, Facebook reports that up to 3\% of its storage devices fail every day~\cite{sathiamoorthy2013xoring}. 
Although the reliability of SSDs is significantly higher than that of HDDs, such frequent failures should be efficiently handled. 
In addition to the failures of disks, hardware and software failures of network switches and storage nodes caused by soft errors, hard errors and/or power outages can prevent accesses to storage devices in storage nodes \cite{greenberg2008cost}.
To keep data available and protected against such software and hardware failures, most conventional distributed storage systems employ replication, which is a simple but effective method for making distributed storage systems resilient. However, replicating a large amount of data can in practice introduce significant storage overheads and performance degradation~\cite{hu2009write, jung2013revisiting, jung2016exploring}.

Erasure coding can be used as an alternative fault tolerance mechanism to replication because it offers the same reliability as or higher reliability than triple replication (here in after denoted as ``3-replication''), with considerably lower storage overheads. The Reed-Solomon (RS) code is one of the most popular erasure codes owing to its optimal storage space utilization~\cite{reed1960polynomial, rashmi2015having, rashmi2016ec, mitra2016partial} and can be easily applied to SSD arrays in distributed storage systems to address the overheads imposed by the traditional replication method. 
When the RS code is employed, all input data are stored as \emph{chunks} of fixed size. 
Generally, RS($k$,$m$) divides target data into $k$ data chunks and computes $m$ coding chunks associated with the data chunks. The system then distributes the total $n$(=$k$+$m$) chunks across different storage devices or nodes. 
RS($k$,$m$) can recover the original data in up to $m$ data chunks among $n$ chunks in case of a failure. 
For example, Google Colossus, which is the successor of the Google File System~\cite{metz2012google, ghemawat2003google, ford2010availability}, uses RS(6,3) to tolerate any failure in up to three chunks with only 1.5$\times$ storage overheads. 
Compared with the 3$\times$ storage overheads of the traditional replication, the 1.5$\times$ storage overheads of RS(6,3) are attractive for distributed storage systems.

However, one of the disadvantages of adopting erasure coding to distributed storage systems is the detrimental effect of their reconstruction operations on the system performance. 
When a node discovers that a data chunk is missing due to a failure, RS($k$,$m$) requires the node to bring $k-1$ remaining chunks over the network, reconstruct the missing chunk, and sends the repaired chunk to the corresponding node. 
Such a network traffic overhead, which is also referred to as \emph{repair traffic}, is a well-known issue behind the adoption of erasure coding~\cite{dimakis2011survey, sathiamoorthy2013xoring}. 
For example, a Facebook cluster deploying erasure coding increases the network traffic by more than 100TB for data reconstruction in a day~\cite{rashmi2013solution}. 
To address this problem, there are many studies on finding optimal trade-offs between the network traffic and storage overheads for repairs~\cite{huang2012erasure, sathiamoorthy2013xoring, esmaili2013core}. 
By exploiting such a trade-off (e.g., lower storage overhead at the cost of higher repair network traffic)~\cite{huang2012erasure, sathiamoorthy2013xoring}, system architects can choose the optimal coding scheme for the given system architecture.

Even though there also exist many studies on the repair traffic, little attention has been paid to the impact of deploying the RS code on SSD-based distributed storage systems. 
The repair traffic occurs only when there is a failure, but encoding, concatenating, decoding data always incur various overheads that consume more CPU cycles, and introduce network traffic and I/O amplification in SSDs.
Furthermore, since SSDs are significantly faster than HDDs, data must be encoded/concatenated/decoded and transferred over network at higher rates to realize their benefits.
This motivates us to study the overheads on the network and storage nodes in detail.
Such an investigation will allow us to more efficiently use distributed storage systems based on SSD arrays. 

To this end, we construct a distributed storage system consisting of 96 processor cores and 52 high-performance SSDs. We leverage an open-source parallel file system, called \emph{Ceph}~\cite{weil2006ceph}, which is widely employed by distributed storage systems; Ceph can also easily support erasure coding as a plug-in module. 
In this study, we first examine the distributed storage system employing the popular RS configuration, RS(6,3) used by Google Colossus \cite{metz2012google}, with synthetic I/O-intensive workloads and compare its behaviors with system employing 3-replication.
While our observations on the synthetic I/O-intensive workloads yield many key insights for tuning distributed storage systems and existing parallel file systems, it is also desired to examine the systems in real application scenarios.  
Therefore, we evaluate the distributed storage system with a Standard Performance Evaluation Corporation (SPEC) suite for measuring the performance of file servers, which is denoted as SFS and was recently released (January 18, 2018) \cite{spec2014sfs}. This SPEC SFS executes a series of ``real'' applications atop the storage system, based on diverse file server usage scenarios, including a database (DB), virtual desktop infrastructure (VDI), electronic design automation (EDA), and video data acquisition (VDA).
 
We classify the key observations made in this study into six categories: 
1) \emph{throughput impact}, 2) \emph{latency impact}, 3) \emph{CPU utilization}, 4) \emph{context switch overheads}, 5) \emph{I/O amplifications} (related to reliability of SSDs) and 6) \emph{network traffic overheads}. 
Compared with 3-replication, RS(6,3) degrades the system throughputs and increases the latency by 42.5\% and 1250\%, respectively, on average. This is because parallel file systems must run a module performing RS coding at the user-level owing to the architectural support for a scale-out option, which consumes more CPU cycles and introduces additional context switches (related to system calls) and network traffic, compared with 3-replication by 91\%, 234\%, and 105\%, respectively, on average. 
More importantly, we observe that, in contrast to common expectations, RS(6,3) incurs 151\% and 387\% more write and read requests than 3-replication, respectively, on average.
This in turn shortens the lifespan and reliability of the underlying SSD arrays. More details regarding our in-depth analyses are presented in Section \ref{sec:dataoverhead}.

To the best of our knowledge, this study is the first examination of various overheads imposed by online erasure coding on a 
physical distributed storage system with an array of SSDs. We also collected all block-level traces 
from the SSD array enabled cluster that we built and made them available to freely download for other studies\footnote{All the traces collected for this paper are available for download from http://trace.camelab.org.}. Lastly, we discuss several system implications which can be applied to future system to reduce the performance degradation and overheads.

\section{Background}
\label{sec:background}

In this section, we explain distributed storage systems that employ SSD arrays and Ceph as their parallel file system. We also describe the fault tolerance mechanisms therein and illustrate online erasure coding in detail. 
\subsection{Distributed SSD Array Systems and Ceph}
\noindent\textbf{Overall architecture.} Figure \ref{fig:ceph_bg}a illustrates the architecture of a distributed SSD array system, which is managed by a Ceph parallel file system \cite{weil2006ceph}. From a high-level viewpoint, the client nodes and storage nodes comprise multiple server computers. Each client node can execute multiple host applications, and these applications can access the underlying storage cluster through a key library of Ceph, called \emph{reliable autonomic distributed object store} (RADOS) \cite{weil2007rados}. This RADOS library module (i.e., libRADOS) connects all the underlying storage nodes as a cluster with client nodes. 
The libRADOS manages the storage cluster via \emph{objects}, 
each including the corresponding data, a variable amount of metadata, and a globally unique identifier. Therefore, the Ceph file system (CephFS) library (i.e., libCephFS) is employed between the libRADOS and applications that use a conventional file system for their I/O services. On the other hand, the clients are connected to the storage nodes through a high speed network, which is referred to as a \emph{public network}. Each storage node consists of several monitor daemons and one or more \emph{object storage device daemons} (OSDs). For using Ceph as a parallel file system, the cluster needs to employ a \emph{metadata server} (MDS). This daemon performs all operations related to metadata. However, it always employs replication for supporting a highest reliability of metadata management; thus, we do not discuss MDS herein. While an OSD handles read/write services from/to a storage device (i.e., an SSD in this study), a \emph{monitor} manages the layout of objects, the access permissions, and the status of multiple OSDs. The storage nodes are also connected through a network, which is referred to as \emph{private network}, and is used for ensuring the reliability of the underlying distributed storage system. Through this network, multiple OSDs and monitors communicate with each other and check the health of their daemons. For example, every six seconds, each OSD checks the heartbeat of other OSDs, thereby providing highly reliable and consistent backend data storage \cite{weil2007rados}.

\begin{figure}
	\centering
	\includegraphics[width=\linewidth]{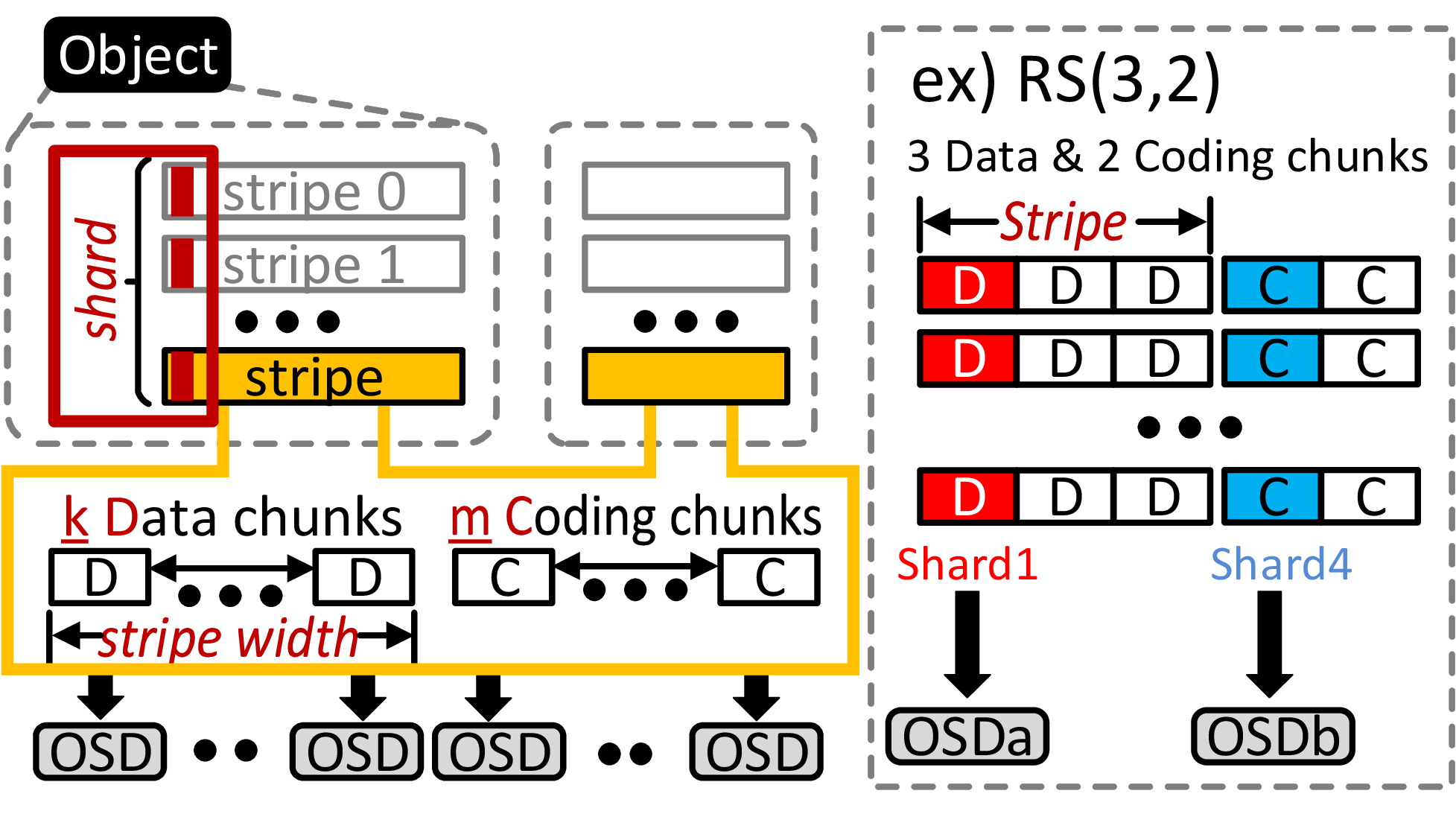}
	\vspace{-15pt}
	\caption{Terminology in erasure coding.}
	\vspace{-10pt}
	\label{fig:term}
\end{figure}

\noindent \textbf{Data Path.} For an application running on a client, libCephFS and libRADOS establish a channel with the underlying storage cluster by retrieving a key and a cluster map from the cluster's monitor. This includes information about the cluster topology related to mapping and authorization. Once the application sends an I/O request with the information of its target file system to libCephFS, libCephFS determines the object associated with the file system and forwards the request to libRADOS. This object is first mapped to one of the \emph{placement groups} (PGs) by determining an object ID with a simple hash function and adjusting the object ID aligned with the total number of PGs that the corresponding pool includes. After determining the corresponding PG ID, \emph{controlled replication under scalable hashing} (CRUSH) assigns an ordered list of OSDs to the PG, based on the cluster topology and ruleset, which represents the information of a given pool such as a type or the number of replicas \cite{weil2006crush}.
Finally, libRADOS issues the I/O request to the primary OSD, which manages all the corresponding I/O services (usually related to resilience and workload balance) by referring to all other OSDs listed in the PG map. This PG exists in the cluster map, which can be also retrieved by communicating with the monitor that resides at the node of the primary OSD.

\begin{figure*}
\centering
\begin{subfigure}[b]{0.35\linewidth}
\includegraphics[width=1\linewidth]{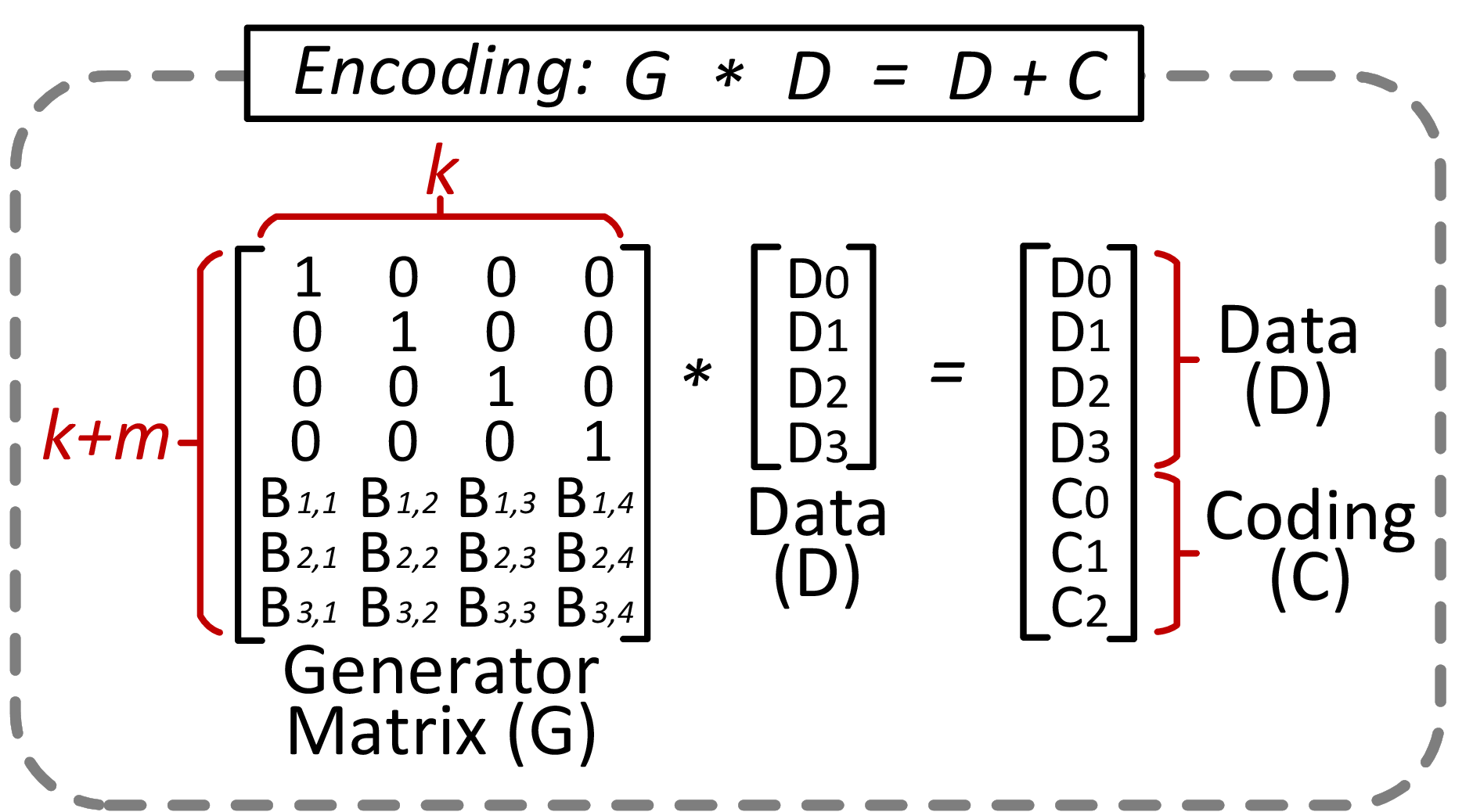}
\vspace{-13pt}
\caption{Encoding.}
\label{fig:encoding}
\end{subfigure}
~
\begin{subfigure}[b]{0.32\linewidth}
\includegraphics[width=1\linewidth]{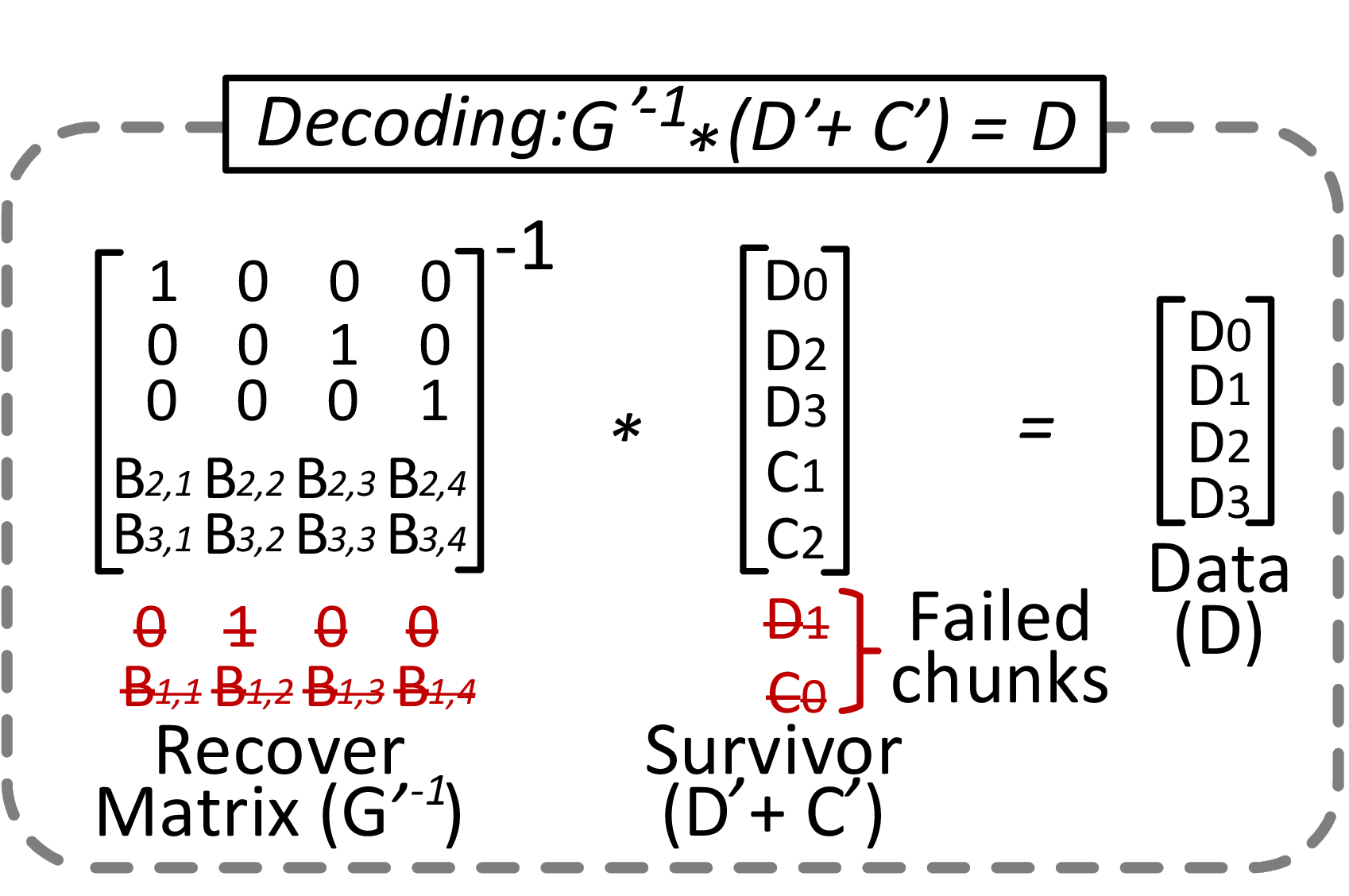}
\vspace{-13pt}
\caption{Decoding.}
\label{fig:decoding}
\end{subfigure}
~
\begin{subfigure}[b]{0.29\linewidth}
\includegraphics[width=1\linewidth]{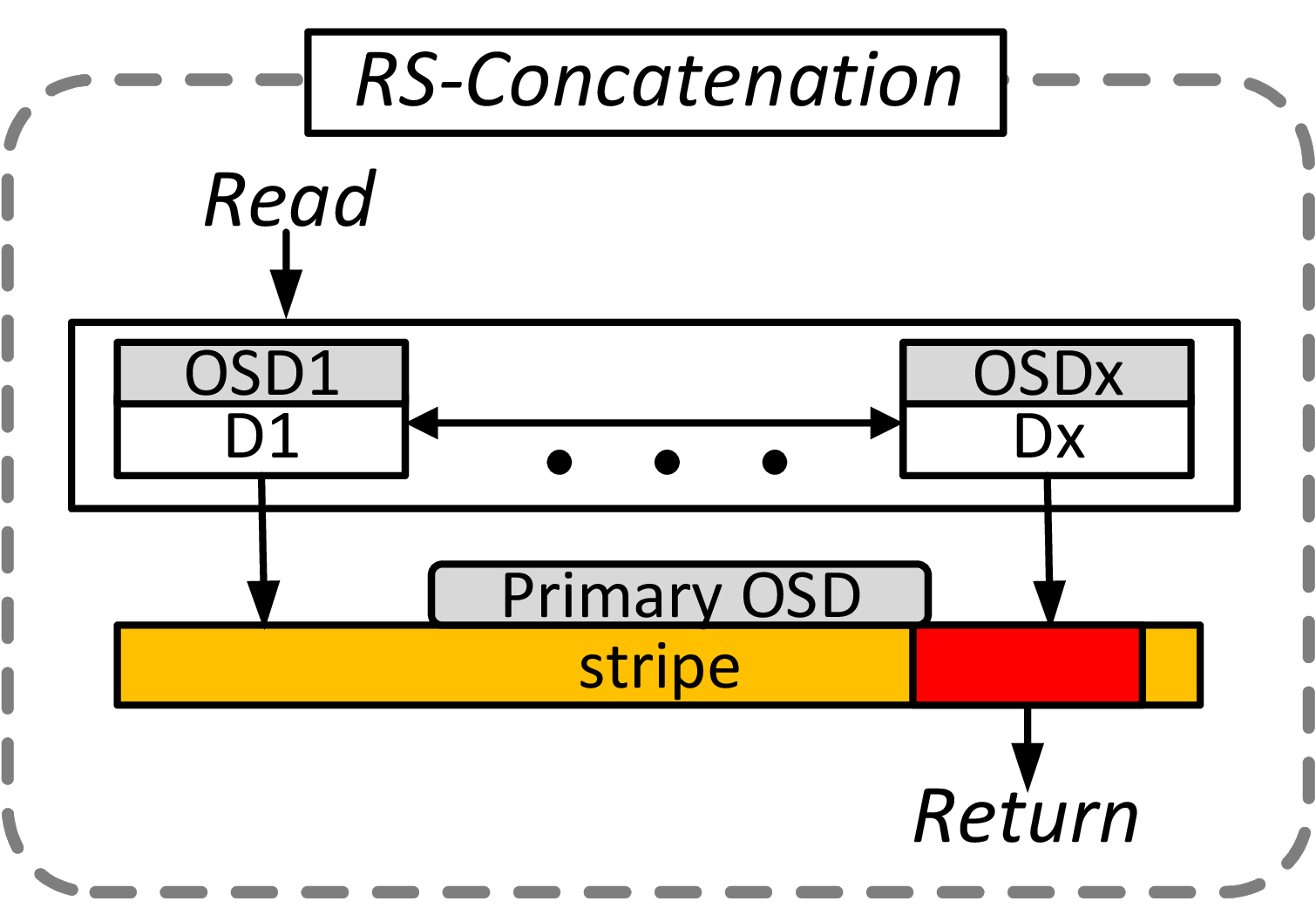}
\vspace{-13pt}
\caption{RS-concatenation.}
\label{fig:rsconcat}
\end{subfigure}
\vspace{-25pt}
\caption{Erasure coding in Ceph parallel file system.\vspace{-17pt}}
\label{fig:overview}
\end{figure*}

\noindent \textbf{Storage stack.} As shown in Figure \ref{fig:ceph_bg}c, a client messenger, which handles all incoming I/O requests, is located at the top of the storage stack of each node. Underneath the client messenger, a dispatcher fetches an I/O request and forwards it to the PG backend module. Because a storage node can fail at any time for many unknown reasons, a log system, called \emph{PrimaryLogPG}, keeps track of all I/O requests fetched by the dispatcher. The PG backend replicates or encodes data based on the given RS code. During this phase, it generates data copies (for replication) or data/coding chunks (for erasure coding), which are managed by a transaction module. This transaction module forwards the copies or chunks to another cluster messenger, which resides on the OSD of the replica or chunk. Note that, all these activities are performed by storage nodes without any intervention of the client-side software. In particular, the data related to replication or erasure coding are transferred across multiple OSDs located in different storage nodes through the private network. Consequently, this data transfer is completely invisible to the client-side applications.

\subsection{Fault Tolerance Mechanisms}
When an application sends an I/O request to a specific target PG, it first acquires a lock, called a \emph{PG lock}, to make sure that the data are successfully stored in the storage cluster. This lock can be released by a commit message, signaling that the corresponding primary OSD will return. During this phase, the primary OSD replicates the data or performs erasure coding on the data to ensure the high resilience of the underlying storage cluster.

\noindent \textbf{Replication.} Traditionally, a target system is made to be fault-tolerant by replicating data in distributed file systems \cite{weil2006ceph}. Ceph applies 3-replication as its default fault-tolerance mechanism \cite{cephpoolconfig}. Similar to a client that retrieves the PG and the corresponding OSD through the cluster map (i.e., CRUSH and PG maps), any primary OSDs can examine the indices of OSD in the PG through a copy of the cluster map, which can be retrieved from the monitor of target node. The PG backend of the primary OSD writes replicated data (object) to the secondary OSD and the tertiary OSD through the transaction modules. If the target system adopts a stronger replication strategy, it can replicate the object to the appropriate PGs for targeting as many OSDs since there are extra replicas. 
Once the replication process is complete, the primary OSD informs the client about whether all the data are successfully stored by responding to a commit. Note that the amount of data that the primary OSD transmits through the private network can be at least 2$\times$ greater than the amount of data received from the public network.

\noindent \textbf{Erasure coding.} Unfortunately, in some cases, a replication method cannot be straightforwardly applied to a storage cluster that employs only SSDs due to the high cost of SSDs per gigabyte. To address this, the distributed storage system community has paid attention to erasure coding, which introduces less storage overhead (less redundant information) for a given level of reliability. A very recent version of Ceph (version 12.2.2 \cite{cephluminousrelease}) employed RS-coding, which is one of the most popular and effective erasure coding techniques, and is now available for a file system interface (libCephFS). In practice, RS coding is classified as a maximum distance separable code \cite{rosenthal1999maximum}, which is an optimal technique that secures the highest possible reliability within a given storage budget. As shown in Figure \ref{fig:term}, RS codes have two types of chunks to manage the fault tolerance for target data of size $N$: (1) \emph{data chunk} and (2) \emph{coding chunk} (also known as parity chunk). While $k$ data chunks are related to the original contents of the target data, $m$ coding chunks maintain the parity data, which can be calculated from the $k$ data chunks. A \emph{stripe} is the unit that an RS code encodes, and it consists of $k$ data chunks. The size of the stripe, referred to as the \emph{stripe width}, is $k\times n$ where $n$ is usually 4KB in Ceph. 
Since failures can occur at the SSD-level and/or the node-level, Ceph stripes the data chunks and coding chunks into $k+m$ different OSDs.
To make the binary data in an object reliable, Ceph manages the data chunks and coding chunks of RS($k,m$) for each object whose default size is 4MB. Thus, for a given RS($k,m$) code, there exist $N/(k*n)$ stripes. All the chunks that exhibit the same offset for the $N/(k*n)$ stripes are referred to as a \emph{shard}, and all the shards are managed by different OSDs. 

\subsection{Online Encoding and Repair for Failures}
\noindent \textbf{Encoding.} Figure \ref{fig:encoding} shows a matrix-vector multiplication for generating $m$ coding chunks from $k$ data chunks. The $k$ words ($w$ bits of each data chunks) are considered as a vector, and $m$ coding chunks can be calculated by multiplying the vector by a matrix, referred to as \emph{generator matrix} (G). This G is $(k+m)*k$ geometric progression matrix, which is constructed by Vandermonde matrix \cite{lacan2004systematic}, and each of its element is calculated via Galois Field arithmetic \cite{omura1986computational}. Each row of the Vandermonde matrix has a form of a geometric sequence that begins with 1. The $(k+m)*k$ extended Vandermonde matrix is needed to construct a generator matrix, whose first and last rows are the same as those of the identity matrix, $k*k$, respectively. On the other hand, the rest of the matrix complies with the general form of the Vandermonde matrix. This extended matrix is converted into a generator matrix that whose size is as same as the size of the extended matrix. The conversion is performed by multiplying a row/column by a constant value and adding a row/column to another row/column. The first $k$ rows comprise the identity matrix $k*k$ and the following $m$ rows compose coding matrix whose first row has 1 for each element.

\noindent \textbf{Decoding.}
To construct original data with $k$ chunks among the $k+m$ chunks, a recover matrix is constructed, which is the inverse matrix of the generator matrix whose rows were calculated with failure chunks, removed in the encoding operation. This inverse matrix is multiplied by remaining chunks, in a similar process to the encoding operation. As a result of the matrix-vector multiplication, we obtain the original chunks as shown in Figure \ref{fig:decoding}. The network traffic and computation overheads are greater than the encoding of RS codes. Since we also have to read remaining chunks and construct the recover matrix, the repairing bandwidth and performance degradation are well-known problems in decoding due to failure \cite{huang2012erasure, sathiamoorthy2013xoring, esmaili2013core}. For example, Facebook uses RS(10,4) with a 256MB chunk size, which generates 2GB of data traffic to reconstruct data for a single data server \cite{sathiamoorthy2013xoring}. However, we observe that it is challenging to encode and decode data online, which has not been observed in prior studies. In the remaining sections, we examine the performance impacts and overheads imposed by online encoding and decoding on an SSD array cluster. 

\noindent \textbf{RS-Concatenation.} In Ceph, while replications no need to read data from the backend storage for encoding and decoding, RS codes require to read multiple data chunks to generate new coding chunks or recover the data in the case of a failure. If a read operation is issued to the underlying storage cluster, the primary OSD first constructs the stripe. Then  the primary OSD serves (returns) the actual data chunk from the stripe. If every data chunk is available and pulled in reasonable time, primary OSD can construct a stripe by concatenating the corresponding data chunks, as shown in Figure \ref{fig:rsconcat}, without any decoding operation. This process is referred to as \emph{RS-concatenation} herein. However, such as an undetected OSD failure or lots of pending requests in OSD can delay data chunks from being served in reasonable time. In these cases, stripe is constructed by decoding the first served $k$ chunks among the $k+m$ chunks.

\section{Evaluation Methodology and Summary}
\label{sec:eval}
In this section, we describe our distributed storage system that employs SSD arrays and its system configuration. For better understanding, we also summarize the in-depth studies and contributions of this paper. 

\begin{figure}
	\centering
	\includegraphics[width=\linewidth]{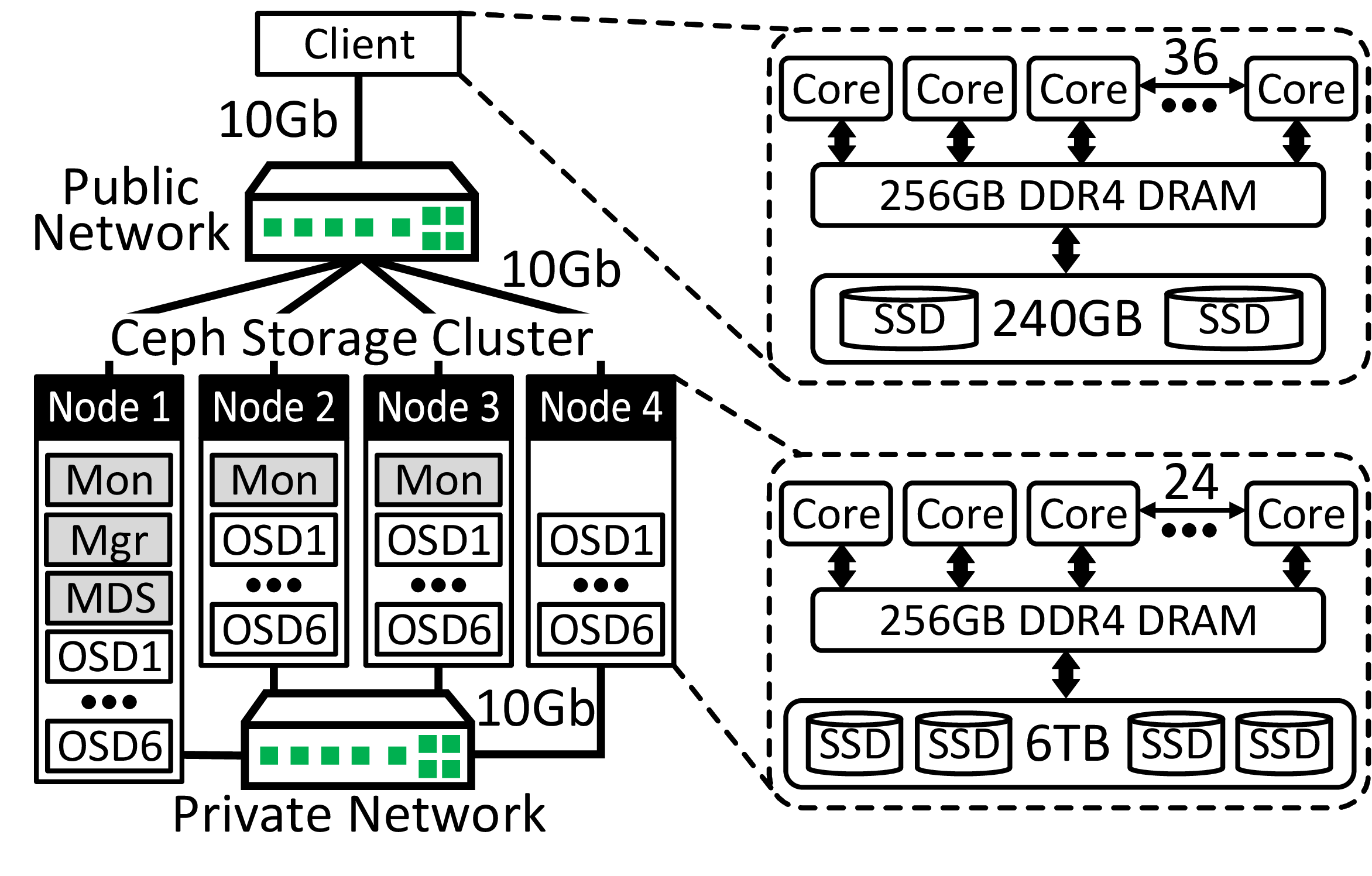}
	\vspace{-20pt}	
	\caption{Configuration of our storage cluster.}
	\label{fig:cluster_set}
	\vspace{-18pt}
\end{figure}

\subsection{Methodology}
\noindent \textbf{Target node architecture.} Figure \ref{fig:cluster_set} illustrates a real system that we built for the evaluation of performance impacts on erasure coding and presents an analysis of the system implications. The client employs 36 2.3 GHz cores (Intel Xeon E5-2669) and 256GB DDR4-2133 DRAM (16 channels, 2 ranks per channel). All the operating system and executable images are booted from and store their local data in the client-side SSD. This client is connected to four different storage nodes of a Ceph storage cluster through a 10Gb public network. Each storage node also employs 24 2.6GHz cores (Intel Xeon E5-2690) and 256GB DDR4-2133 DRAM. For the storage, a high-performance SSD (600GB) is employed for the local OS image, while OSD daemons consist of 12 Intel SSD 730 (6TB) units, each grouping two Intel 730 units through a hardware striping RAID controller \cite{refraid}. In total, the Ceph storage cluster that we built comprises 1TB DRAM, 96 cores, 52 SSDs (26TB). The storage nodes are connected over another 10Gb private network, which is separated from the public network. The important characteristics and hardware configuration of our system are shown in Table \ref{eval_sum}.

\begin{table}[]
\begin{tabular}{|c||c|l|l|l|}
\hline
\multicolumn{5}{|c|}{\textbf{Client Node}}                                                                                                                                              \\ \hline
Processors                            & \multicolumn{4}{c|}{36 cores, each 2.3GHz (Intel Xeon E5-2669)}                                                                        \\ \hline
DRAM                                  & \multicolumn{4}{c|}{256GB DDR4-2133}                                                                                                   \\ \hline
Public Network                        & \multicolumn{4}{c|}{10Gb}                                                                                                              \\ \hline
\multicolumn{5}{|c|}{\textbf{Storage Node (x4)}}                                                                                                                                        \\ \hline
Processors                            & \multicolumn{4}{c|}{24 cores, each 2.6GHz (Intel Xeon E5-2690)}                                                                        \\ \hline
DRAM                                  & \multicolumn{4}{c|}{256GB DDR4-2133}                                                                                                   \\ \hline
Storage                               & \multicolumn{4}{c|}{\begin{tabular}[c]{@{}c@{}}OS image: 600GB SSD\\ OSD daemons: 12 480GB SSDs \\ (RAID w/ each 2 SSDs)\end{tabular}} \\ \hline
\multicolumn{1}{|l||}{Private Network} & \multicolumn{4}{c|}{10Gb}                                                                                                              \\ \hline
\end{tabular}
\caption{Target node hardware specification.}
\label{eval_sum}
\end{table}

\noindent \textbf{Software and workload.}
For the parallel file system, we install Ceph 12.2.2 Luminous, which is the most recently released version (December 2017) \cite{cephluminousrelease}. We use OS kernel version 4.15.2 for the client node and 3.10.0 for the storage nodes, which are the recommended kernel versions for the Ceph file system. We employ Ceph Luminous uses the Jerasure plugin module \cite{plank2014jerasure} and Bluestore, optimized for modern SSDs. To evaluate the impact of the most common usage scenario of storage clusters, we use a flexible I/O tester, FIO \cite{axboe2015flexible}, on our mini cluster. Specifically, we create a data pool and a metadata pool separately, and measure the overall performance of the file system employing these two pools. We set the data pool size as 512 and 256 PGs for 3-replication and RS(6,3), respectively, and set the size of metadata pool as 128 PGs for both configurations. Note that metadata is stored in a replicated pool both for 3-replication and RS(6,3). Our Ceph file system is mounted on the kernel driver. We create a 1TB dummy file, and read/write a specific offset from this file. This allows us to examine the impact of the data pool type while excluding the impact of the metadata. We also disable the client cache during the FIO tests to exclude variables such as locality. We collect the block traces for both the data and metadata pools by using blktrace \cite{brunelle2006block}. We performed a pre-evaluation to determine the number of queue entries that exhibit the best performance of the underlying storage cluster. We observed that the 256 queue depth is the best for diverse workloads, and therefore, we apply this queue depth to all evaluation scenarios. When there is no data written upon the target storage, the read performance can differ significantly from the actual storage performance since the underlying SSD in practice serves the reads with garbage values from the DRAM if there is no target data written (rather than accessing actual Flash media). Thus, we sequentially write the whole 1TB file before performing each set of evaluations.

\subsection{Summary of In-Depth Analyses}
\begin{figure}
	\centering
	\includegraphics[width=\linewidth]{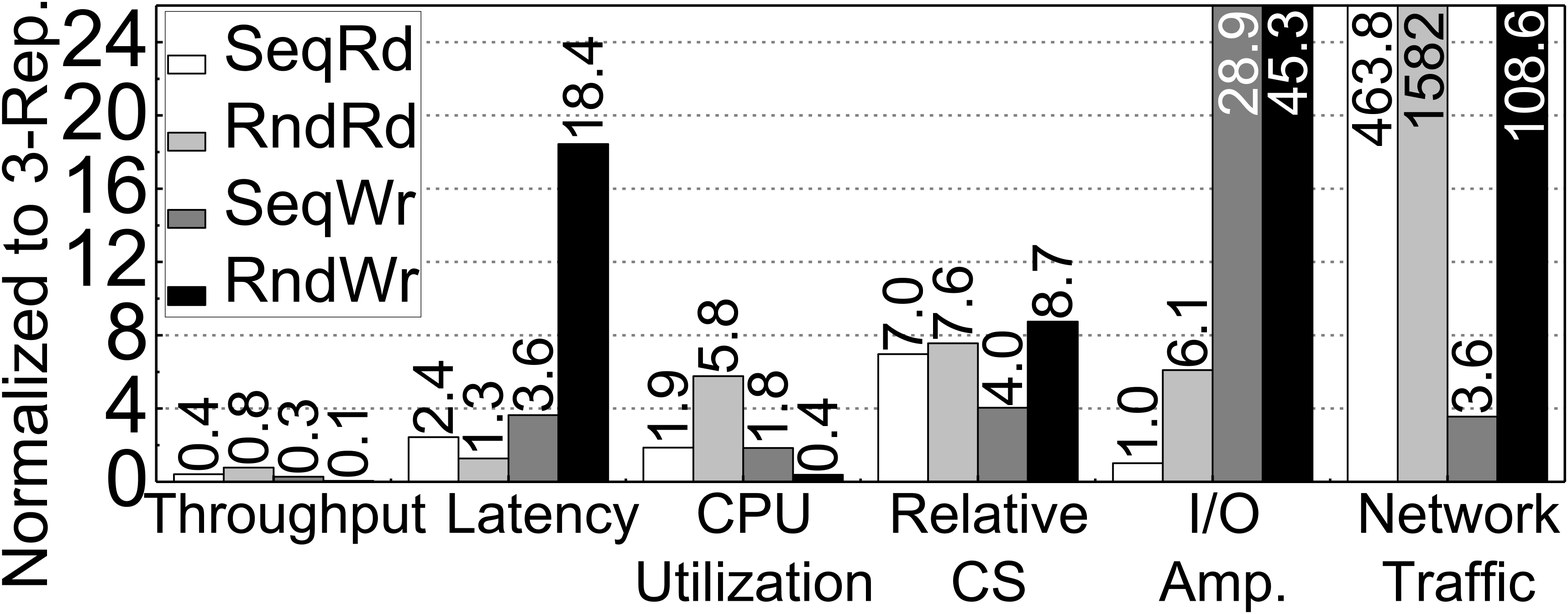}
	\vspace{-10pt}
	\caption{Summary of comparison between replication and erasure coding mechanism synthetic workloads.}
	\label{fig:introduction_syn}
	\vspace{-10pt}
\end{figure}

The observations and findings obtained through our in-depth analyses are summarized as follows:

\noindent \textbf{1. Storage system I/O performance overhead.} This work comprehensively analyzes the impacts of online RS coding on the I/O performance of storage systems, and compares them with those of the default fault-tolerance scheme (i.e., 3-replication). 
As shown in Figure \ref{fig:introduction_syn}, even without any failure, RS(6,3) coding yields a bandwidth 44\% lower and latency 1.8$\times$ longer than those of 3-replication for read operations. 
Furthermore, RS(6,3) coding exhibits bandwidth 83\% lower and latency 8.1$\times$ longer than 3-replication for write operations. We consider this performance degradation imposed by erasure coding a significant challenge that makes it difficult to deploy SSD arrays for HPC and DC systems. (cf. Section \ref{sec:overall})

\noindent \textbf{2. Computing and software overheads.} 
In each node, erasure coding is mostly implemented with software as part of a RAID virtual device driver in the kernel space or with hardware as part of RAID controllers \cite{refraid}. 
However, since most distributed file systems, including a fault tolerant mechanism, are implemented in the user space, I/O services always need to pass through all the modules of the storage stack in both the user and kernel spaces. Consequently, this user-space implementation of erasure coding can increase the computing overheads, including context switching, encoding data through a generator matrix, and data placement. Figure \ref{fig:introduction_syn} compares the overall CPU utilizations of 3-replication with those of RS(6,3) coding. 
We observe that RS(6,3) coding incurs 7.3$\times$ and 5.9$\times$ more context switches per I/O operation than 3-replication for read and write operations, respectively. 
We performed an in-depth study on the computing and software overheads of online erasure coding by decomposing them into user- and kernel-level activities. (cf. Section \ref{sec:computation})

\noindent \textbf{3. I/O amplification overheads.} We observe that online erasure coding significantly increases the amount of I/O data served by storage nodes compared with 3-replication, even without failure. This differs completely from the common expectation for erasure coding (e.g., significantly lower storage overheads). We define these additional reads/writes as \emph{read/write amplification}. As shown in Figure \ref{fig:introduction_syn}, RS(6,3) coding shows 2.5$\times$ more read amplification and hundreds of times more write amplification than 3-replication during read and write operations, respectively. This is because RS coding manages all I/O requests at the stripe level, which is inefficient from a storage viewpoint. However, it incurs severe problems especially for low-level flash in SSD arrays.
Since flash does not allow an overwrite to a physical block (without erasing), SSD firmware in practice forwards the incoming write to another reserved block, which was erased in advance, and invalids the overwritten data by remapping the corresponding address with new one \cite{zhang2015opennvm, shahidi2016exploring}. Therefore, the increased amount of data that needs to be written to the underlying flash shortens the life time of SSD regardless of the storage overhead of the erasure coding. (cf. Section \ref{sec:dataoverhead})

\noindent \textbf{4. Private network overheads.} 
In contrast to erasure coding implemented for a single node, erasure coding for distributed storage systems pulls or pushes a large amount of data over its private network which connects storage nodes and is invisible to client nodes. 
Figure \ref{fig:introduction_syn} summarizes the network traffic volume that erasure coding generates, normalized to the volume that 3-replication generates. 
Due to the characteristics of erasure coding, a significant amount of read traffic is generated for write operations in contrast to 3-replication. Importantly, erasure coding congests the private network because of the stripes and data regions that erasure coding must manage for coding chunks, whereas 3-replication hardly congests the network. 
Even for reads, the private network of a system deploying erasure coding is significantly busier than that of a system using 3-replication because erasure coding needs to pull chunks to construct a stripe. 
Such high private network traffic is necessary for recovery from errors in one or more storage nodes, but we believe that the higher network overheads of erasure coding compared with replication should be optimized for future distributed storage systems while especially considering that the SSDs have far shorter latency than HDDs. (cf. Section \ref{private_network})

\begin{figure}
	\centering
	\includegraphics[width=\linewidth]{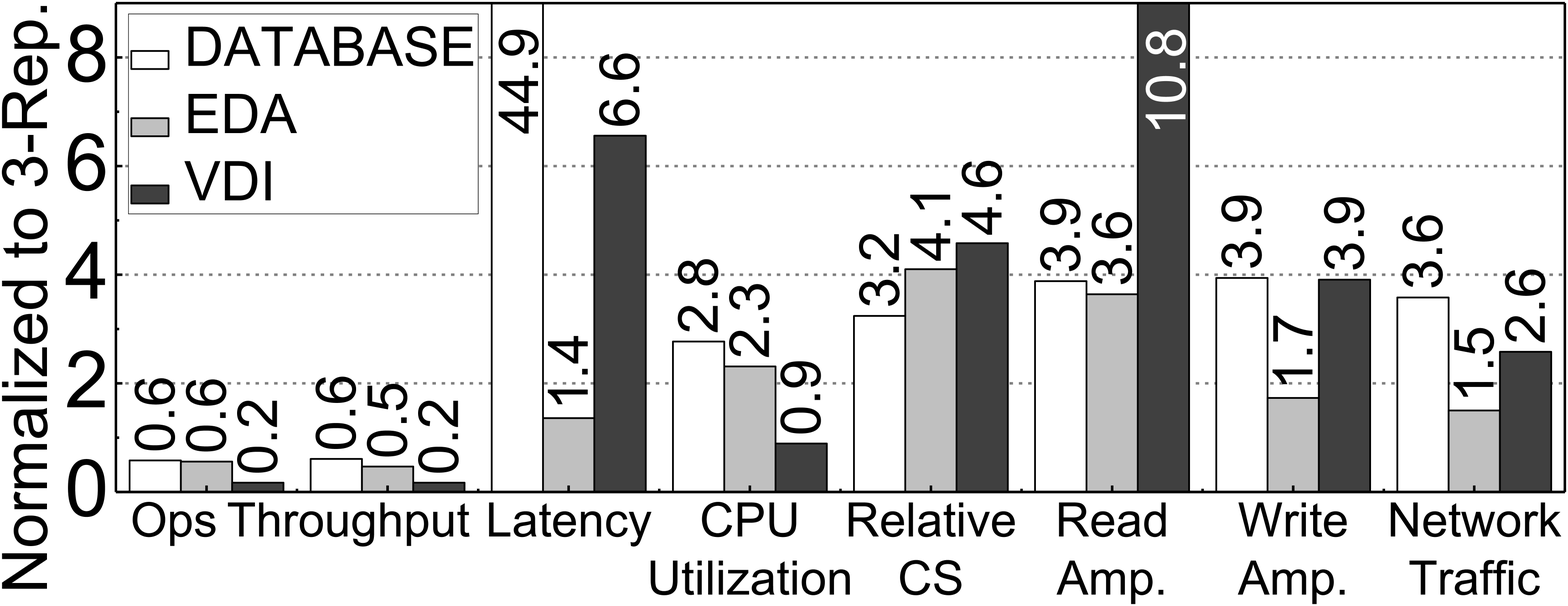}
	\vspace{-10pt}
	\caption{Summary of comparison between replication and erasure coding mechanism under real workloads.}
	\label{fig:introduction}
	\vspace{-10pt}
\end{figure}

\noindent \textbf{5. Physical data layout on RS-coded SSD array.} 
Since there are many SSD devices in a distributed storage system and all of them are connected by a network, the performance of erasure coding significantly varies depending on the layout of the data across the SSD devices. In practice, the performance of an SSD degrades if reads and writes are interleaved and/or there are many random accesses \cite{chen2009understanding}. 
In contrast to this device-level characteristic, we observe that the throughput of data updates in erasure coding (each of which consists of reads and writes due to the parity chunks) is 3.6$\times$ better than that of a new write (encoding). In this work, we will examine the system overheads and performance characteristics of RS codes by considering the physical data layout that the parallel file system manages. (cf. Section \ref{sec:distribution})

\noindent \textbf{6. Real application usages.}
We examine all the aforementioned characteristics and observations by re-evaluating our system with different real application scenarios. The results are summarized in Figure \ref{fig:introduction}. This figure indicates severe performance degradation, and significant overheads are also observed in the real application usages. 
Specifically, the throughput and latency with online erasure coding degrades by 61\% and 7.5$\times$, respectively, on average, compared with the system that employs 3-replication. Furthermore, we observe that erasure coding consumes 1.8$\times$ more CPU cycles than 3-replication since it introduces context switches more than 3-replication by 3.9$\times$ on average. Note that significant data overheads are also observed; overall, 2.4$\times$ more data are transferred via the private network compared with 3-replication, which in turn makes the network overly crowded, and 4$\times$ more data are served from the underlying SSDs in erasure coding. Unfortunately, this heavy I/O amplification can significantly reduce the lifetime and worsen the reliability of SSD arrays. (cf. Section \ref{sec:realworkload})

\section{Performance Comparison}
\label{sec:overall}
In this section, we compare the performance of the major erasure coding configuration, RS(6,3), with that of 3-replication at a high level.

\begin{figure}
	\centering
	\begin{subfigure}{0.49\linewidth}
		\includegraphics[width=\linewidth]{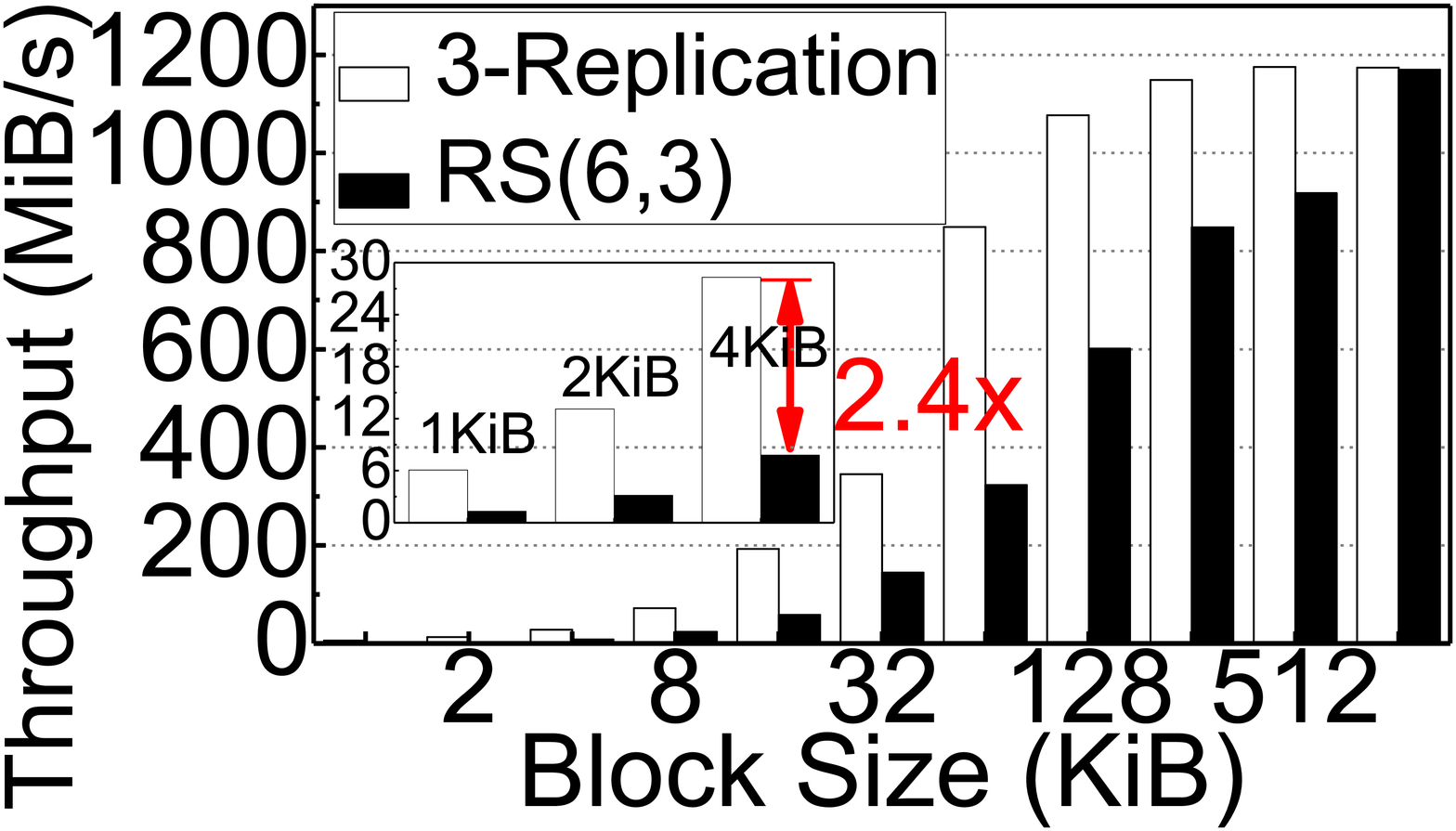}
		\caption{Throughput.}
		\label{fig:SW_thr}
	\end{subfigure}
	\begin{subfigure}{0.49\linewidth}
		\includegraphics[width=\linewidth]{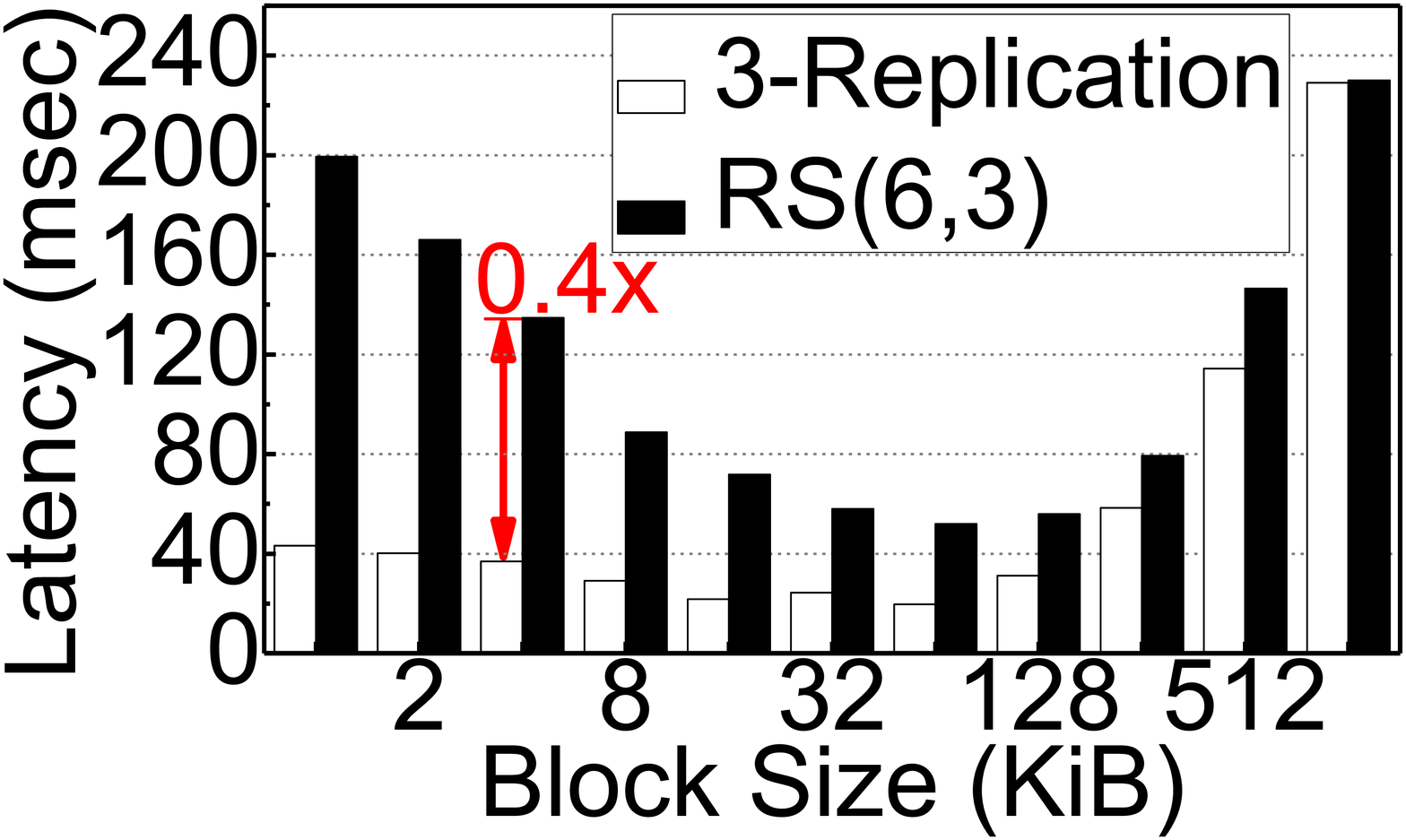}
		\caption{Latency.}
		\label{fig:SW_lat}
	\end{subfigure}
	\vspace{-5pt}
	\caption{Sequential write performance.}
	\vspace{-10pt}
	\label{fig:SW_perf}
\end{figure}

\begin{figure}
	\centering
	\begin{subfigure}{0.49\linewidth}
		\includegraphics[width=\linewidth]{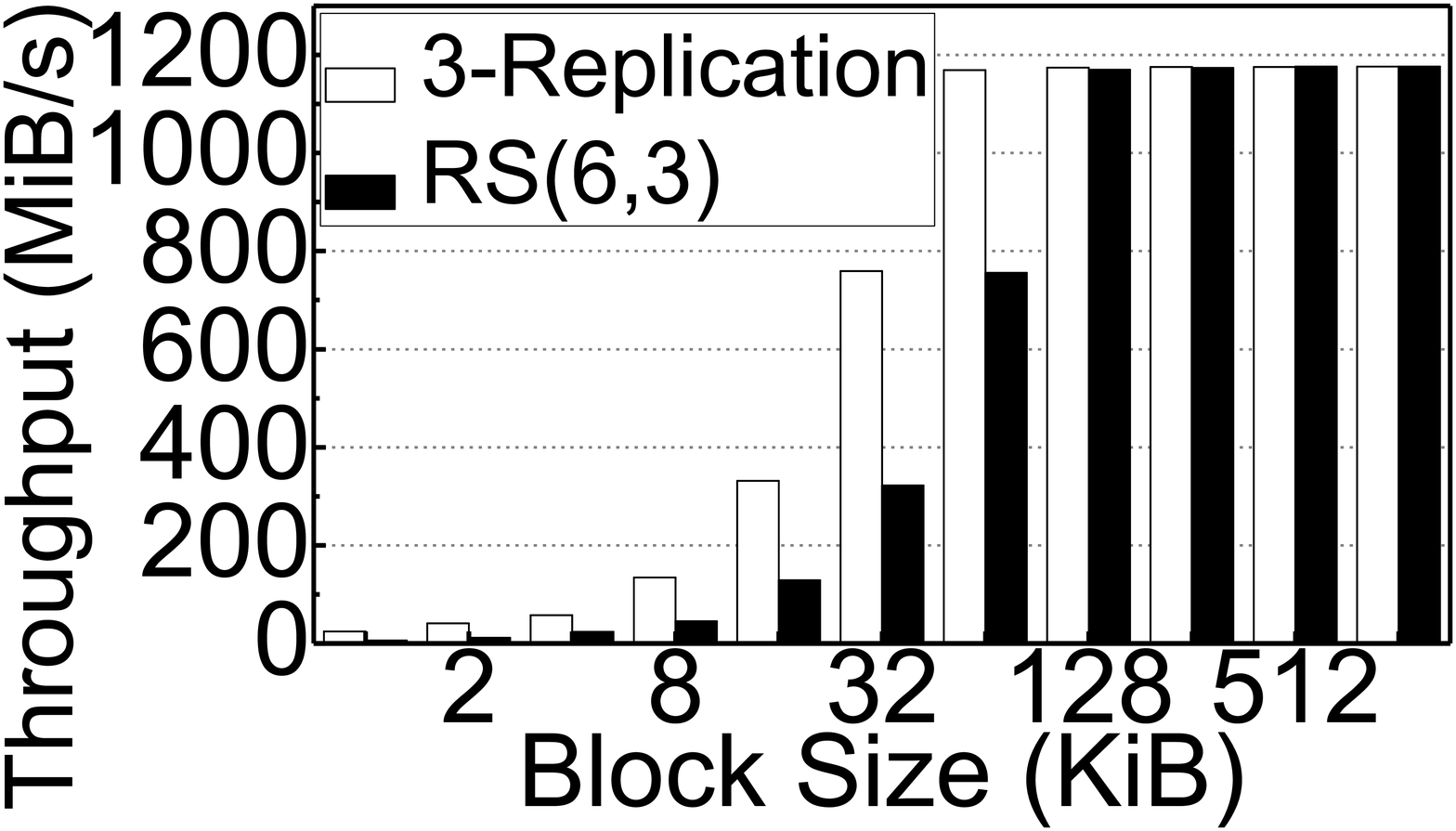}
		\caption{Throughput.}
		\label{fig:SR_thr}
	\end{subfigure}
	\begin{subfigure}{0.49\linewidth}
		\includegraphics[width=\linewidth]{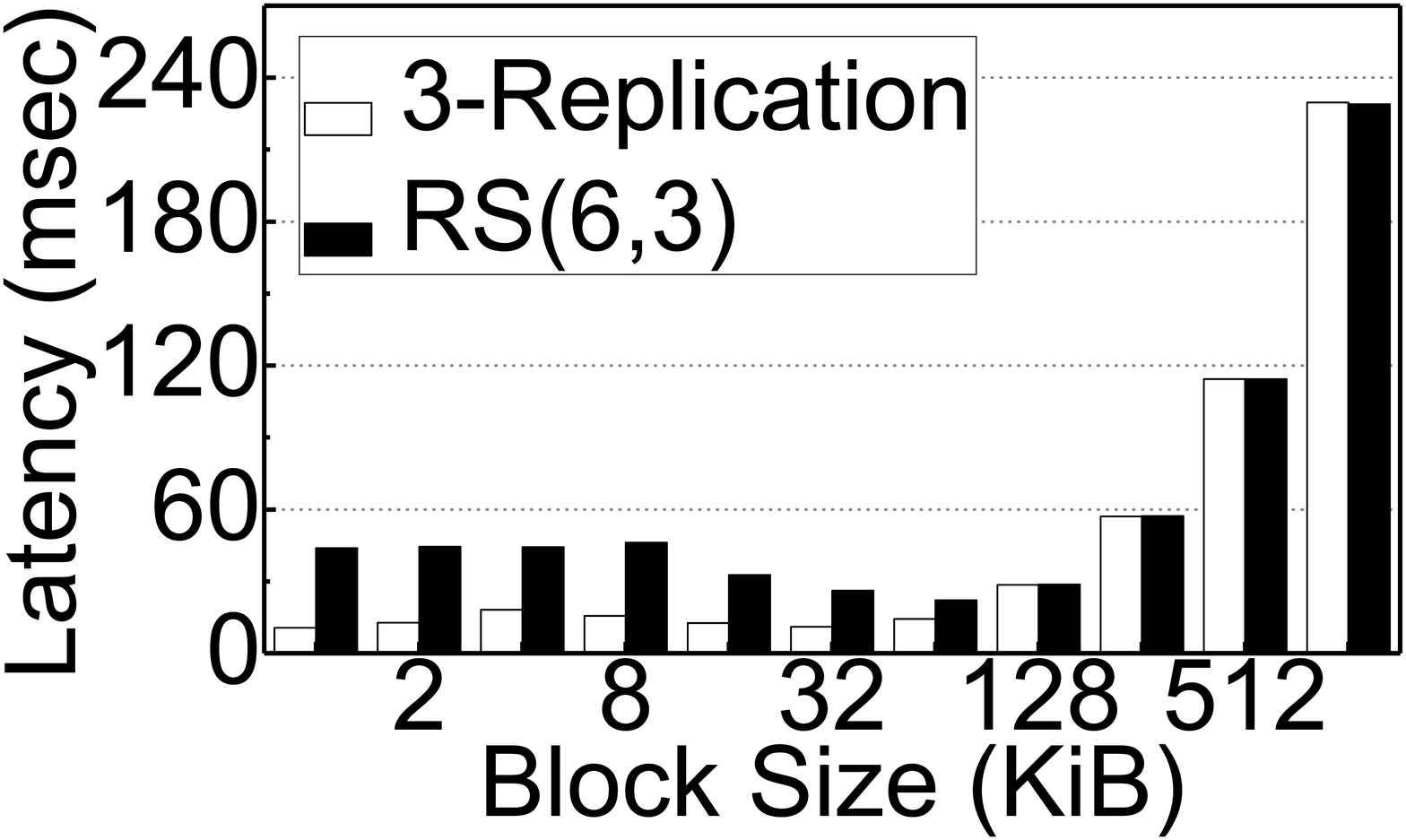}
		\caption{Latency.}
		\label{fig:SR_lat}
	\end{subfigure}
	\vspace{-5pt}
	\caption{Sequential read performance.}
	\vspace{-10pt}
	\label{fig:SR_perf}
\end{figure}

\begin{figure}
	\centering
	\begin{subfigure}{0.48\linewidth}
		\includegraphics[width=\linewidth]{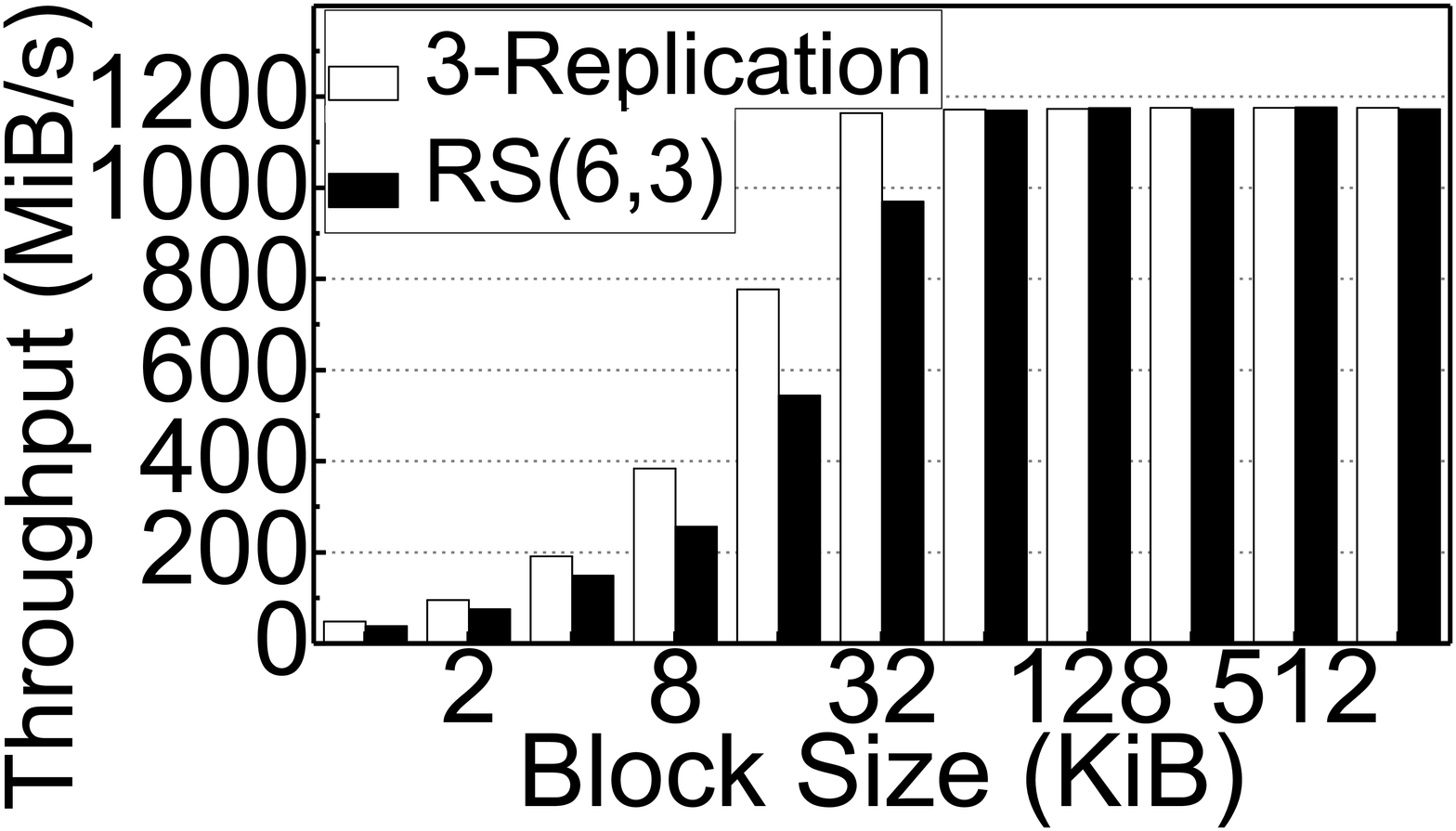}
		\caption{Throughput.}
		\label{fig:RR_thr}
	\end{subfigure}
	~
	\begin{subfigure}{0.48\linewidth}
		\includegraphics[width=\linewidth]{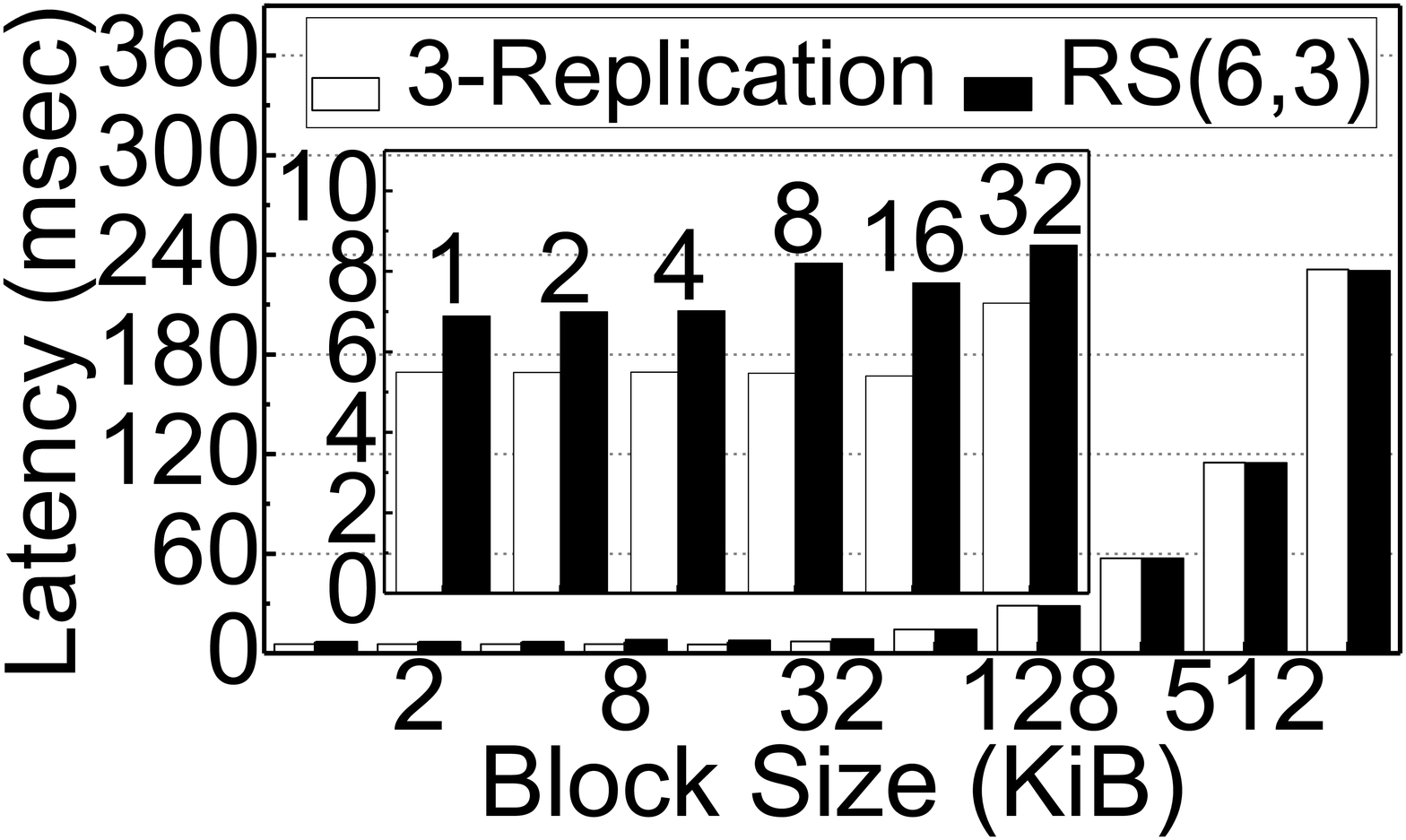}
		\caption{Latency.}
		\label{fig:RR_lat}
	\end{subfigure}
	\vspace{-5pt}
	\caption{Random read performance.}
	\vspace{-10pt}
	\label{fig:RR_perf}
\end{figure}
\begin{figure}
	\centering
	\begin{subfigure}{0.49\linewidth}
		\includegraphics[width=\linewidth]{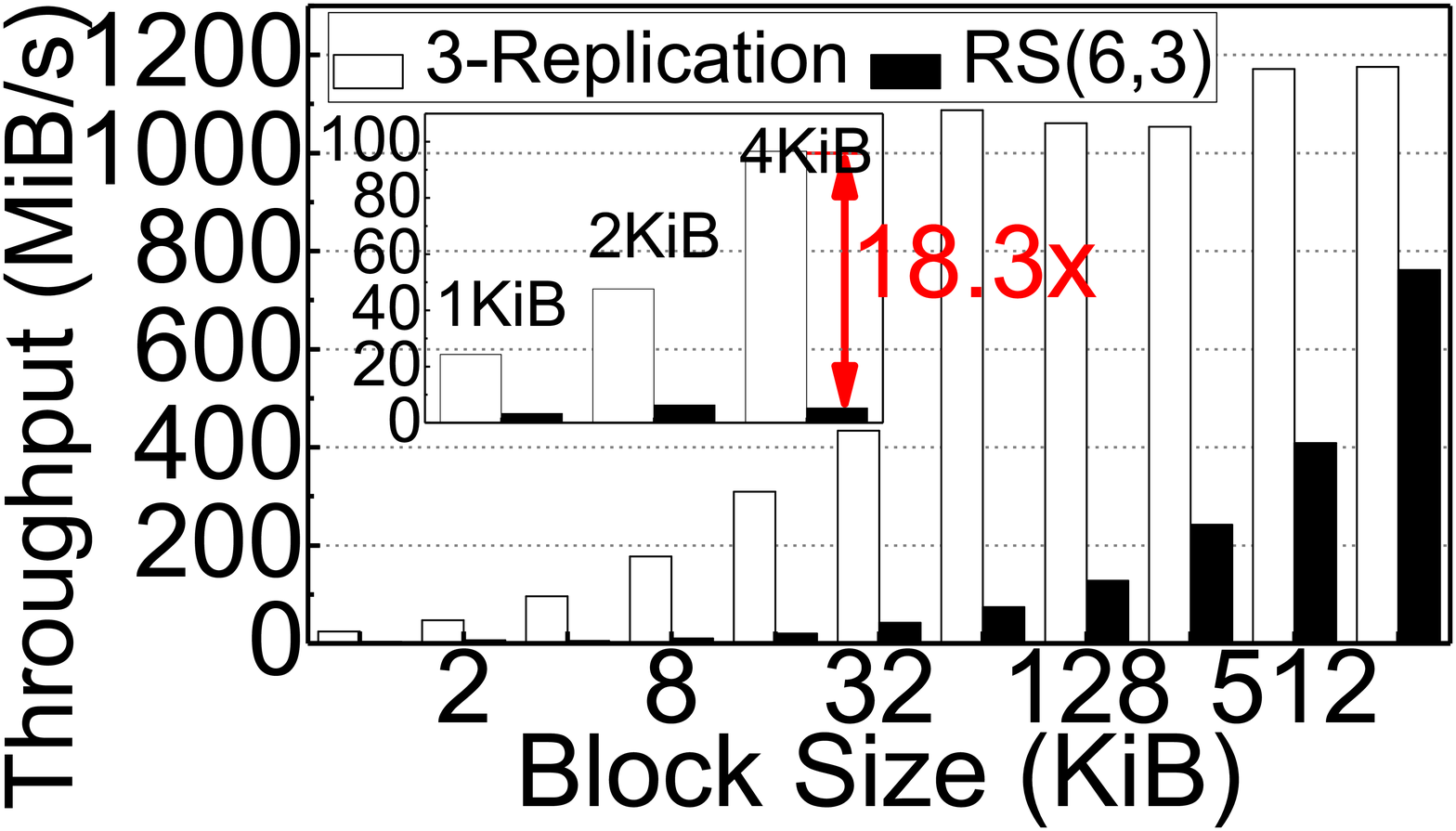}
		\caption{Throughput.}
		\label{fig:RW_thr}
	\end{subfigure}
	\begin{subfigure}{0.49\linewidth}
		\includegraphics[width=\linewidth]{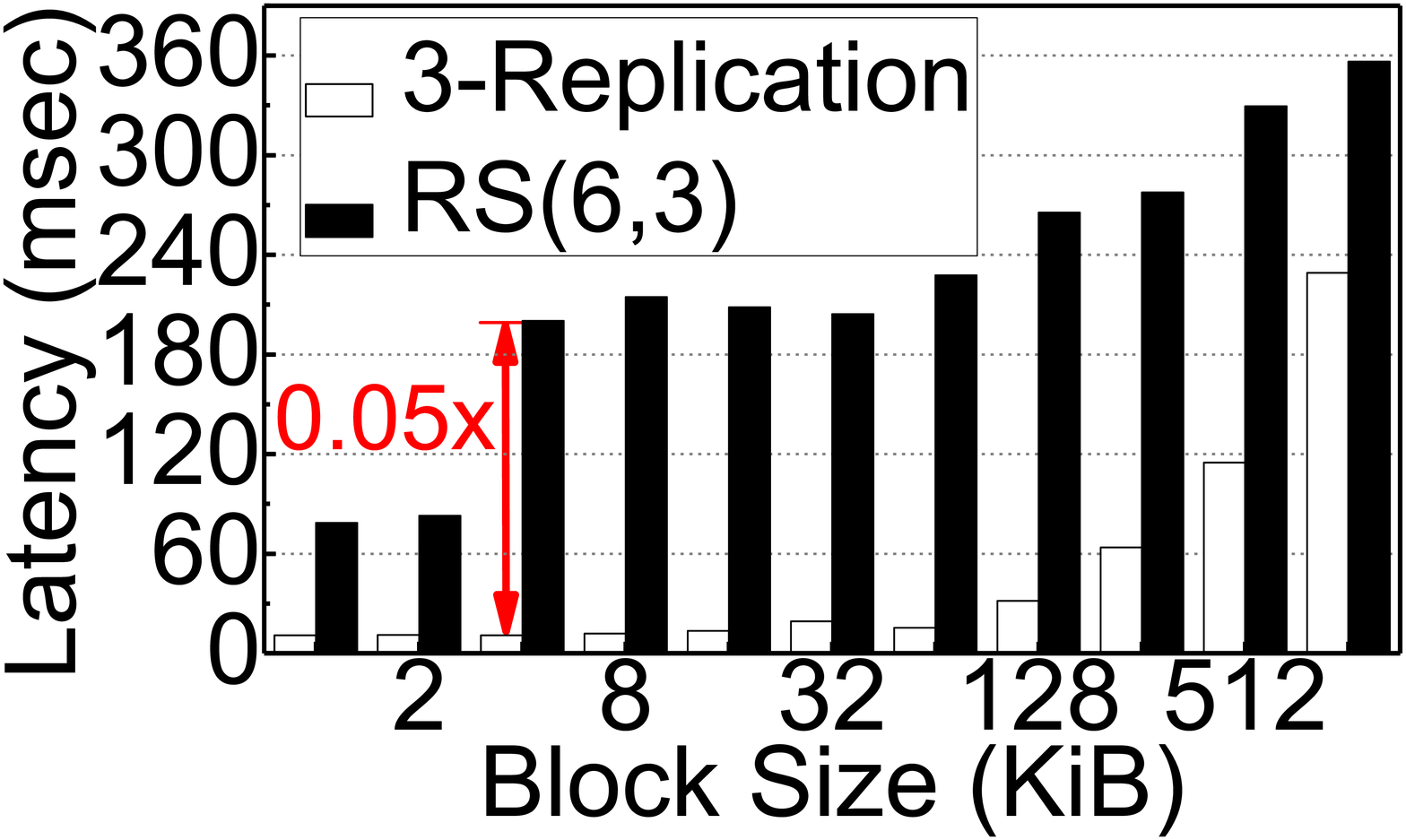}
		\caption{Latency.}
		\label{fig:RW_lat}
	\end{subfigure}
	\vspace{-5pt}
	\caption{Random write performance.}
	\vspace{-10pt}
	\label{fig:RW_perf}
\end{figure}

\subsection{Sequential Performance}
\noindent \textbf{Writes.} Figures \ref{fig:SW_thr} and \ref{fig:SW_lat} show the throughput and latency of sequential writes with various block sizes ranging from 1KB to 1MB, respectively. As shown in Figure \ref{fig:SW_thr}, 3-replication offers around 553 MB/s, on average, whereas RS(6,3) provides 373 MB/s for the sequential writes. Overall, the throughput of RS(6,3) is 70\% worse than that of 3-replication for 4KB$\sim$16KB request sizes. Considering that the most popular request size of diverse applications and native file systems is 4KB$\sim$16KB, the performance degradation of online RS coding may not be acceptable in many computing domains. As shown in Figure \ref{fig:SW_lat}, the latency of RS(6,3) is 2$\times$ longer than that of 3-replication on average. The latency of a conventional Ceph configuration (with 3-replication) is less than 45 ms for most block sizes of I/O accesses. We believe that the latency of 3-replication is in a reasonable range. However, RS(6,3) requires 103 ms, on average, which is not in a reasonable range.
Such long latencies can be a serious issue for many latency-sensitive applications. The reason behind the long latency and low bandwidth of RS(6,3) is that online erasure coding requires computation for encoding, data management, and additional network traffic. 
Therefore, even though 3-replication stores 2$\times$ more data than RS(6,3), 3-replication shows better performance in our sequential write tests. 
We will analyze the contribution of the performance degradation in more detail in Section \ref{sec:syschar}. Note that the latency of sequential writes decreases as the block size increases until 64KB. This is because a small request actually needs to read out more chunks from the stripe because fewer chunks are modified. Therefore, the consecutive requests are more delayed by the PG backend with a small request. We examine this phenomenon in detail in Section \ref{sec:distribution}.

\noindent \textbf{Reads.} Figure \ref{fig:SR_thr} compares the throughput of RS(6,3) with that of 3-replication by performing sequential read operations. In contrast to the writes (cf. Figure \ref{fig:SW_thr}), RS(6,3) degrades the throughput by only 51\% on average, compared with 3-replication. This is because in the most cases, the decoding process (i.e., repair bandwidth) is not involved during read operations. Nevertheless, RS(6,3) gives 2.7$\times$ longer latency than 3-replication. Even though the decoding and repair overheads occur only when the data chunks are slowly pulled, the reads of online erasure coding always require composing the data chunks into a stripe, which in turn introduces the overheads not observed by any replication method. This process associated with reads is herein referred to as \emph{RS-concatenation}, and it generates extra data transfers over the private network, further increasing the latency of RS(6,3). As shown in Figure \ref{fig:SR_lat}, the latency of traditional configuration (with 3-replication) is less than 22 ms, whereas RS(6,3) requires a latency as high as 45 ms. We examine the performance and system impacts of RS-concatenation in Section \ref{sec:computation}.

\begin{figure}
	\centering
	\begin{subfigure}{0.49\linewidth}
		\includegraphics[width=\linewidth]{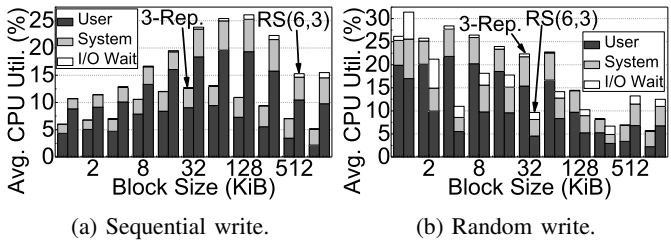}
		\caption{Sequential write.}
		\label{fig:SW_cpu}
	\end{subfigure}
	\begin{subfigure}{0.49\linewidth}
		\includegraphics[width=\linewidth]{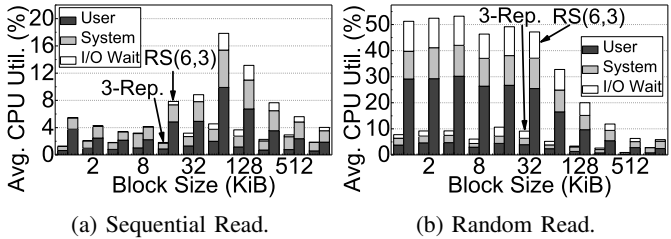}
		\caption{Random write.}
		\label{fig:RW_cpu}
	\end{subfigure}
	\caption{CPU utilization.}
	\label{fig:write_cpu}
	\vspace{-10pt}
\end{figure}
\begin{figure}
	\centering
	\begin{subfigure}{0.49\linewidth}
		\includegraphics[width=\linewidth]{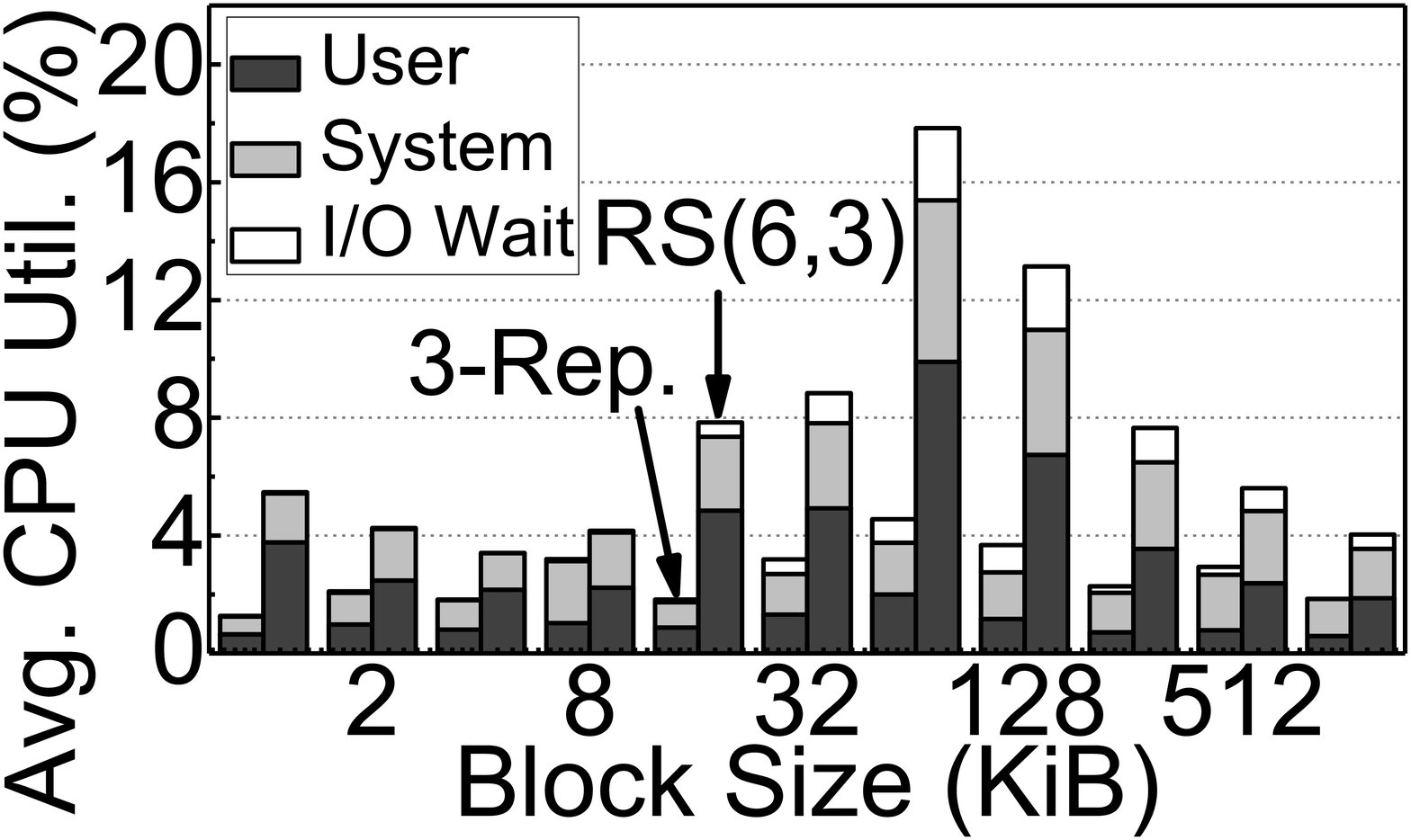}
		\caption{Sequential Read.}
		\label{fig:SR_cpu}
	\end{subfigure}
	\begin{subfigure}{0.49\linewidth}
		\includegraphics[width=\linewidth]{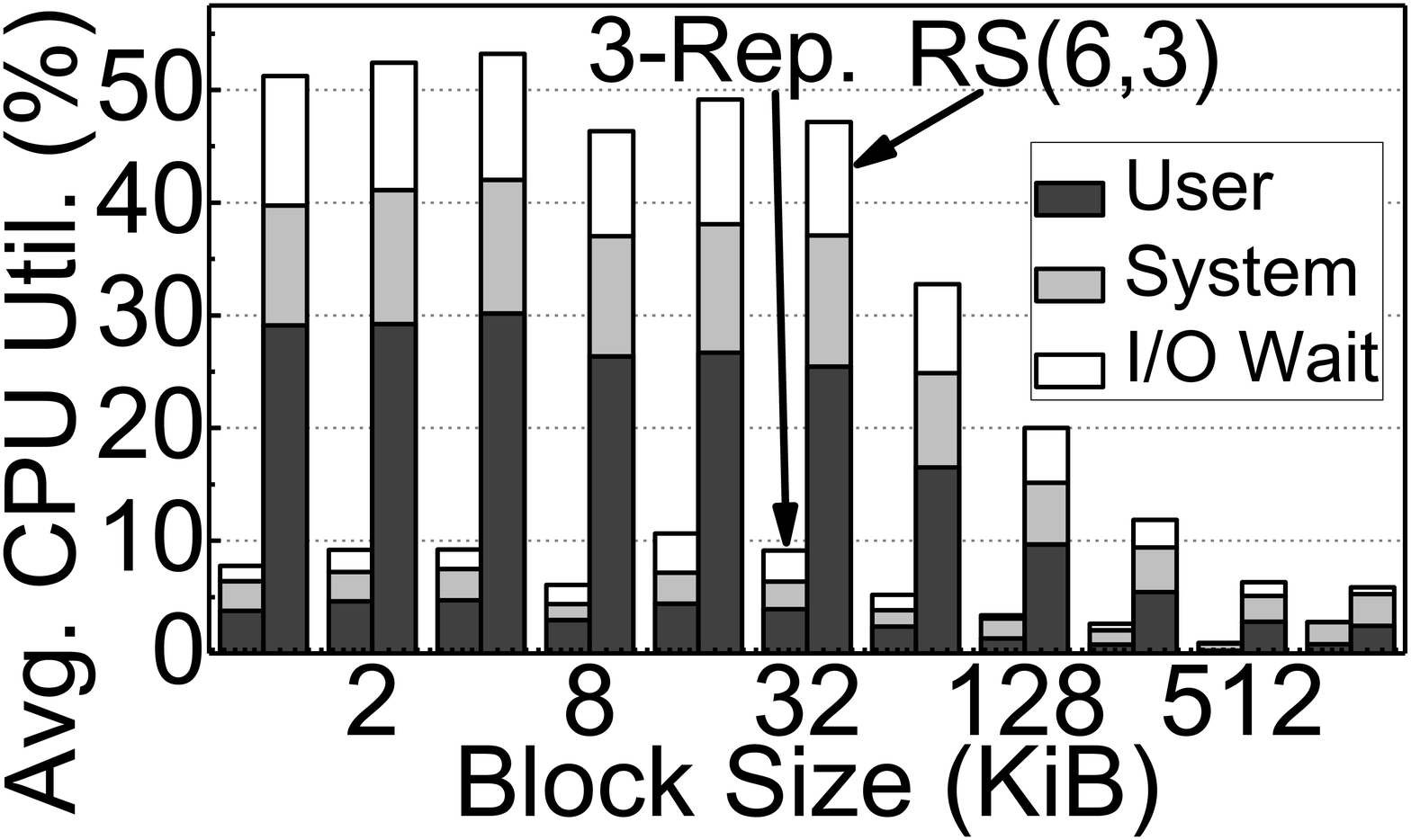}
		\caption{Random Read.}
		\label{fig:RR_cpu}
	\end{subfigure}
	\vspace{-5pt}
	\caption{CPU utilization.}
	\vspace{-10pt}
	\label{fig:read_cpu}
\end{figure}

\subsection{Random Performance}

\noindent \textbf{Reads.} Figure \ref{fig:RR_perf} shows the read performance of 3-replication and RS(6,3) with random I/O accesses. In contrast to sequential reads, the difference between 3-replication and RS(6,3) with regard to the bandwidth and latency is less than 16\% and 22\%, respectively, on average. RS-concatenation, which is a process for composing data chunks into a stripe, not only wastes computation but also requires pulling the data chunks from the other underlying OSDs. Since the primary OSD does not send the acknowledgement signal to the client until all data chunks have arrived from the other RS-associated OSDs, the target read request can be delayed as long each request as delayed. While the impact of this overhead on the performance is in a reasonable range, the sequential reads are further delayed. A block size smaller than an object mostly targets the same PG, and especially primary OSD. Hence many resource conflicts are caused at the PG level. As explained in Section \ref{sec:background}, since the dispatcher of the target (underneath its client messenger) locks the target PG to offer strong consistency of the storage cluster, the underlying PG backend is not available, which decreasing the throughput and increasing the latency. In contrast, since I/O requests under random accesses can be distributed across different OSDs (in our case 24 OSDs), such lock contentions can be addressed to some extent, which in turn increases the performance. This phenomenon is also observed in 3-replication, but the number of OSDs (and nodes) that the PG backend needs to handle through its cluster messenger is less than that for online erasure coding.  Because of this higher degree of PG-level parallelism, the performance of random reads is better than that of sequential reads.

\begin{figure}
	\centering
	\begin{subfigure}{0.49\linewidth}
		\includegraphics[width=\linewidth]{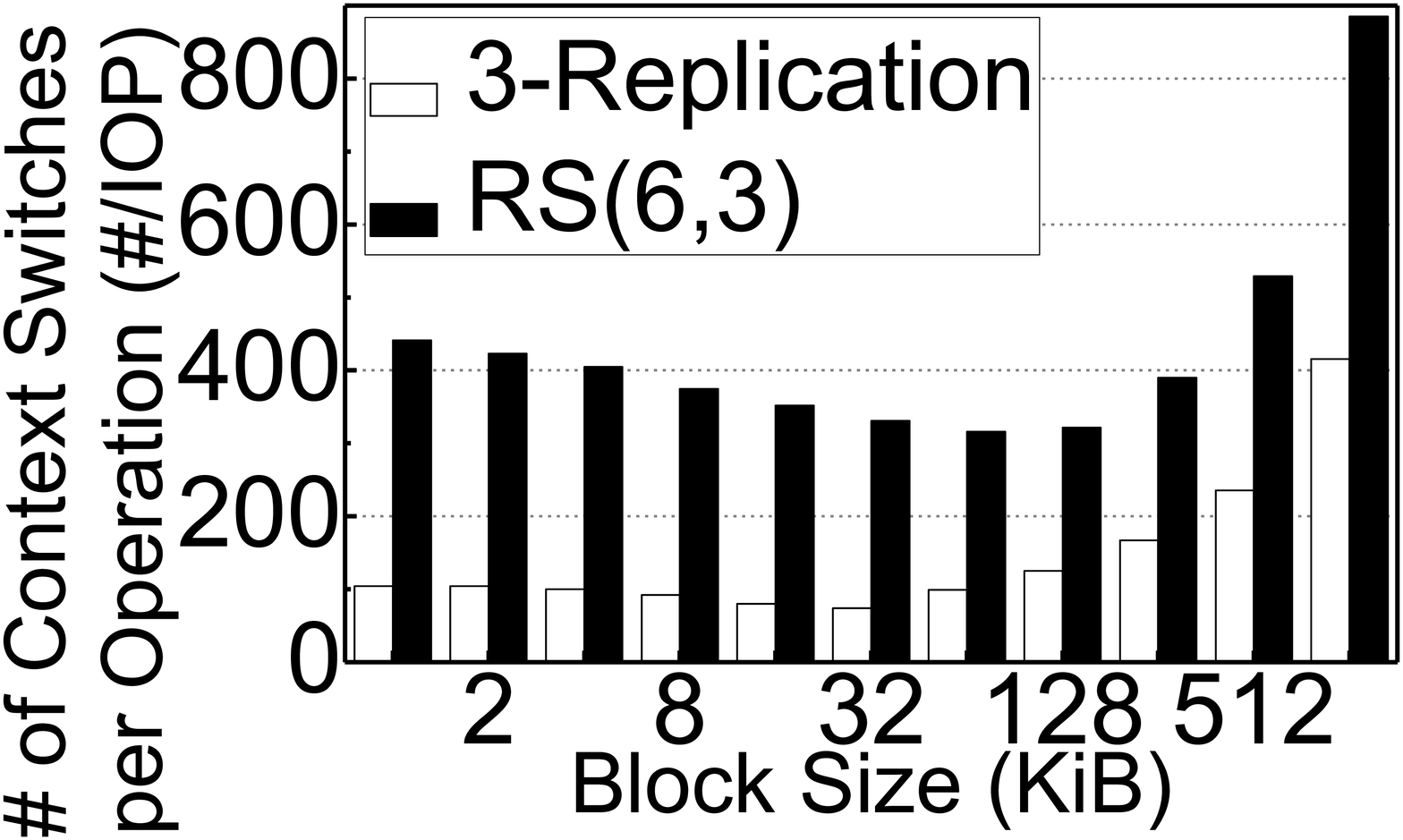}
		\caption{Sequential write.}
		\label{fig:SW_ctx}
	\end{subfigure}
	\begin{subfigure}{0.49\linewidth}
		\includegraphics[width=\linewidth]{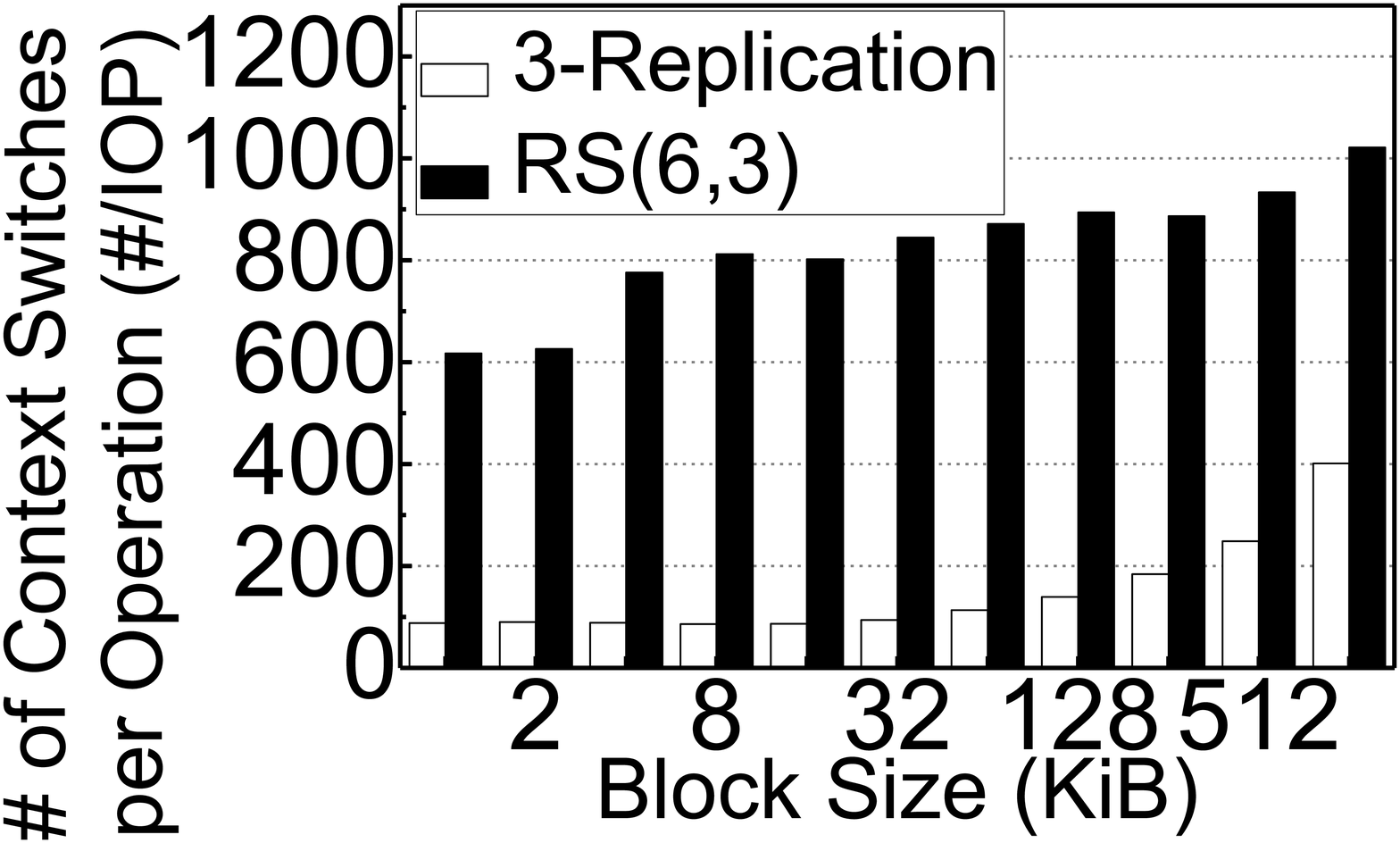}
		\caption{Random write.}
		\label{fig:RW_ctx}
	\end{subfigure}
	\vspace{-5pt}
	\caption{Relative number of context switches under writes.}
	\vspace{-10pt}
	\label{fig:write_ctx}
\end{figure}
\begin{figure}
	\centering
	\begin{subfigure}{0.49\linewidth}
		\includegraphics[width=\linewidth]{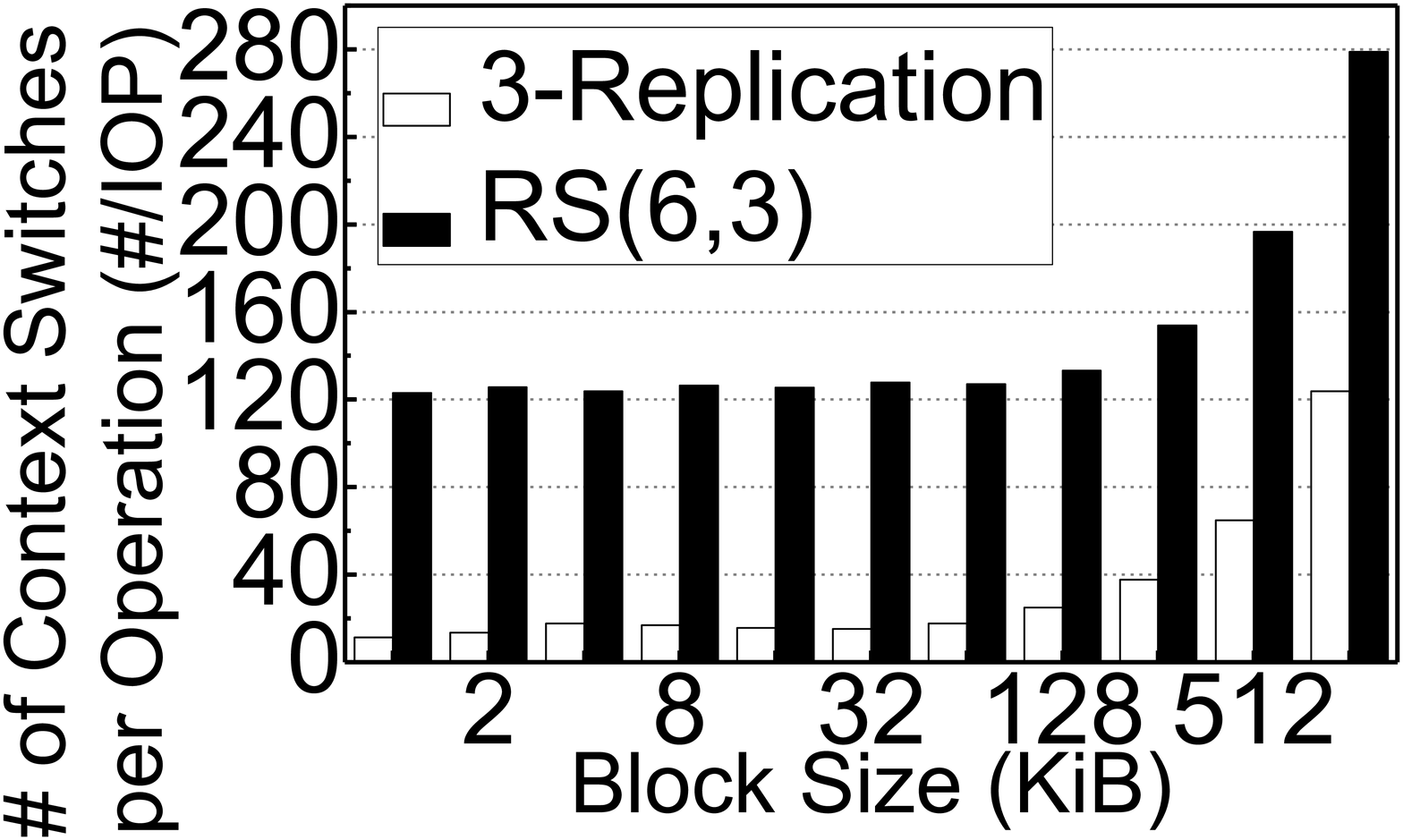}
		\caption{Sequential Read.}
		\label{fig:SR_ctx}
	\end{subfigure}
	\begin{subfigure}{0.49\linewidth}
		\includegraphics[width=\linewidth]{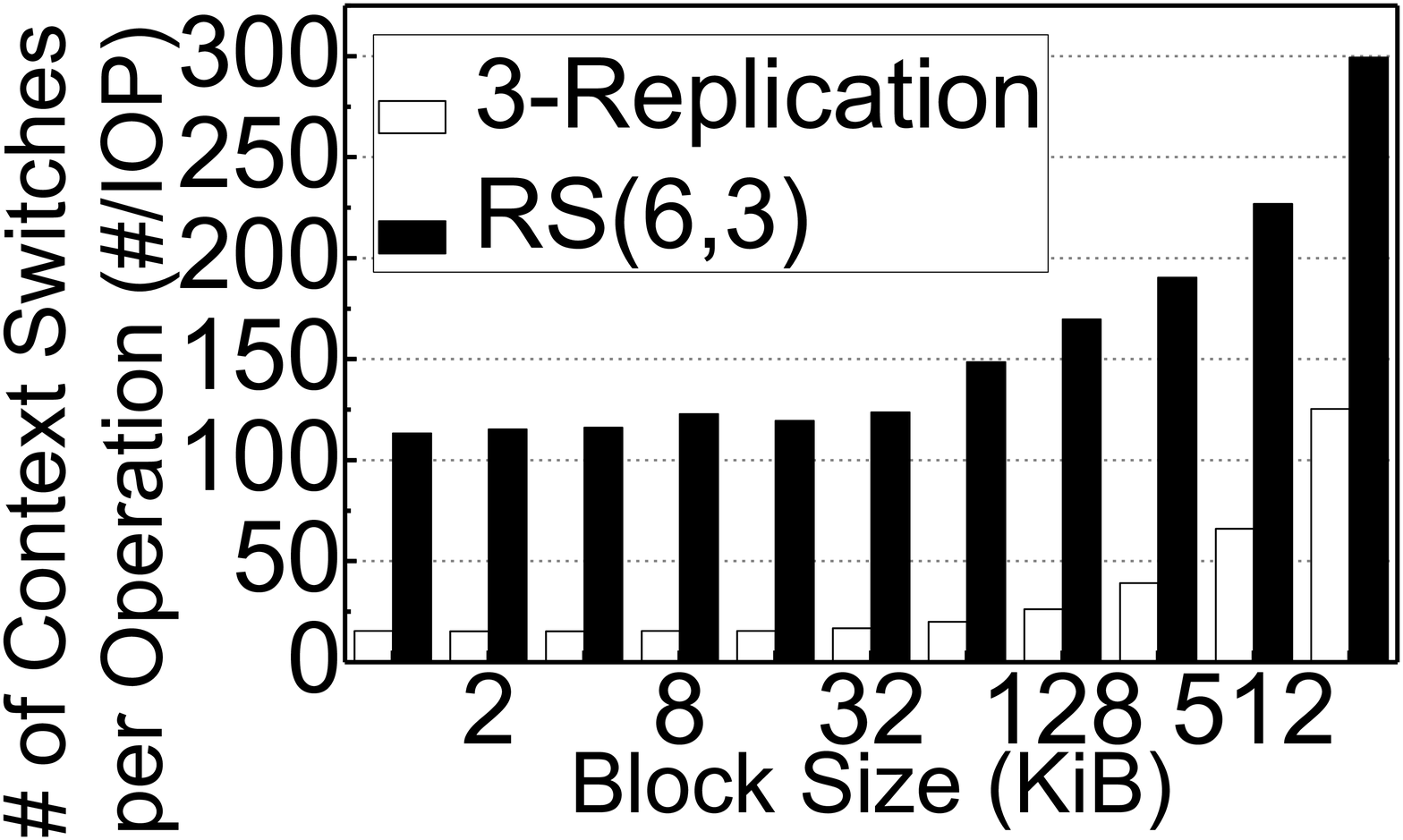}
		\caption{Random Read.}
		\label{fig:RR_ctx}
	\end{subfigure}
	\vspace{-5pt}
	\caption{Relative number of context switches under reads.}
	\vspace{-10pt}
	\label{fig:read_ctx}
\end{figure}

\noindent \textbf{Writes.} As shown in Figure \ref{fig:RW_perf}, RS(6,3) gives 5.6$\times$ worse write throughput than 3-replication for random I/O accesses. The trend of the performance difference is similar to that of sequential writes, but the random write bandwidth of RS(6,3) offers a 2.4$\times$ lower write bandwidth for random writes than for sequential writes, even though the sequential writes are delayed due to the PG lock contention. This is because of object initialization, which must be completed at the first write to the object in erasure coding. Object initialization introduces more significant performance degradation and overheads compared to object update. Random writes are more likely to access an object that has never been initialized, whereas sequential writes are more likely to update the initialized object because the requests after initialization target the same object. We closely examine these performance issues related to writes in Section \ref{sec:distribution}. 

\section{System Characterizations of Online RS}
\label{sec:syschar}

\subsection{Computing and System Analysis}
\label{sec:computation}

In this section, we analyze the computing overheads (in terms of CPU utilizations and context switches) imposed by 3-replication and RS(6,3). We performed a pre-evaluation to measure the computing overheads involved with serving I/O subsystems even when the system is idle and excluded such overheads from our results. The computing overheads reported in this section are average values for the 96 cores employed by our mini-storage cluster. Note that the stripe widths of RS(6,3) are 24KB and 40KB, respectively.

\noindent \textbf{Writes.} Figure \ref{fig:write_cpu} shows the CPU utilization of 3-replication and RS(6,3) for writes. In these CPU utilization tests, we measure the utilization of the kernel side (``system'') and user side (``user'') separately. For sequential writes, erasure coding consumes around 2$\times$ more total CPU execution cycles than replication, on average. 
A notable aspect of this measurement is that the user-mode CPU utilizations account for 63\%$\sim$82\% of the total CPU cycles (96 cores of the storage cluster), which is not observed in conventional multiple driver layers in Linux. Since all OSD daemons, including the PG backend and fault-tolerance modules (such as erasure coding and 3-replication), are implemented at the user level, user-mode operations require more CPU cycles than kernel-mode operations to secure storage resilience. On the other hand, as shown in Figure \ref{fig:RW_cpu}, RS(6,3) requires 15\% of the total CPU cycles, which is approximately 20\% less than the amount of CPU usages that 3-replication needs (19\% of the total CPU cycles).

\begin{figure}
	\centering
	\begin{subfigure}{0.49\linewidth}
		\includegraphics[width=\linewidth]{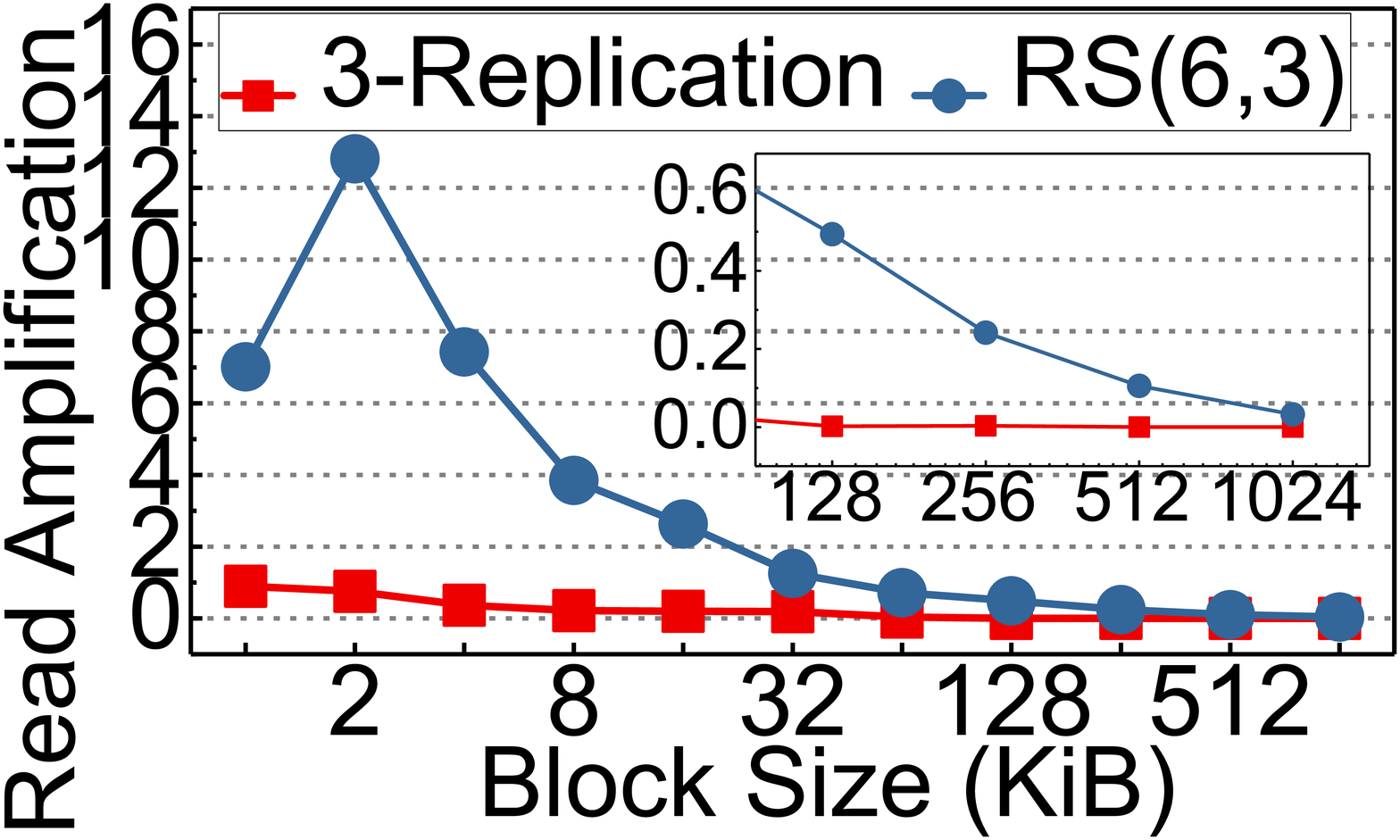}
		\caption{Read amplification.}
		\label{fig:SW_read}
	\end{subfigure}
	\begin{subfigure}{0.49\linewidth}
		\includegraphics[width=\linewidth]{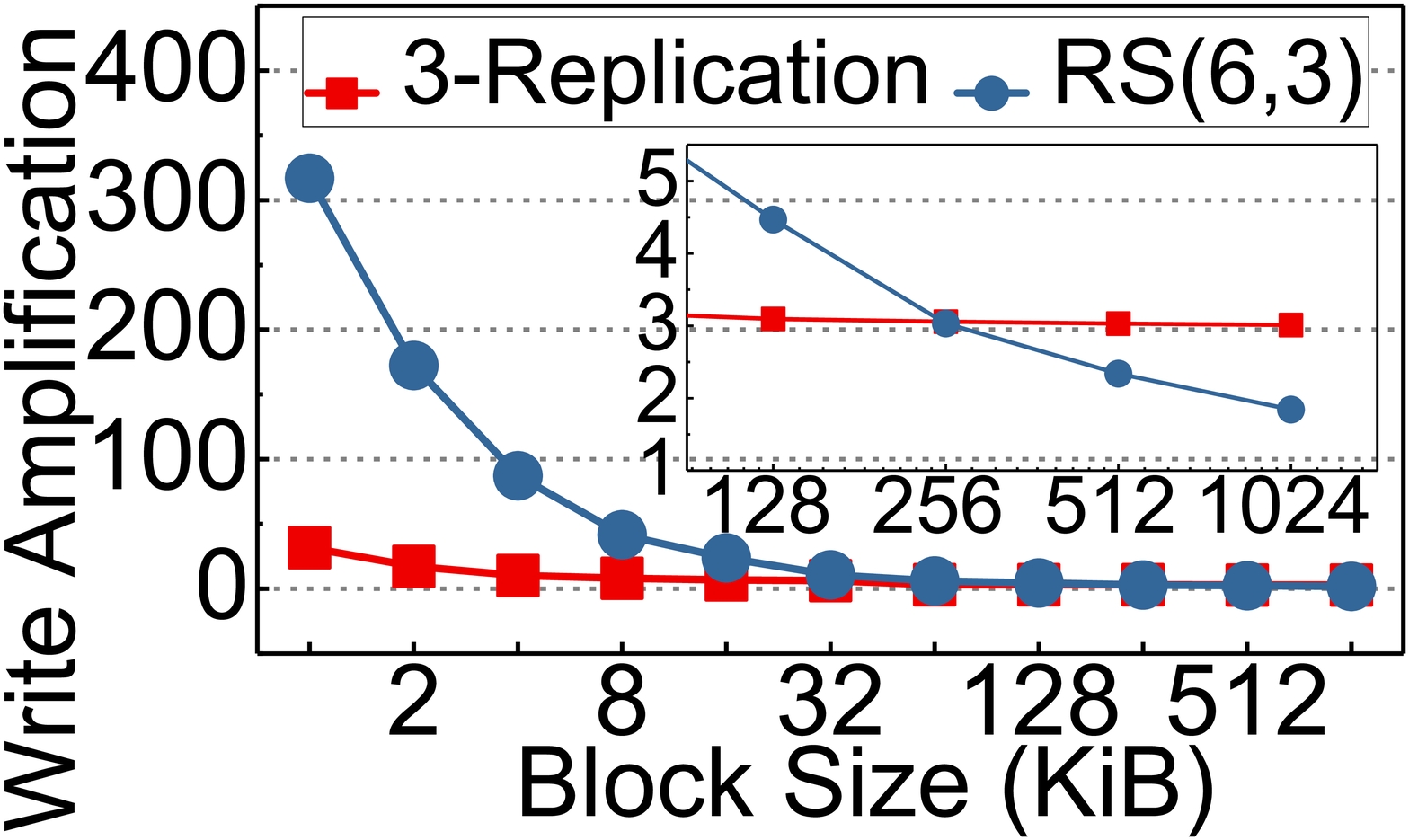}
		\caption{Write amplification.}
		\label{fig:SW_write}
	\end{subfigure}
	\vspace{-7pt}
	\caption{I/O amplifications under sequential writes.}
	\vspace{-10pt}
	\label{fig:sw_io}
\end{figure}

\begin{figure}
	\centering
	\begin{subfigure}{0.49\linewidth}
		\includegraphics[width=\linewidth]{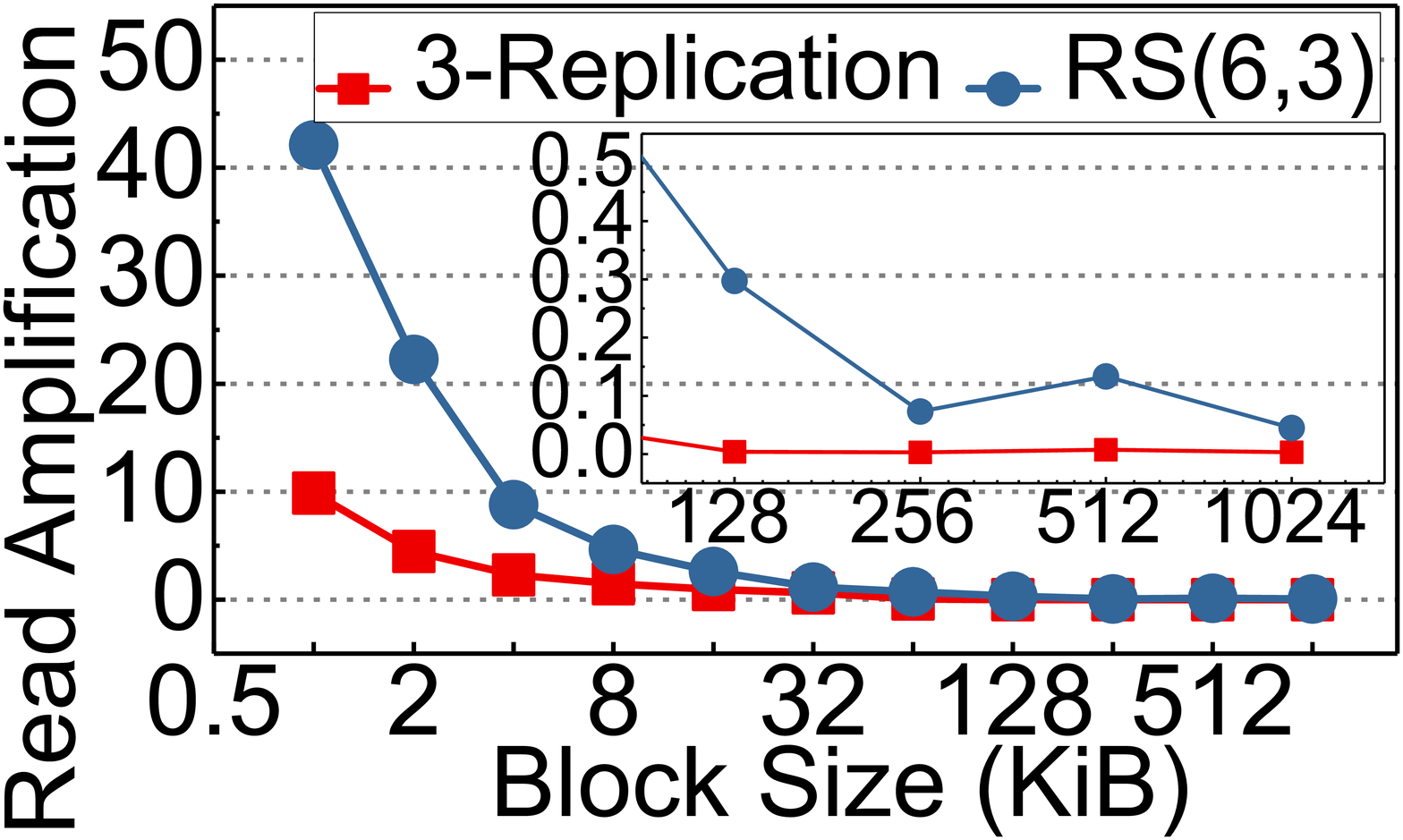}
		\caption{Read amplification.}
		\label{fig:RW_read}
	\end{subfigure}
	\begin{subfigure}{0.49\linewidth}
		\includegraphics[width=\linewidth]{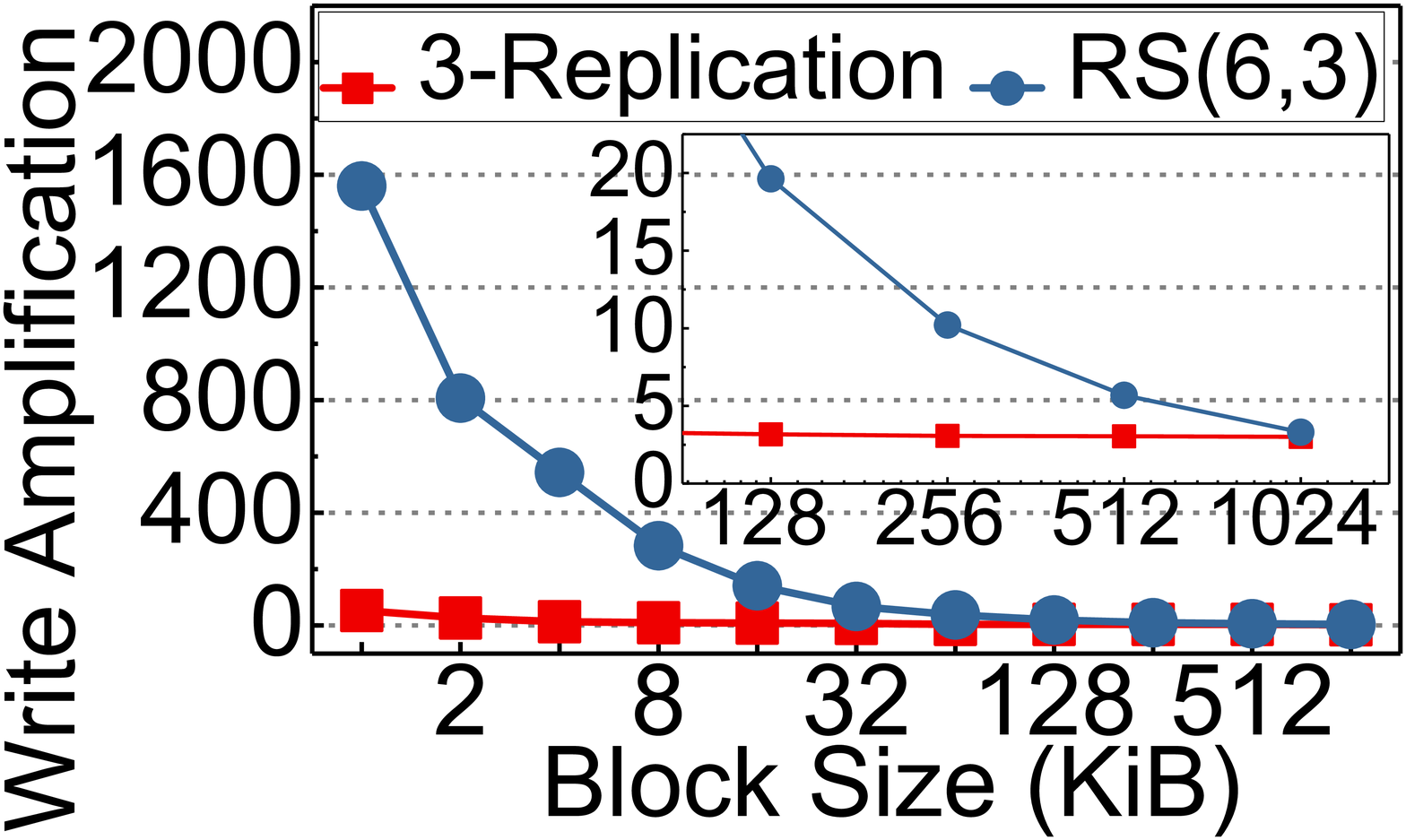}
		\caption{Write amplification.}
		\label{fig:RW_write}
	\end{subfigure}
	\vspace{-7pt}
	\caption{I/O amplifications under random writes.}
	\vspace{-10pt}
	\label{fig:rw_io}
\end{figure}

\noindent \textbf{Reads.} In contrast to writes, read operations mostly do not require encoding and/or decoding processes as there is no failure in our evaluation. However, we observe that CPU cycles are wasted by the fault tolerance modules of the PG backend in the storage nodes. Figure \ref{fig:SR_cpu} shows the CPU cycles consumed by 3-replication and RS(6,3) under sequential reads. As we expected, 3-replication uses only 2.6\% of the total CPU cycles, but RS(6,3) consumes up to 18\%. This is because, while 3-replication does not need to manipulate replicas to serve block I/O requests, erasure coding should concatenate data chunks distributed across multiple OSDs and compose them into a stripe. This RS-concatenation consumes CPU cycles not only for the stripe composition but also for the transaction module to handle data transfers. This phenomenon is more prominent when serving random I/O requests. As shown in Figure \ref{fig:RR_cpu}, 3-replication only consumes 6.1\% of the CPU cycles, whereas RS(6,3) consumes 34\% of the total CPU cycles, on average. The reason why erasure coding consumes more CPU cycles for random accesses is that relative less requests are processed at the same time with sequential accesses since consecutive requests are blocked by previously issued I/O request with sequential accesses due to the PG locks described in Section \ref{sec:background}. Without a failure, erasure coding consumes 34\% of the total CPU cycles, and user-mode operations (54\% of the total CPU cycles) are always involved with it during serving read requests. However, this is not acceptable in terms of power, efficiency, and scalability in many computing domains.

\begin{figure}
	\centering
	\begin{subfigure}{0.49\linewidth}
		\includegraphics[width=\linewidth]{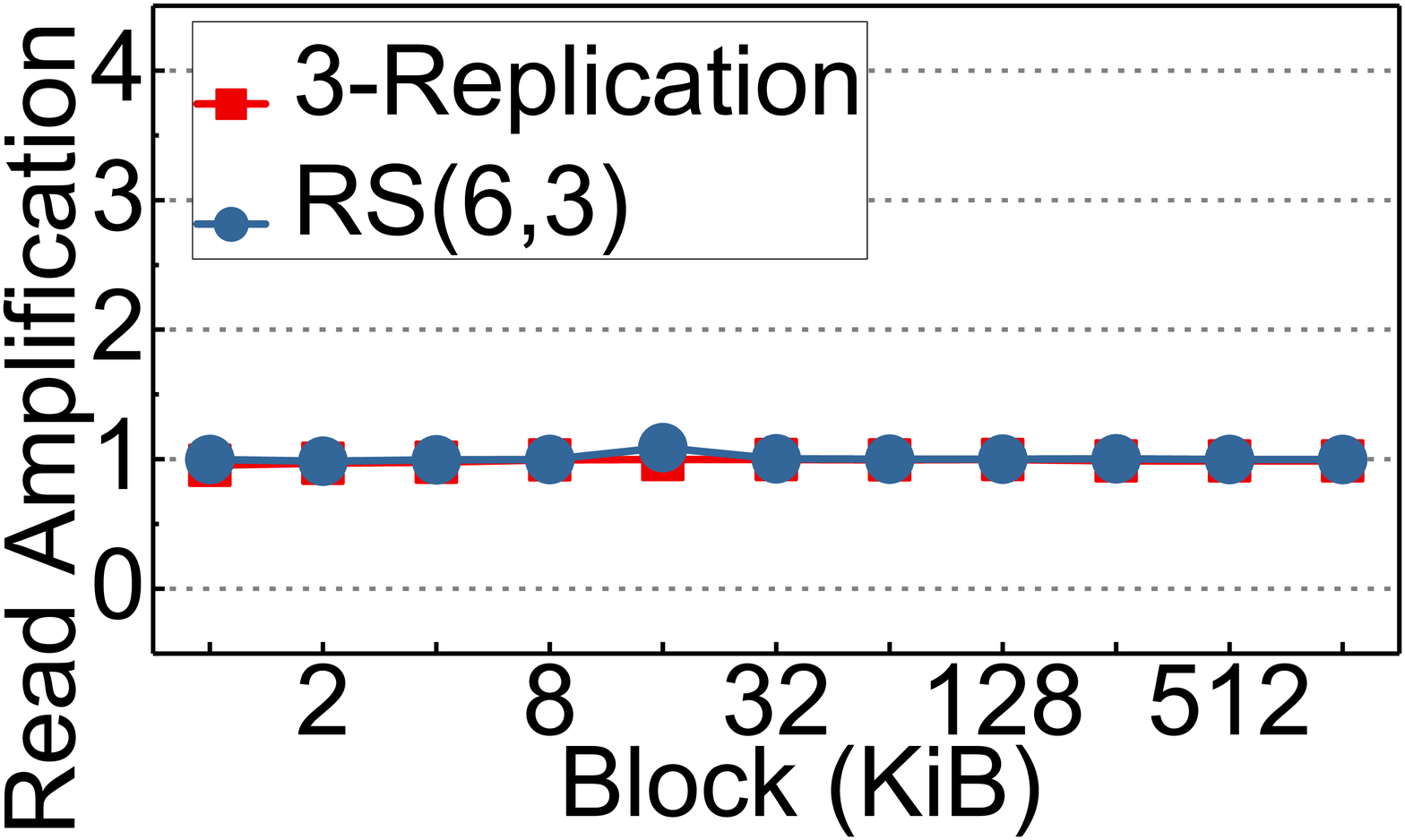}
		\caption{Sequential read.}
		\label{fig:SR_read}
	\end{subfigure}
	\begin{subfigure}{0.49\linewidth}
		\includegraphics[width=\linewidth]{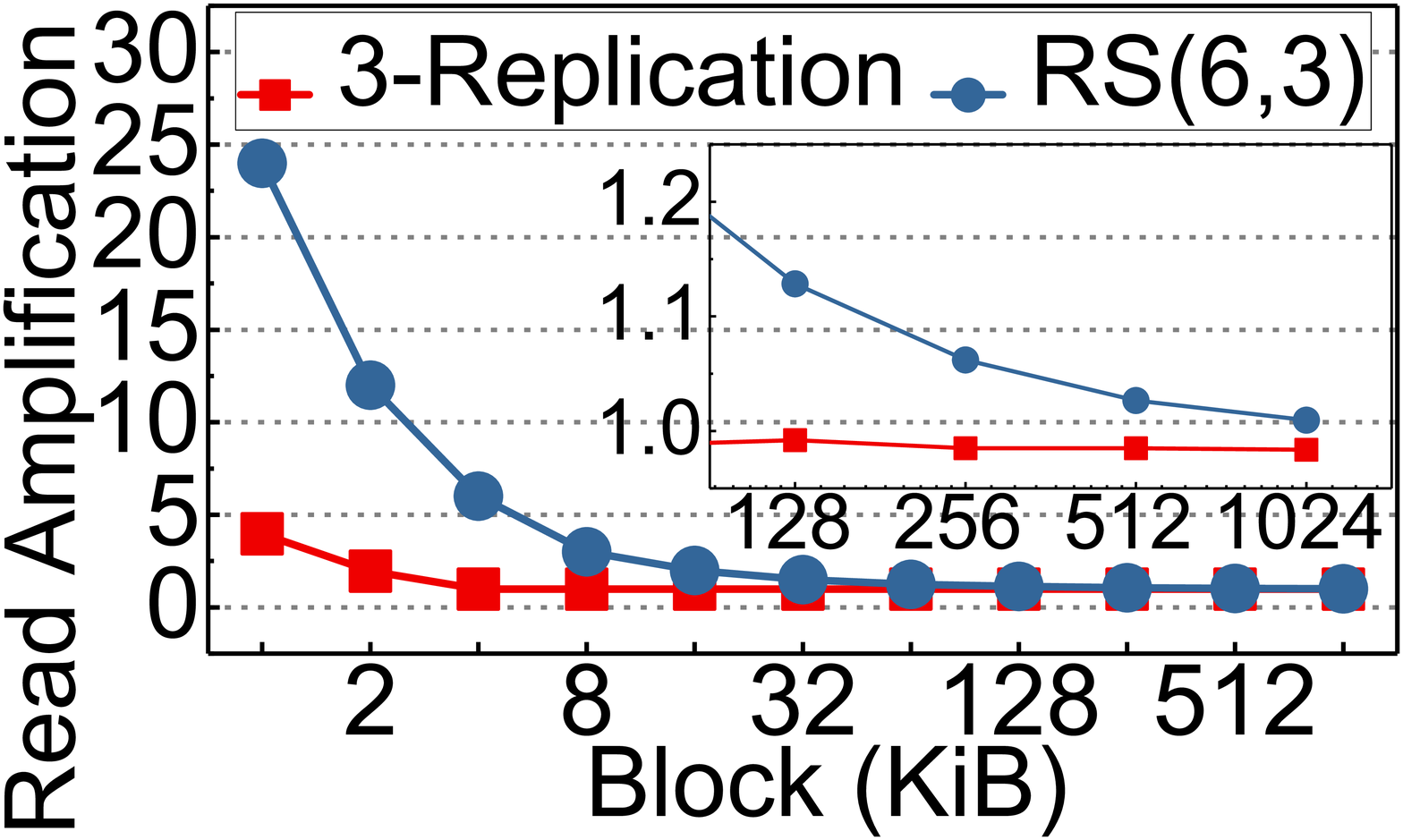}
		\caption{Random read.}
		\label{fig:RR_read}
	\end{subfigure}
	\vspace{-7pt}
	\caption{I/O amplifications under reads.}
	\vspace{-10pt}
	\label{fig:read_io}
\end{figure}

One of the challenges in deploying online erasure coding in distributed storage systems is user-level implementations. As described in the previous section, user-mode operations consume about 58\% of the total CPU cycles for reads and writes, on average. In this section, we examine context switching overheads associated with interleaving worker threads and context switches between the user and kernel modes. Figures \ref{fig:write_ctx} and \ref{fig:read_ctx} illustrate the relative number of context switches incurred by writes and reads. 
Since the number of requests processed varies according storage cluster performance, we use a different metric; the relative number of context switches, i.e., the number of context switches imposed by the storage cluster per I/O operation.

As shown in the Figure \ref{fig:write_ctx}, RS(6,3) exhibits 4.4$\times$ more relative context switches than 3-replication for writes, on average.
We ascribe this to two reasons. First, there is a significant amount of computing for encoding per object at the initial phase. Because Ceph manages the storage cluster with objects, even though libCephFS offers block interfaces, the erasure coding module at the PG backend creates dummy data chunks and coding chunks at the initial phase and writes them to the underlying SSDs. Second, the small size of writes (smaller than the stripe width) is treated as updates. This requires reading the underlying data chunks, regenerating coding chunks and updating the corresponding stripe causing many activities to be introduced at the cluster messenger and PG backend, which reside at the user level, thereby increasing the relative number of context switches. Note that RS(6,3) offers fewer relative context switches for reads than for writes by 75\%, on average, but still requires 4.6$\times$ more context switches than 3-replication, on average. This is mostly because RS-concatenation-related data transfers and computations are also performed at the user level. We will analyze why there are many context switches in online erasure coding with I/O amplification analysis and investigate the network traffic in depth.

\subsection{Data Overhead Analysis}
\label{sec:dataoverhead}
\noindent \textbf{Replication vs. erasure coding.} Figures \ref{fig:sw_io}, \ref{fig:rw_io} and \ref{fig:read_io} show the read and write I/O amplification calculated by normalizing the amount of data that 3-replication and erasure coding at the PG backend generate with the actual amount of data that the application requests. To obtain a high-level insight on the difference between 3-replication and erasure coding, we first select the overheads observed by I/O services with sequential accesses; other ones will be analyzed shortly. The 3-replication generally does not require reading data when it writes, but it performs a read-and-modify operation when the data are smaller than 4KB which is the minimum unit of I/O.

This in turn introduces read amplification 9$\times$ greater than that of the original request size under a sequential write with 1KB blocks. We also observe that the I/O amplification caused by 3-replication is just the same amount of replicas which necessarily needs to be written under the writes with the blocks larger than 4KB (Figures \ref{fig:SW_write} and \ref{fig:RW_write}). 
In contrast, RS(6,3) introduces severe I/O amplification. This is because, as described in Section \ref{sec:background}, each OSD handles block requests managed by libCephFS for each object. Thus, even though writes occur as a sequential order (but their block size is smaller than the stripe), the PG backend requires reading the data chunks for each write and updating them with new coding chunks. As shown in the plots, this increases the amount of data to read and write by up to 1561$\times$ and 142$\times$ compared with the total volume of all I/O requests.

\noindent \textbf{Writes.} Figures \ref{fig:RW_read} and \ref{fig:RW_write} show the I/O amplification under the execution of client-level random writes. As indicated in the previous subsection, RS(6,3) introduces up to 29$\times$ more data to write than 3-replication as well as many more reads. As previously mentioned, the random accesses lead to up to 29$\times$ more write amplification than 3-replication, which differs completely from the common expectation for the effect of erasure coding on the storage overheads. Even though users request small size writes, the OSD requires the creation of an object. When users request data as random accesses, such small requests are distributed across many OSDs and create/initialize the corresponding objects, which in turn leads to greater write amplification. We examine this in detail in the private network analysis (Section \ref{private_network}).

\begin{figure}
	\centering
	\begin{subfigure}{0.49\linewidth}
		\includegraphics[width=\linewidth]{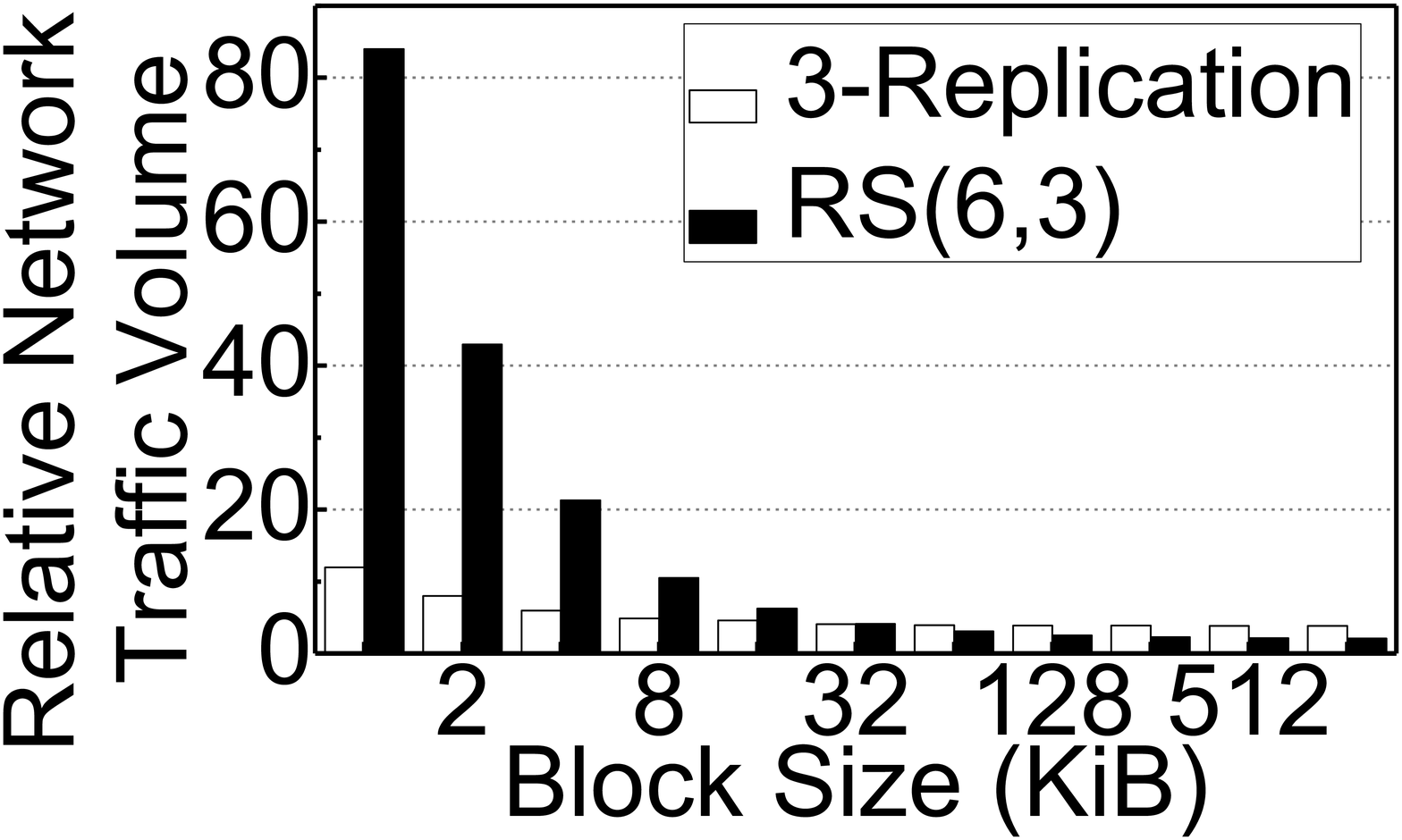}
		\caption{Sequential write.}
		\label{fig:SW_net}
	\end{subfigure}
	\begin{subfigure}{0.49\linewidth}
		\includegraphics[width=\linewidth]{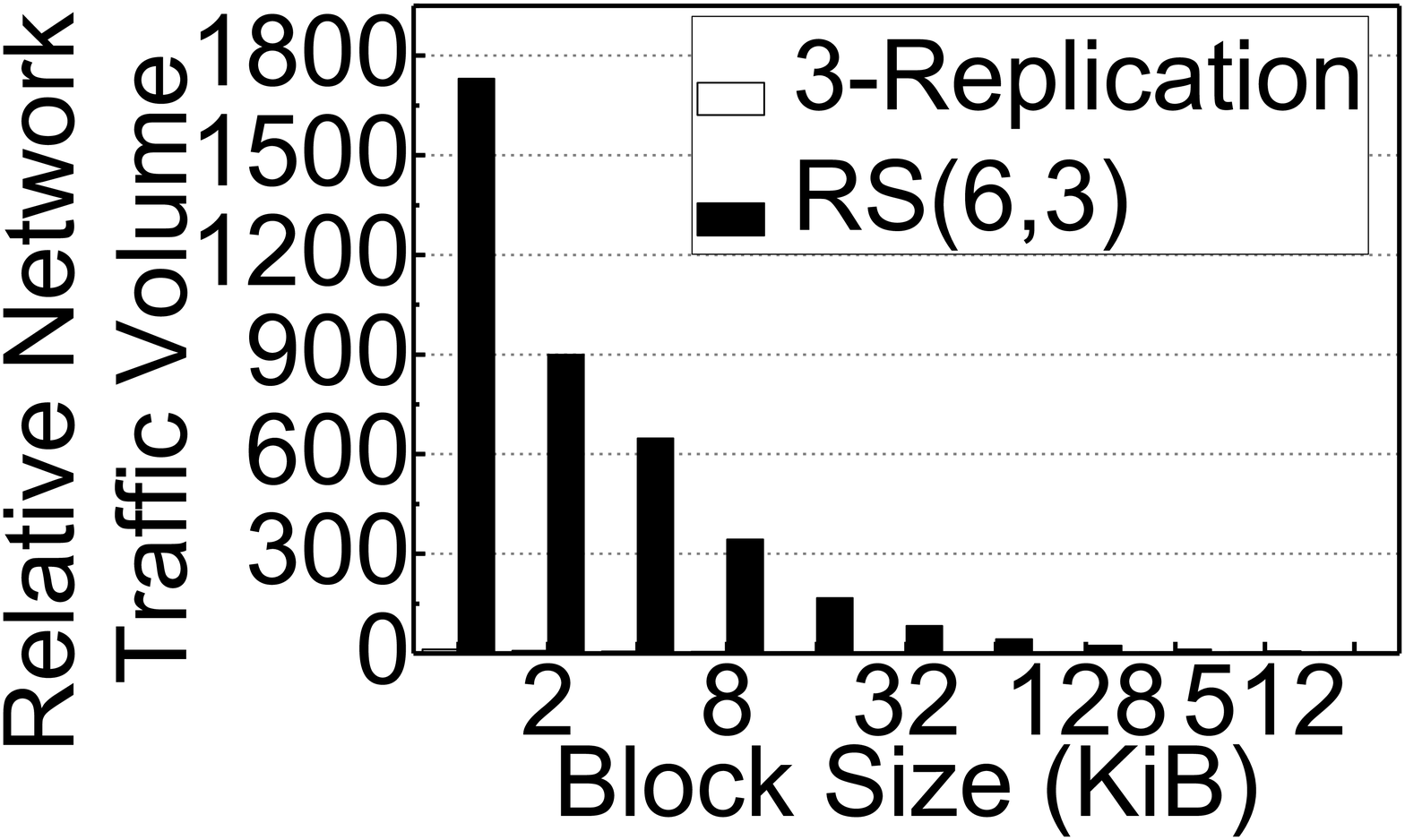}
		\caption{Random write.}
		\label{fig:RW_net}
	\end{subfigure}
	\vspace{-5pt}
	\caption{Relative network traffic volume under writes.}
	\vspace{-5pt}
	\label{fig:write_network}
\end{figure}

\begin{figure}
	\centering
	\begin{subfigure}{0.49\linewidth}
		\includegraphics[width=\linewidth]{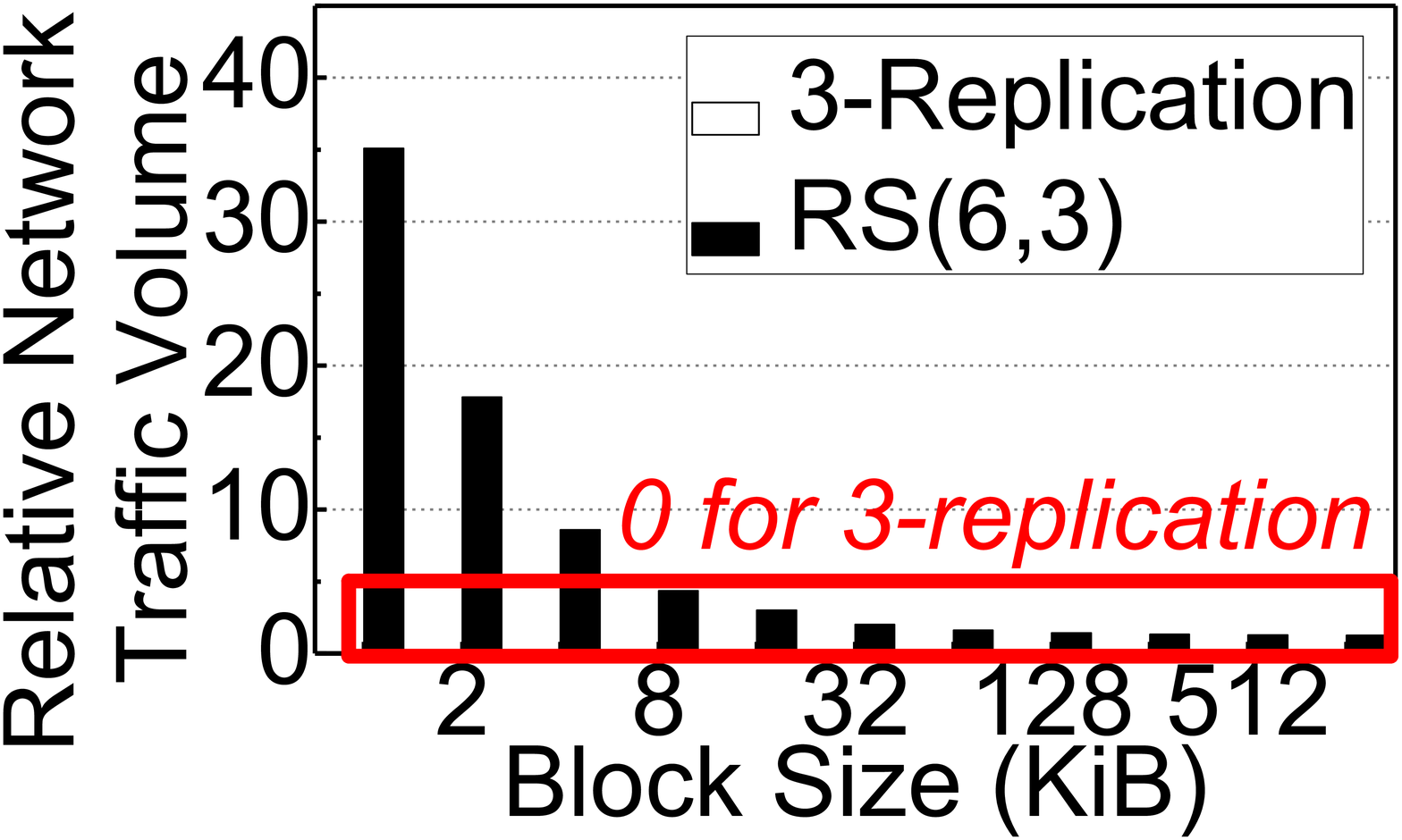}
		\caption{Sequential read.}
		\label{fig:SR_net}
	\end{subfigure}
	\begin{subfigure}{0.49\linewidth}
		\includegraphics[width=\linewidth]{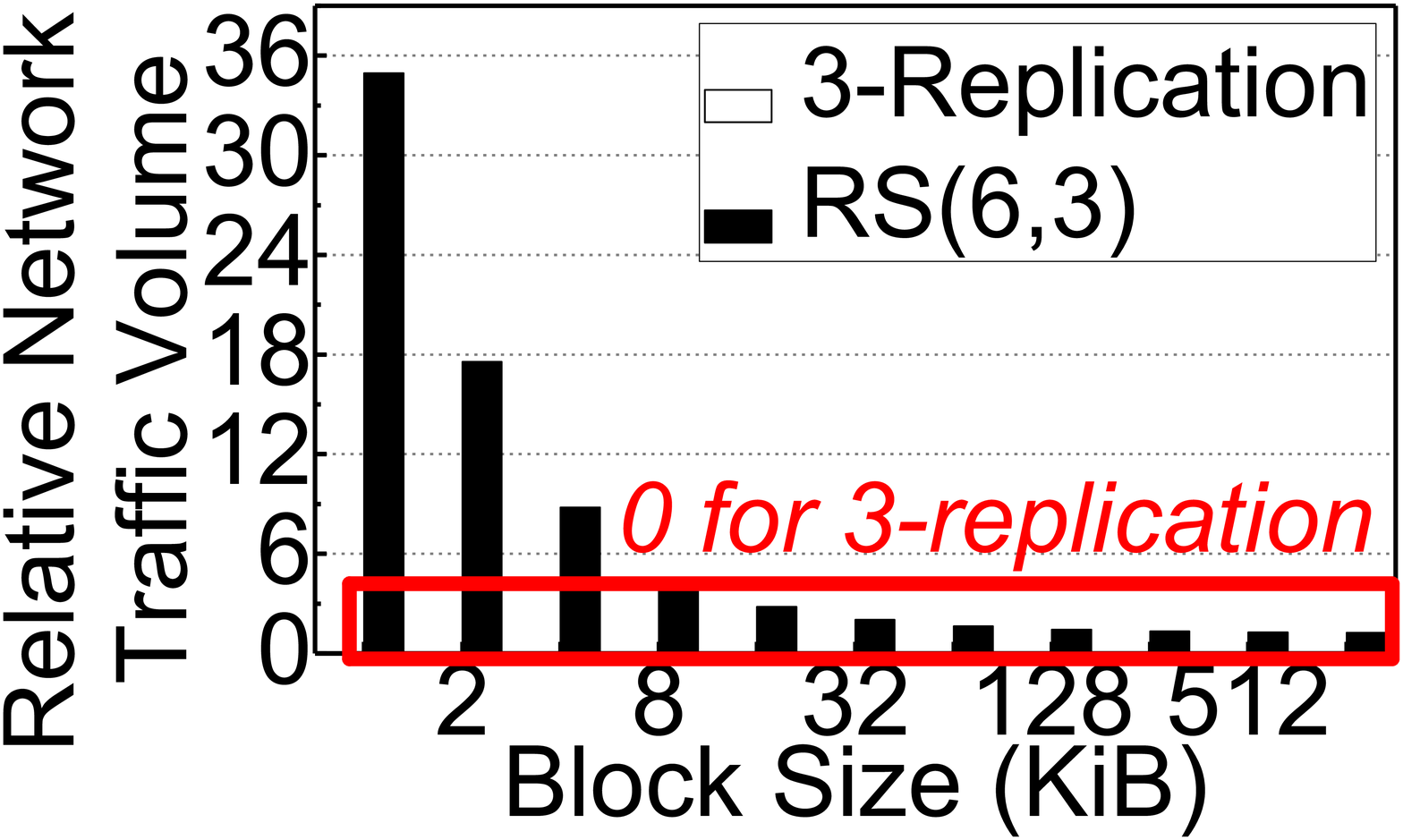}
		\caption{Random read.}
		\label{fig:RR_net}
	\end{subfigure}
	\vspace{-5pt}
	\caption{Relative network traffic volume under reads.}
	\vspace{-5pt}
	\label{fig:read_network}
\end{figure}

\noindent \textbf{Reads.} Figures \ref{fig:SR_read} and \ref{fig:RR_read} show reads with sequential and random accesses, respectively. In sequential reads, since whole data in stripe is requested by consecutive reads, the PG backend for all fault-tolerance mechanisms reads exactly as many data chunks as the client requests, which in turn makes the I/O amplification for all varying block sizes almost once. Even though the block size of a request is smaller than the minimum unit of I/O, consecutive I/O requests leverage the data; therefore, there is no read amplification. However, read requests with a random pattern can be distributed and served by different OSDs. As erasure coding at the PG backend performs I/O service based on the stripe width, if the request is smaller than the stripe width, the data transfers are wasted. If a small block size request spans two stripes, greater amplification can occur. As shown in Figure \ref{fig:RR_read}, 
RS(6,3) imposes 6.1$\times$ greater I/O amplification than 3-replication at 4KB. When there is a request whose block size is slightly greater than the stripe width (e.g., 32KB) and its data spans across two or three stripes, RS(6,3) imposes about 1.2$\times$ greater I/O amplification than 3-replication.

\subsubsection{Private Network Traffic}
\label{private_network}
\noindent \textbf{Writes.} One may expect that 3-replication exhibits more data transfers than RS(6,3) as the amount of replica data exceeds the amount of coding chunks in RS(6,3). However, 
under writes with various block sizes and patterns, the I/O amplification imposed by RS(6,3) congests the private network. Specifically, as shown in Figure \ref{fig:SW_net}, RS(6,3) generates 3.4$\times$ more data transfers over the actual request size than 3-replication, on average, if the block size is smaller than 32KB. 
As the block size increases, the portion of extra stripes decreases, reducing the private network traffic. Importantly, under the random accesses, the private network is overly congested by RS(6,3). As shown in Figure \ref{fig:RW_net}, 
RS(6,3) transfers 82$\times$ more data than 3-replication, which should be optimized for a future system. This is because erasure coding of the PG backend requires initializing the objects, which can cause significant overheads when the requests are distributed across multiple OSDs.

\begin{figure}
	\centering
	\includegraphics[width=\linewidth]{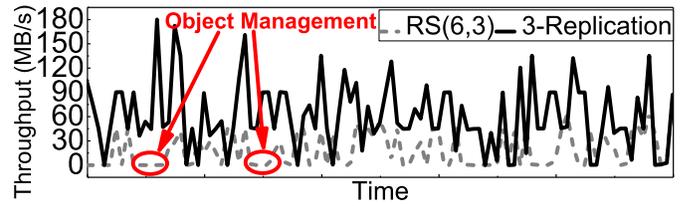}
	\vspace{-5pt}
	\caption{Object management in 16KB.}
	\vspace{-5pt}
	\label{fig:object_manage}
\end{figure}
\begin{figure}
	\centering
	\begin{subfigure}{0.49\columnwidth}
		\includegraphics[width=\columnwidth]{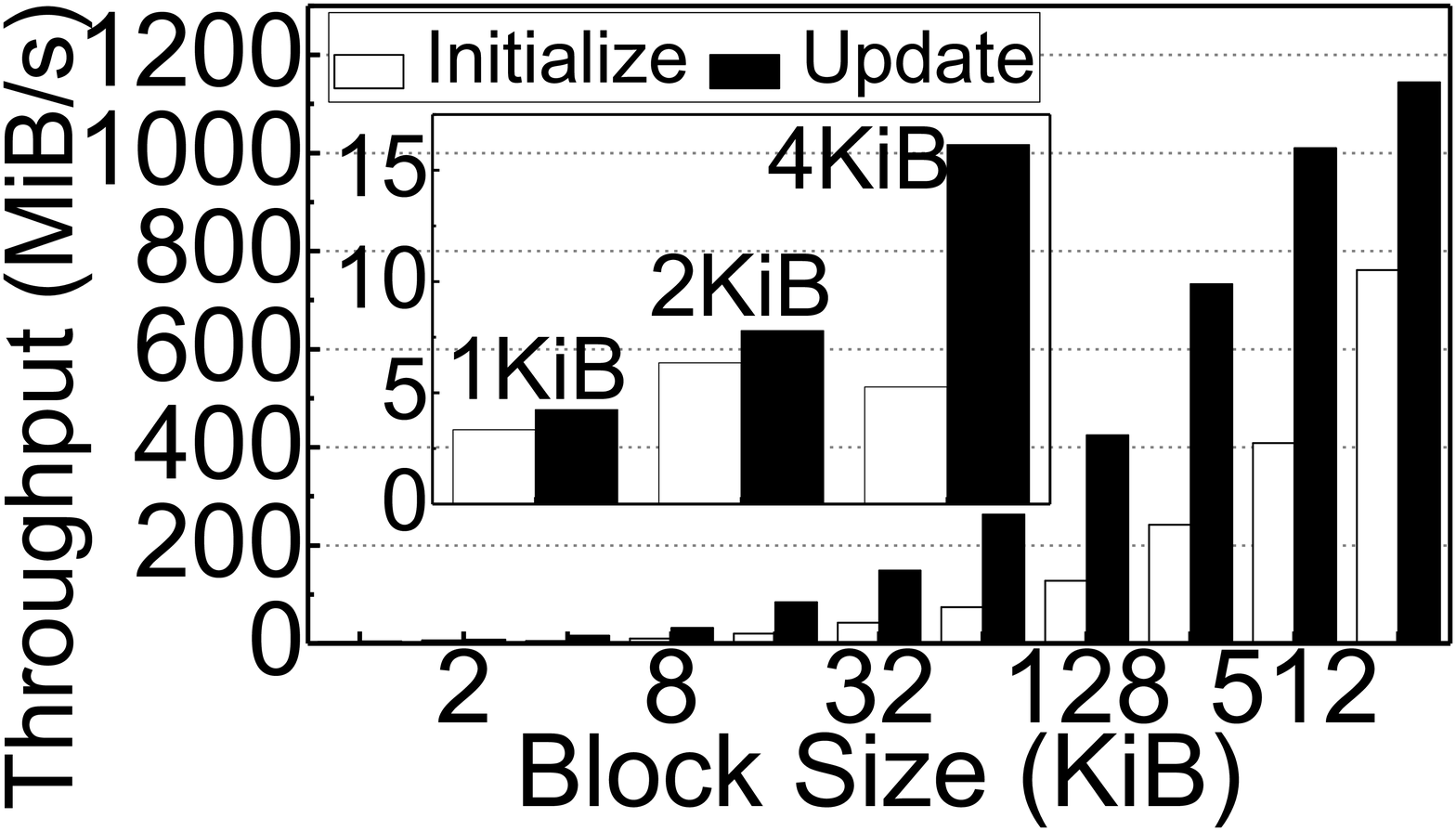}
		\caption{Throughput.}
		\label{fig:ow_thr}
	\end{subfigure}
	\begin{subfigure}{0.49\columnwidth}
		\includegraphics[width=\columnwidth]{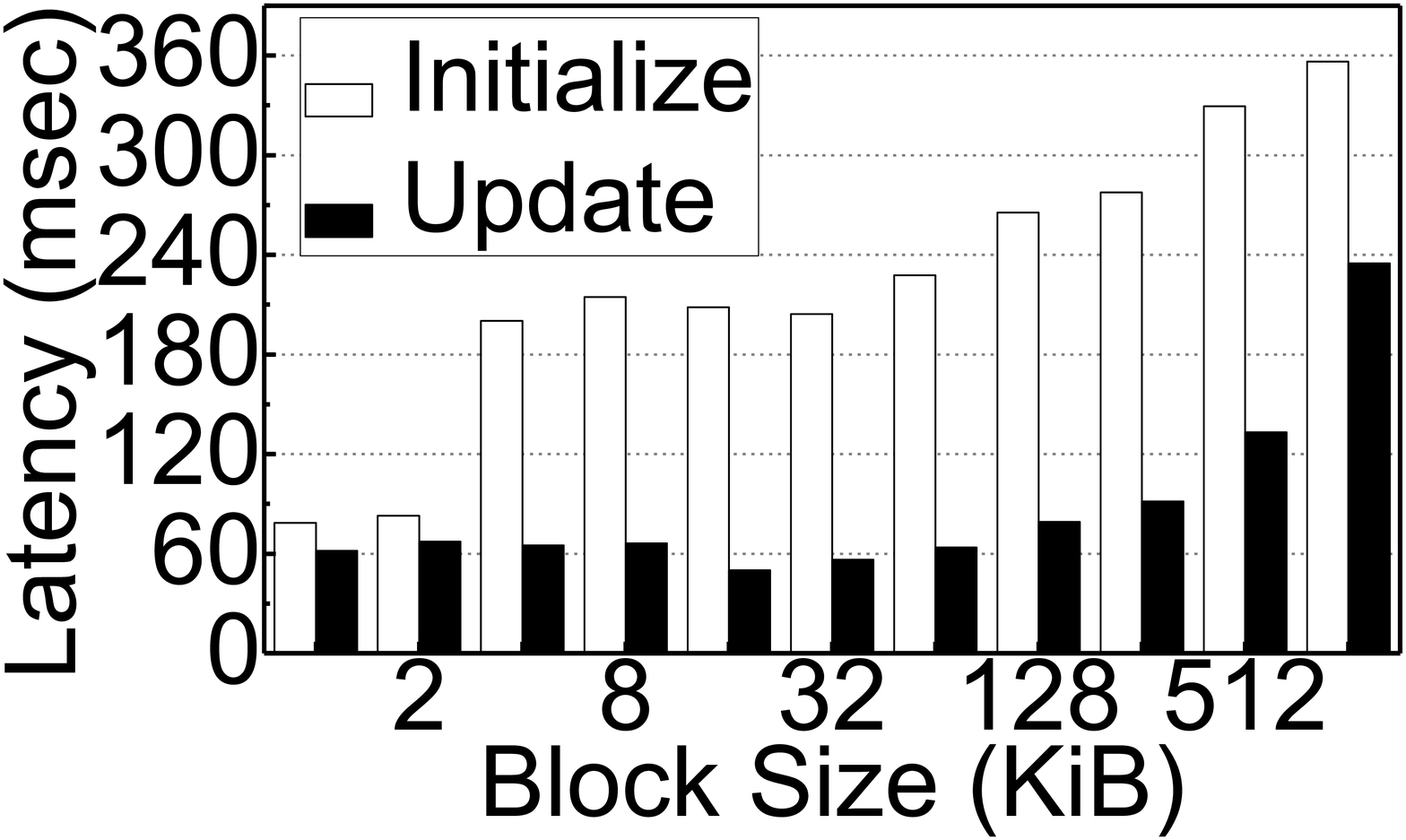}
		\caption{Latency.}
		\label{fig:ow_lat}
	\end{subfigure}
	\vspace{-5pt}
	\caption{Comparing overall performance.}
	\vspace{-5pt}
	\label{fig:overwrite}
\end{figure}

\begin{figure}
	\centering
	\begin{subfigure}{0.49\columnwidth}
		\includegraphics[width=\columnwidth]{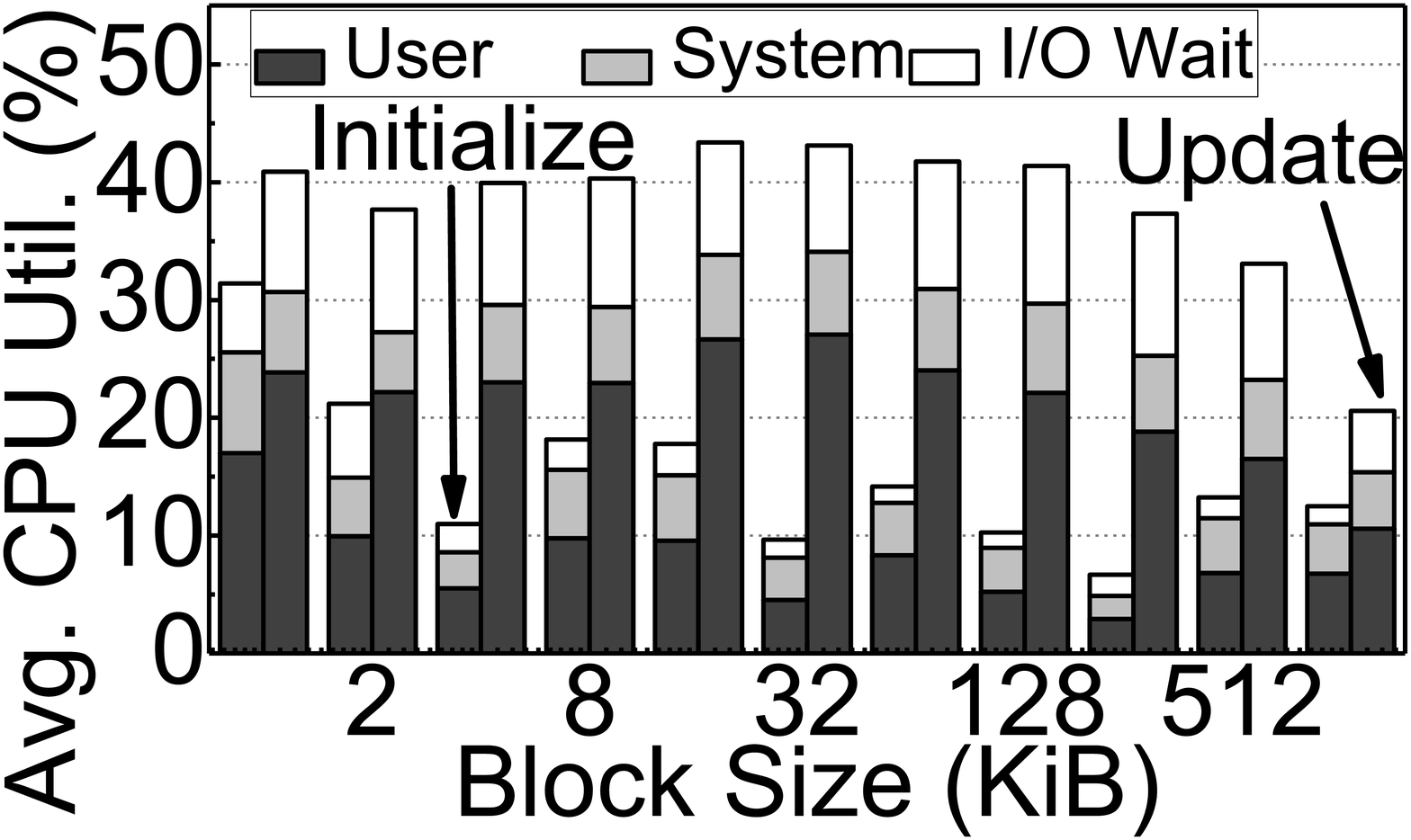}
		\caption{CPU Utilization.}
		\label{fig:row_cpu}
	\end{subfigure}
	\begin{subfigure}{0.49\columnwidth}
		\includegraphics[width=\columnwidth]{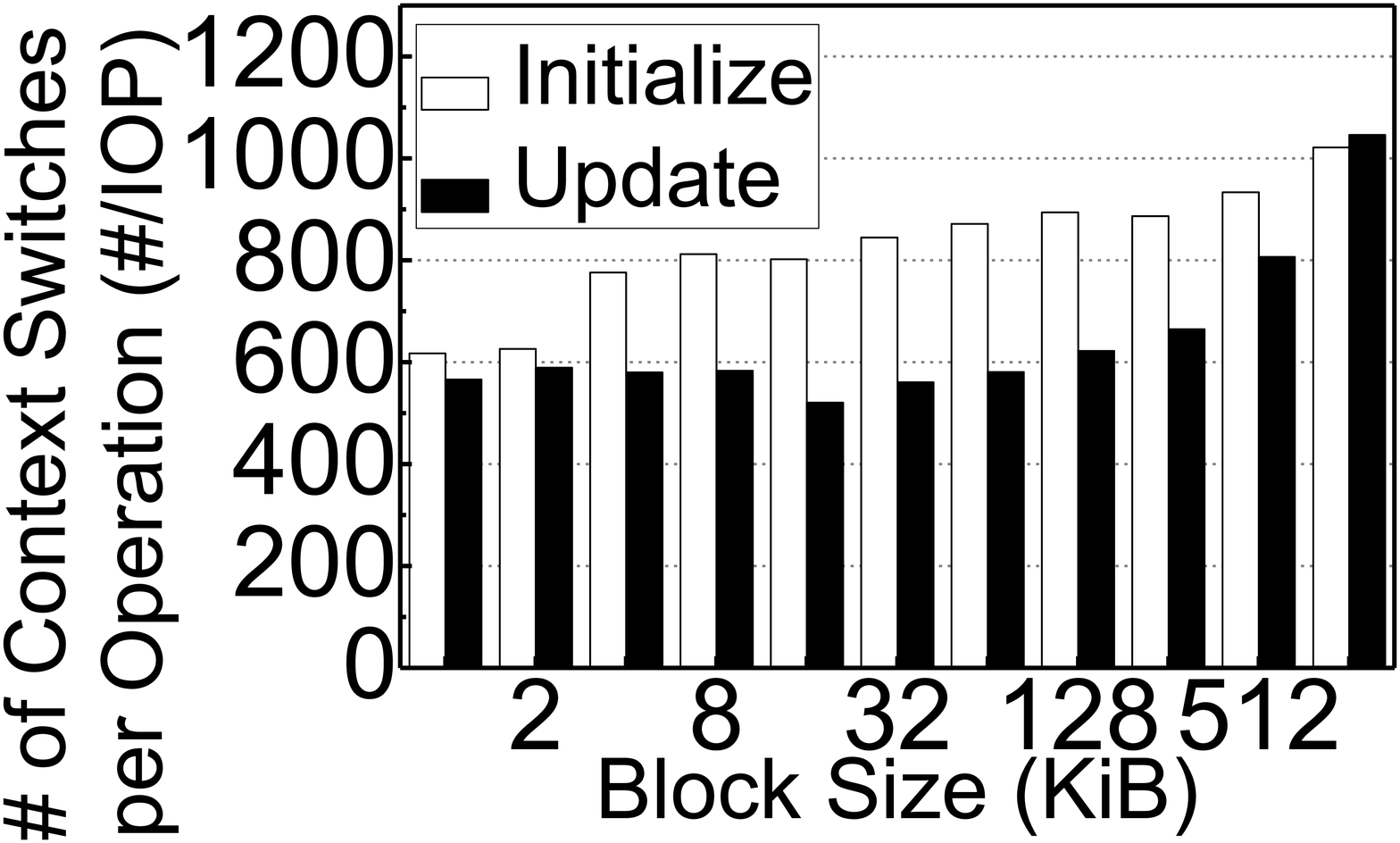}
		\caption{Context Switches.}
		\label{fig:row_ctx}
	\end{subfigure}
	\vspace{-5pt}
	\caption{Comparing computation and software overheads.}
	\vspace{-5pt}
	\label{fig:overwrite_comp}
\end{figure}

\noindent \textbf{Reads.} Figure \ref{fig:read_network} shows the data transfer overheads imposed by the 
fault tolerance mechanisms of the PG backend during read services. 
This indicates that 3-replication exhibits only minimal data transfers related to necessary communications among OSD daemons. This OSD interaction is used for monitoring the status of each OSD, which is referred to as the OSD heartbeat. In this study, it generates 280KB/s data transfers, which has almost no impact for the private network traffic. In contrast, RS(6,3) introduces upto
25$\times$ more data transfers than the total volume of requested reads, which is associated with concatenating the data chunks into a stripe. 
Note that the amount of data transferred (for both reads and writes) over the network cannot be statistically calculated using the relationship between the $k$ data chunks and $m$ coding chunks because the data are often transferred among OSDs within a node.

\begin{figure}
	\centering
	\begin{subfigure}{0.49\columnwidth}
		\includegraphics[width=\columnwidth]{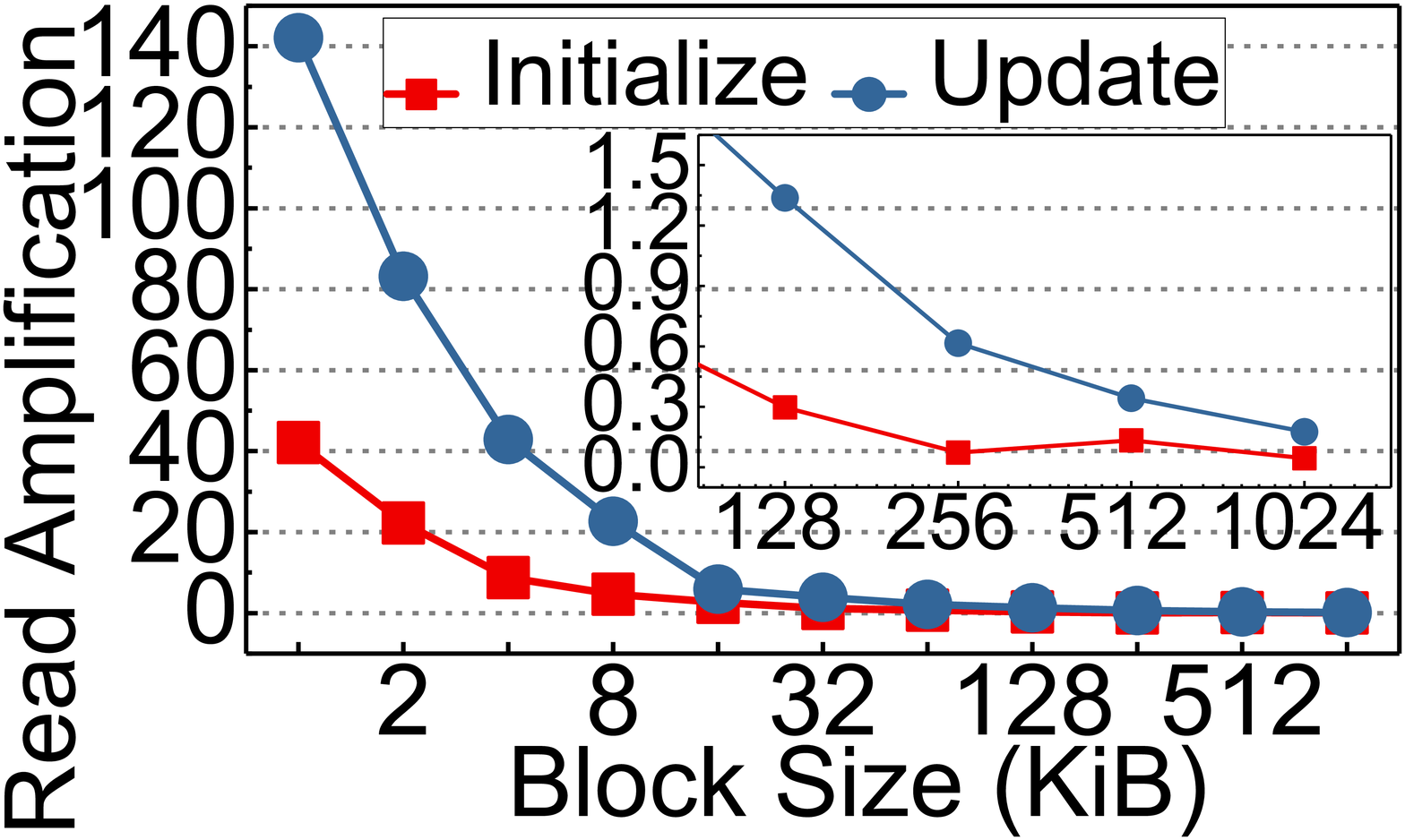}
		\caption{Read Amplification.}
		\label{fig:row_read}
	\end{subfigure}
	\begin{subfigure}{0.49\columnwidth}
		\includegraphics[width=\columnwidth]{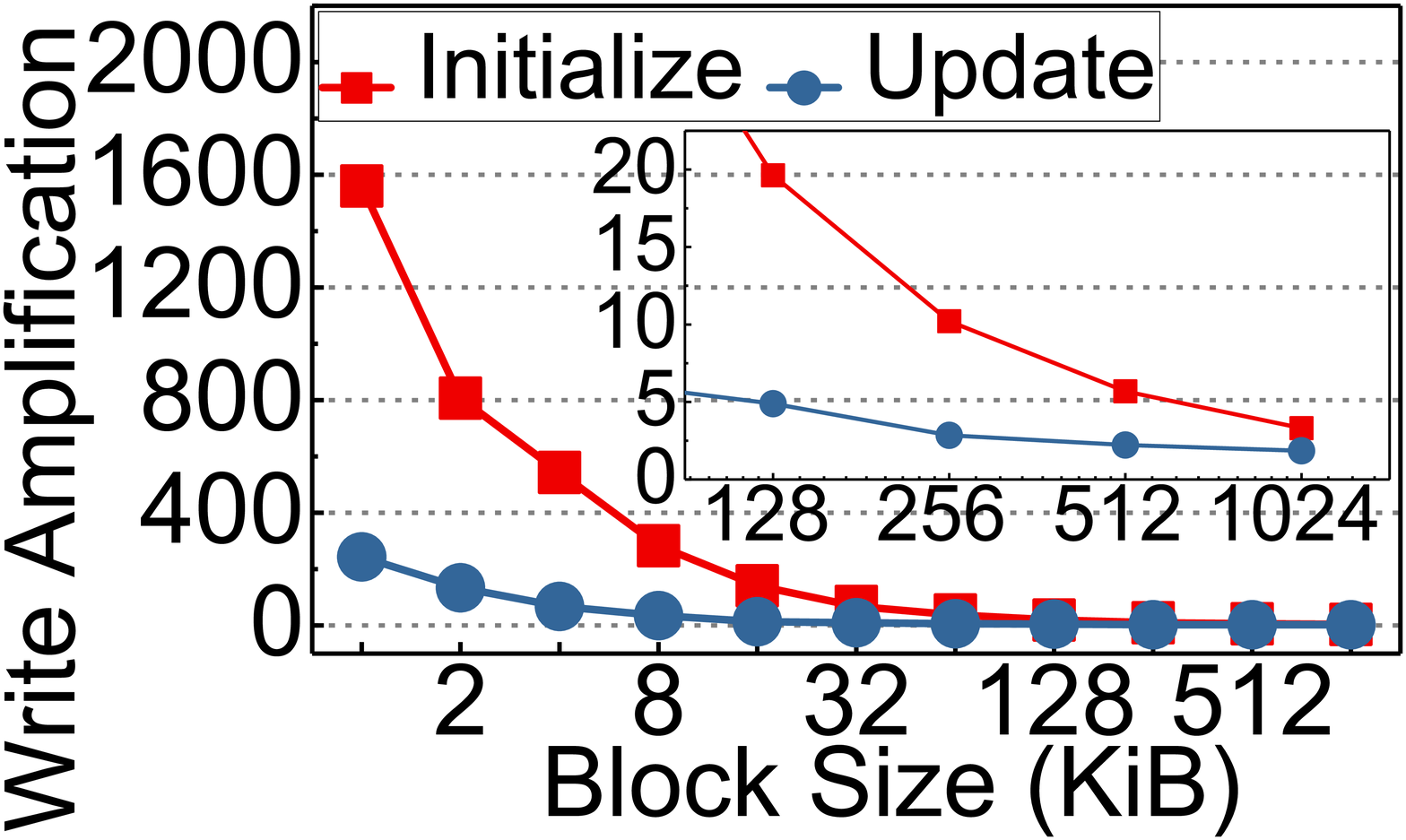}
		\caption{Write Amplification.}
		\label{fig:row_write}
	\end{subfigure}
	\vspace{-5pt}
	\caption{Comparing I/O amplification.}
	\vspace{-5pt}
	\label{fig:overwrite_ioamp}
\end{figure}

\subsection{Data Distribution Characteristics}
\label{sec:distribution}

In this section, we will examine the performance impacts of erasure coding imposed by the object management of the underlying parallel file systems, which were briefly discussed in previous sections. Since the storage cluster manages data over an object, it places a burden on erasure coding. Specifically, while libCephFS offers file system operations, the data are managed by storage nodes as an object. Thus, whenever an OSD daemon receives a new write that heads a target address in the range of an object, it creates the object and fills the data shards and coding shards. This object initialization degrades the overall system performance. Figure \ref{fig:object_manage} shows a time series analysis for the initial performance obtained by issuing 16KB of data as a sequential pattern. The figure indicates that, while 3-replication does not cause performance degradation at the I/O initial phase, RS(6,3) periodically shows a near-zero throughput owing to the object initialization that fills the data and coding shards.

\begin{figure}
	\centering
	\includegraphics[width=0.49\linewidth]{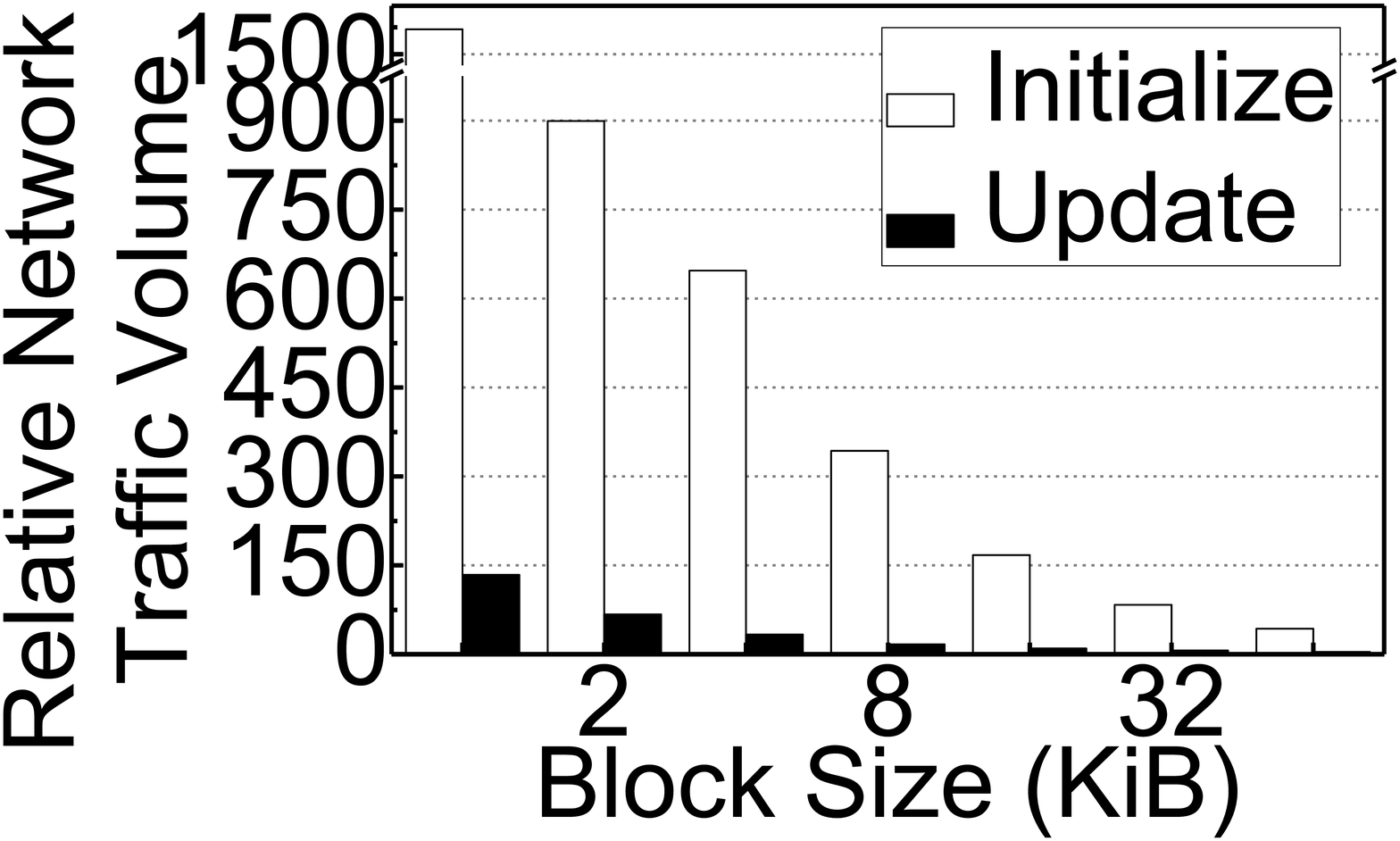}
	\includegraphics[width=0.49\linewidth]{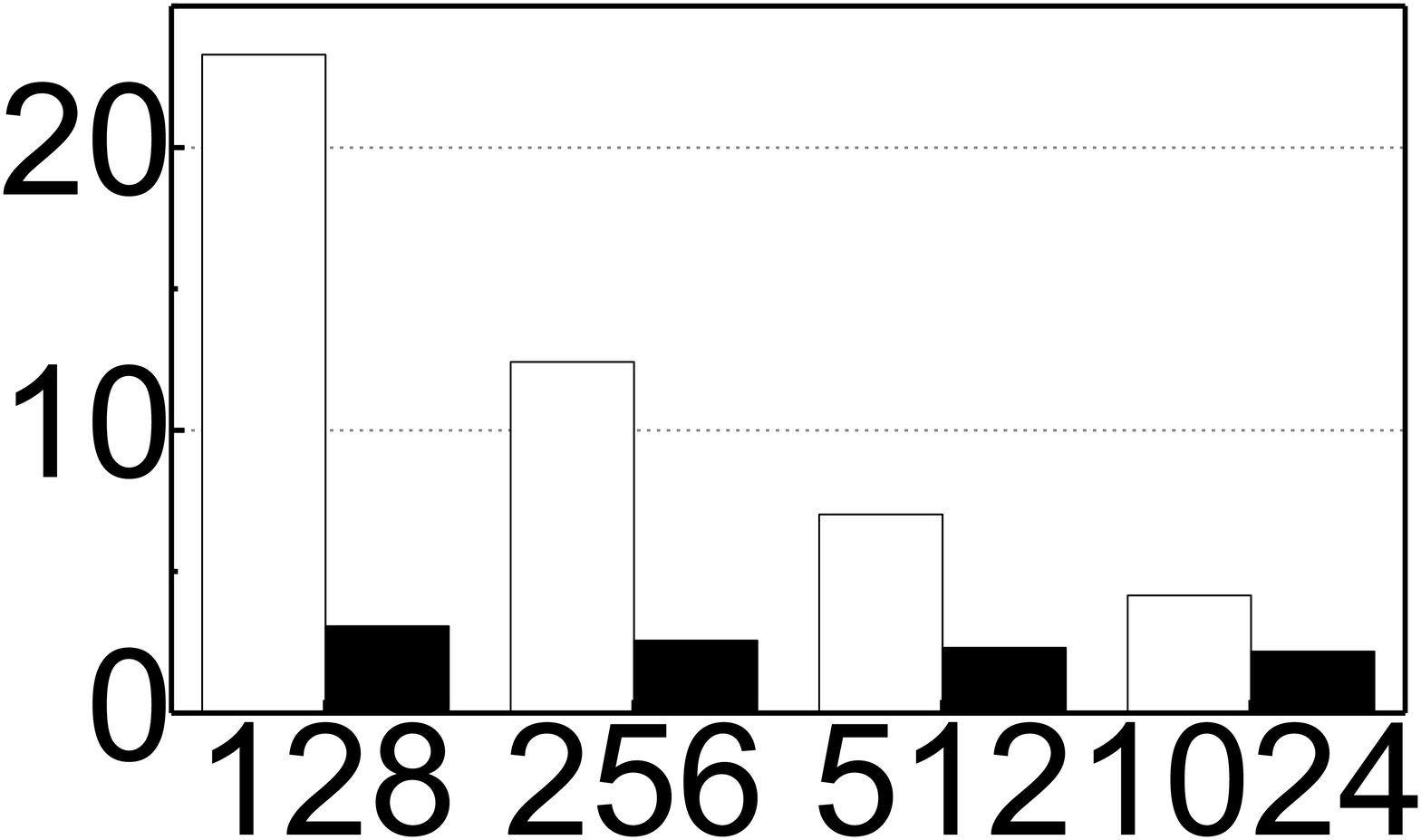}
	\vspace{-5pt}
	\caption{Comparing network traffic.}
	\vspace{-5pt}
	\label{fig:overwrite_network}
\end{figure}

To obtain precise results, we further analyze the performance degradation and overheads imposed by initializing data and coding shards. Figure \ref{fig:overwrite} compares the performance of writes on a pristine image and overwrites. This figure shows that the throughput and latency of updates are 2.8$\times$ and 0.4$\times$ better, respectively, than initial writes on objects, on average. This degradation can be supported by several overheads as shown in Figures \ref{fig:overwrite_comp}$\sim$\ref{fig:overwrite_network}, each of which compares writes on pristine objects and overwrites in terms of the CPU utilizations and context switch overheads, I/O amplification, and private network traffic. As shown in these figures, 4.1$\times$ more I/O amplification is incurred under writes on the clean objects; therefore, the private network is 11.3$\times$ busier than that of overwrites to transfer/receive data from each other. However, the CPU utilization during overwrites is 3$\times$ higher than that for writes on pristine objects, whereas the relative number of context switches is 21\% smaller. We believe that the relative number of context switches is larger in initialization because more reads and writes are issued from the PG backend per write request from client. Nevertheless, the CPU utilization is higher because far more requests are served under overwrites in the same amount of time.

\subsection{RS Settings}
\label{sec:configuration}
\begin{figure}
	\centering
	\begin{subfigure}{0.49\linewidth}
		\includegraphics[width=\linewidth]{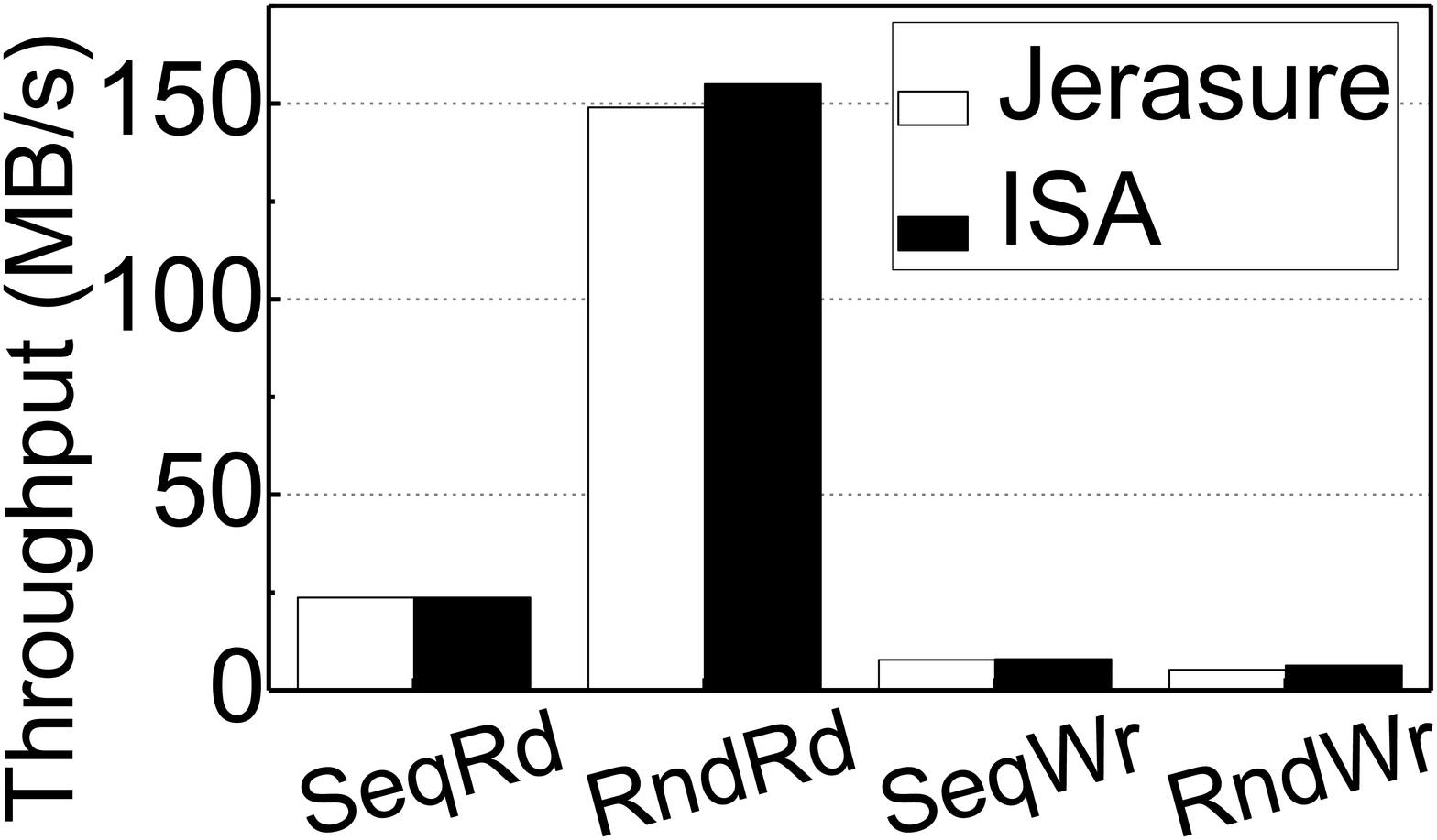}
		\caption{Throughput.}
		\label{fig:isa_thr}
	\end{subfigure}
	\begin{subfigure}{0.49\linewidth}
		\includegraphics[width=\linewidth]{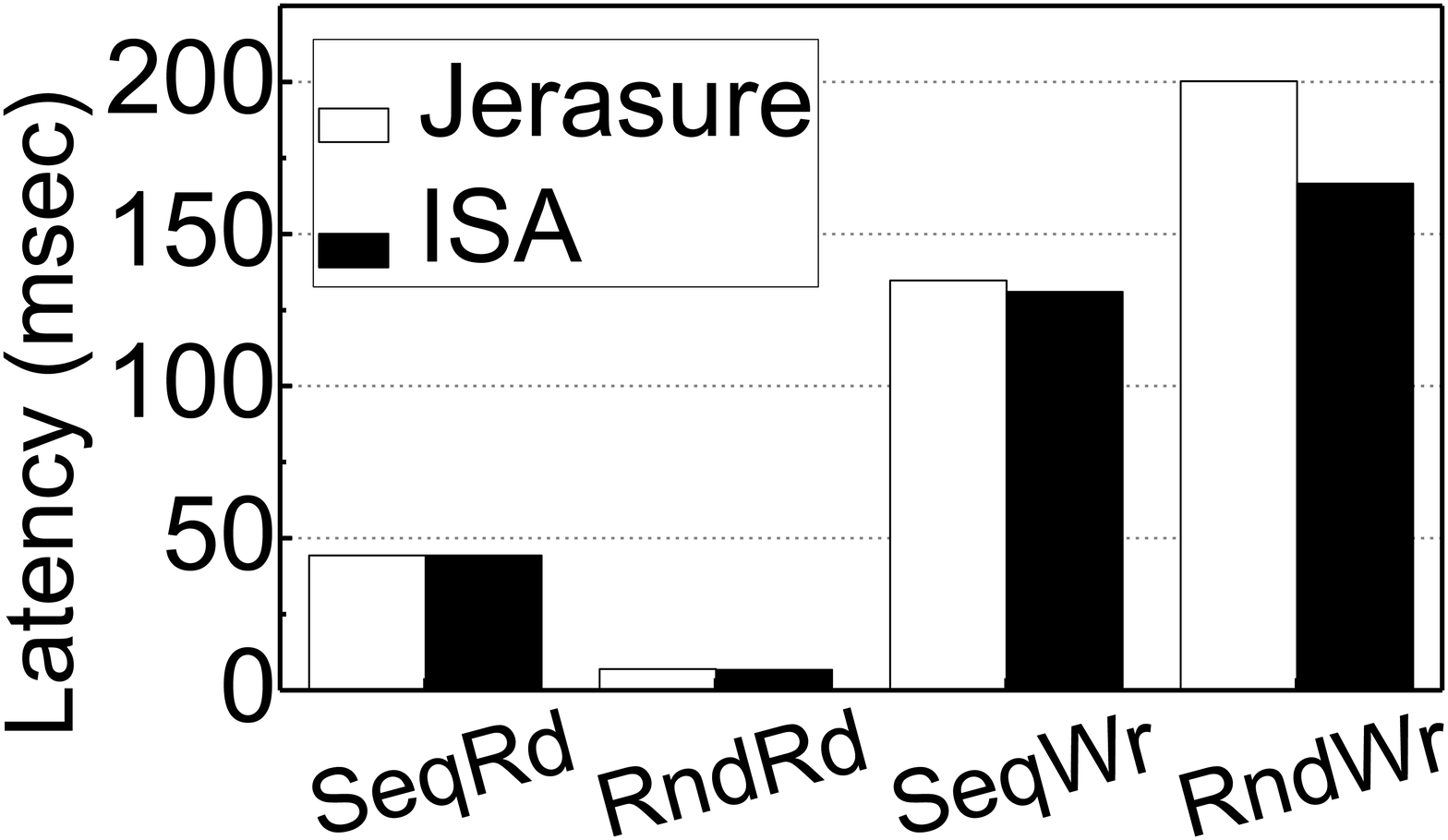}
		\caption{Latency.}
		\label{fig:isa_lat}
	\end{subfigure}
	\vspace{-5pt}
	\caption{Performance comparison between Jerasure and ISA.}
	\vspace{-5pt}
	\label{fig:isa_perf}
\end{figure}

\begin{figure}
	\centering
	\begin{subfigure}{0.49\linewidth}
		\includegraphics[width=\linewidth]{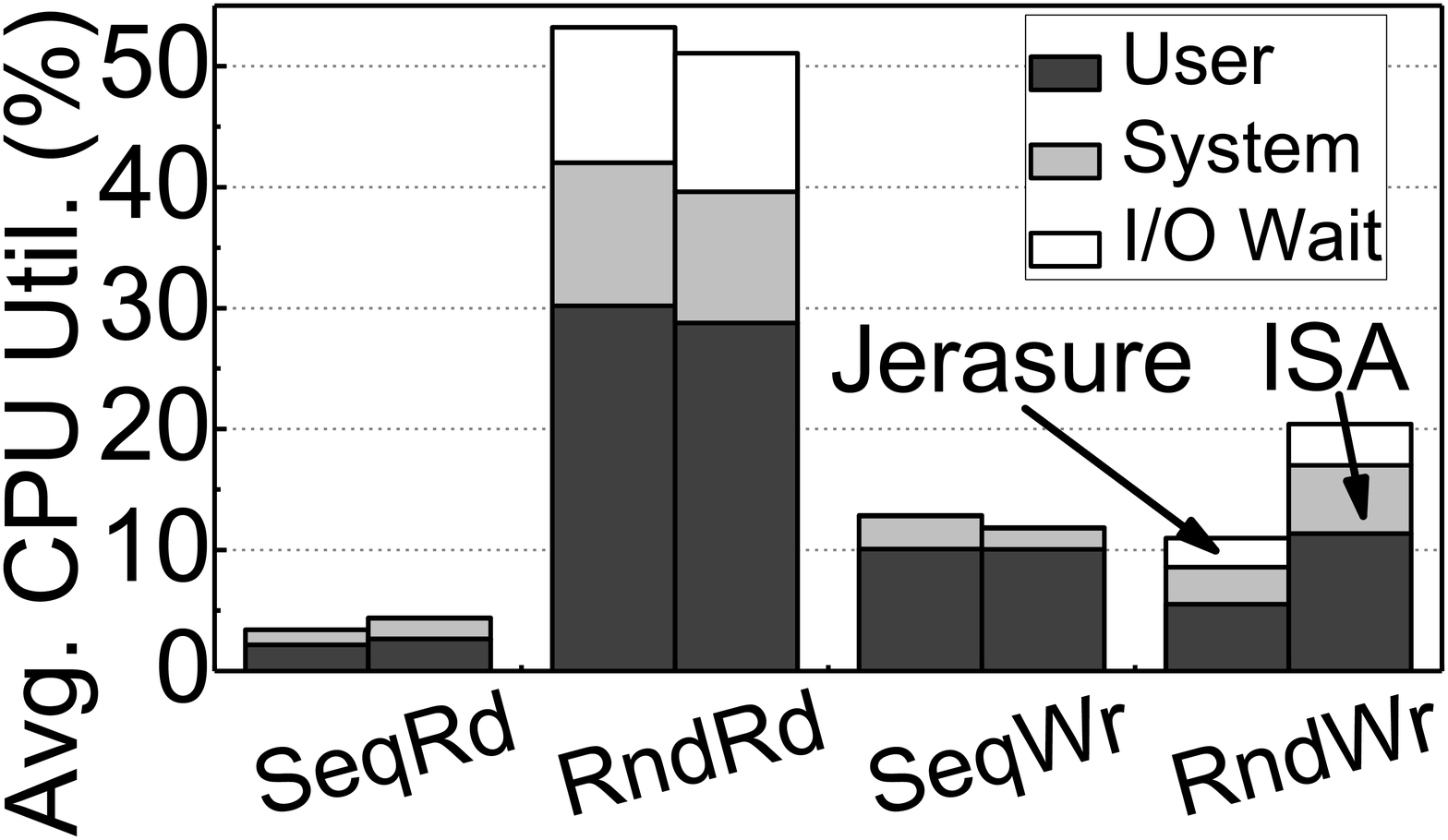}
		\caption{CPU utilization.}
		\label{fig:isa_cpu}
	\end{subfigure}
	\begin{subfigure}{0.49\linewidth}
		\includegraphics[width=\linewidth]{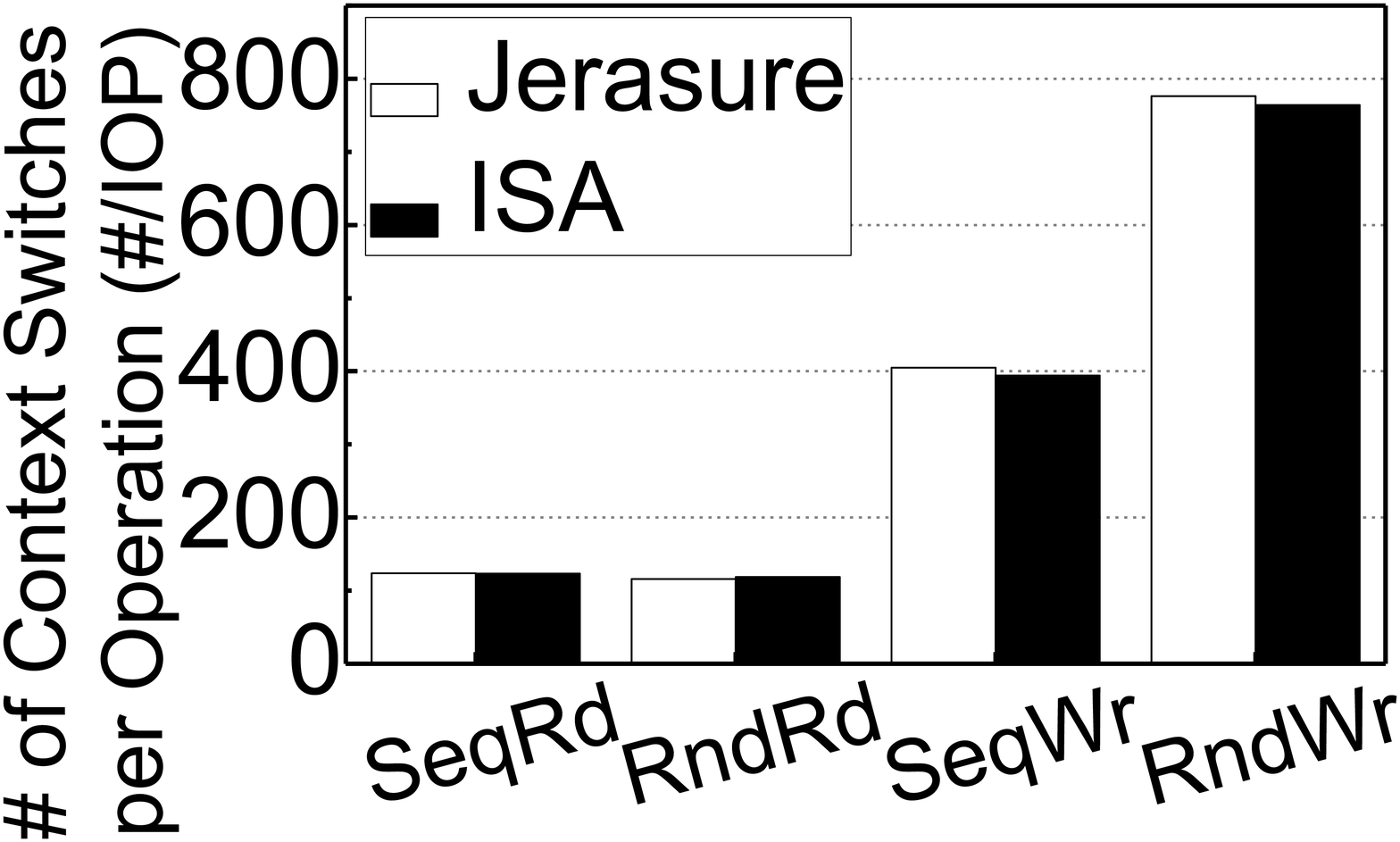}
		\caption{Context switches.}
		\label{fig:isa_ctx}
	\end{subfigure}
	\vspace{-5pt}
	\caption{Computation overheads comparison between Jerasure and ISA.}
	\vspace{-5pt}
	\label{fig:isa_comp}
\end{figure}

In this section we will discuss about the impacts of several Reed-Solomon settings on performance and overheads.

\noindent \textbf{Plugin.} We firstly examined the impact of plugins. Two different libraries implement Reed-Solomon algorithm. These libraries are Jerasure library and Intel Intelligent Storage Acceleration libaray (ISA-L) and these can be adopted to Ceph via Jerasure plugin and ISA plugin. ISA-L only runs on Intel processors while Jerasure is platform independent library. Figure \ref{fig:isa_perf} shows the 4KB read/write performance of the systems with Jerasure and ISA plugin. As we can see from the figure, ISA plugin shows similar performance with both read and write requests. Figure \ref{fig:isa_comp} represents the computation overheads incurred in two different systems during serving 4KB read and write requests. Both systems incur similar amount of CPU usages and the number of context switches per request. This is because the change of the library does not impact overall data path of erasure coding.

\begin{figure}
	\centering
	\begin{subfigure}{0.49\linewidth}
		\includegraphics[width=\linewidth]{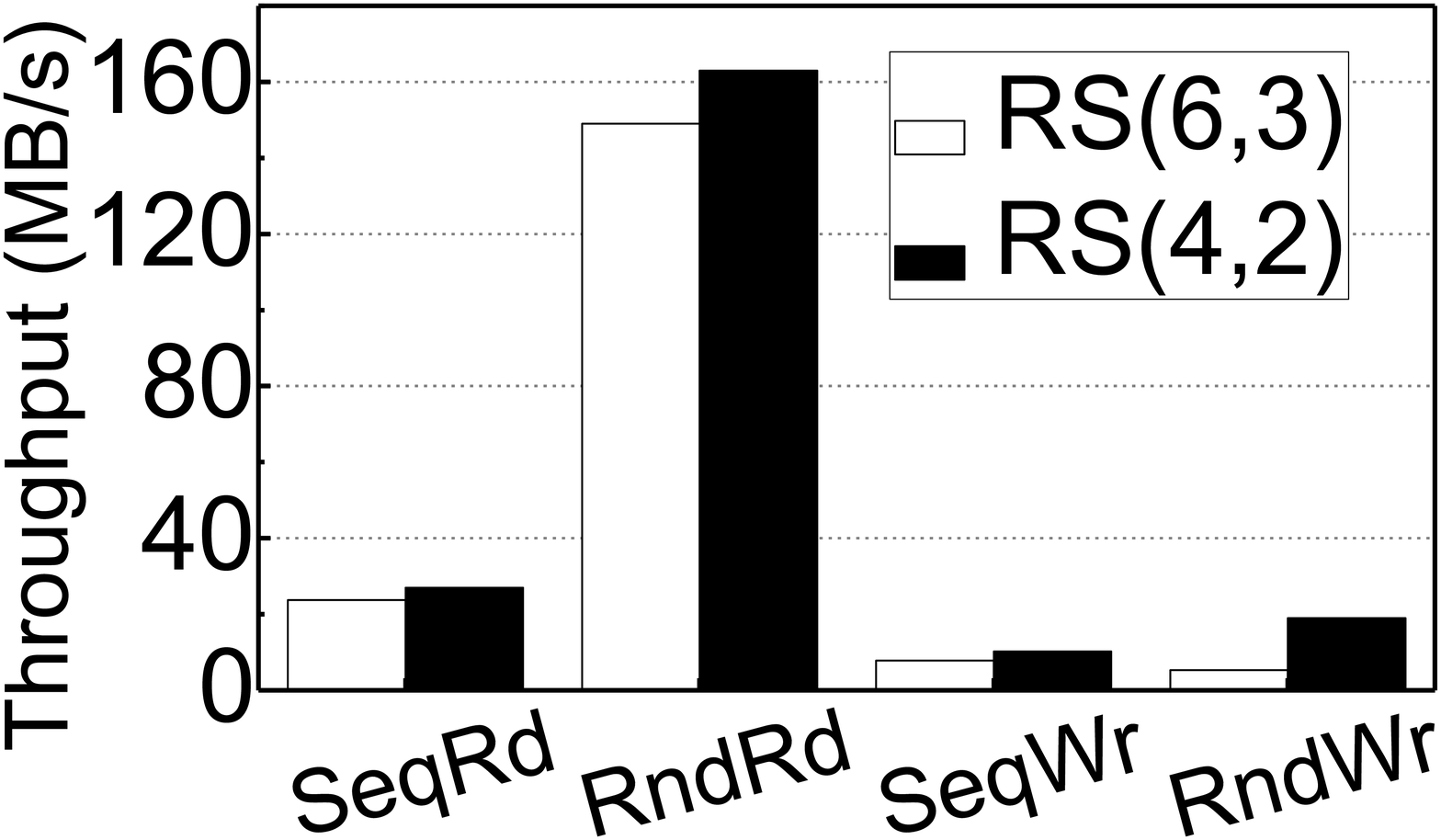}
		\caption{Throughput.}
		\label{fig:42_thr}
	\end{subfigure}
	\begin{subfigure}{0.49\linewidth}
		\includegraphics[width=\linewidth]{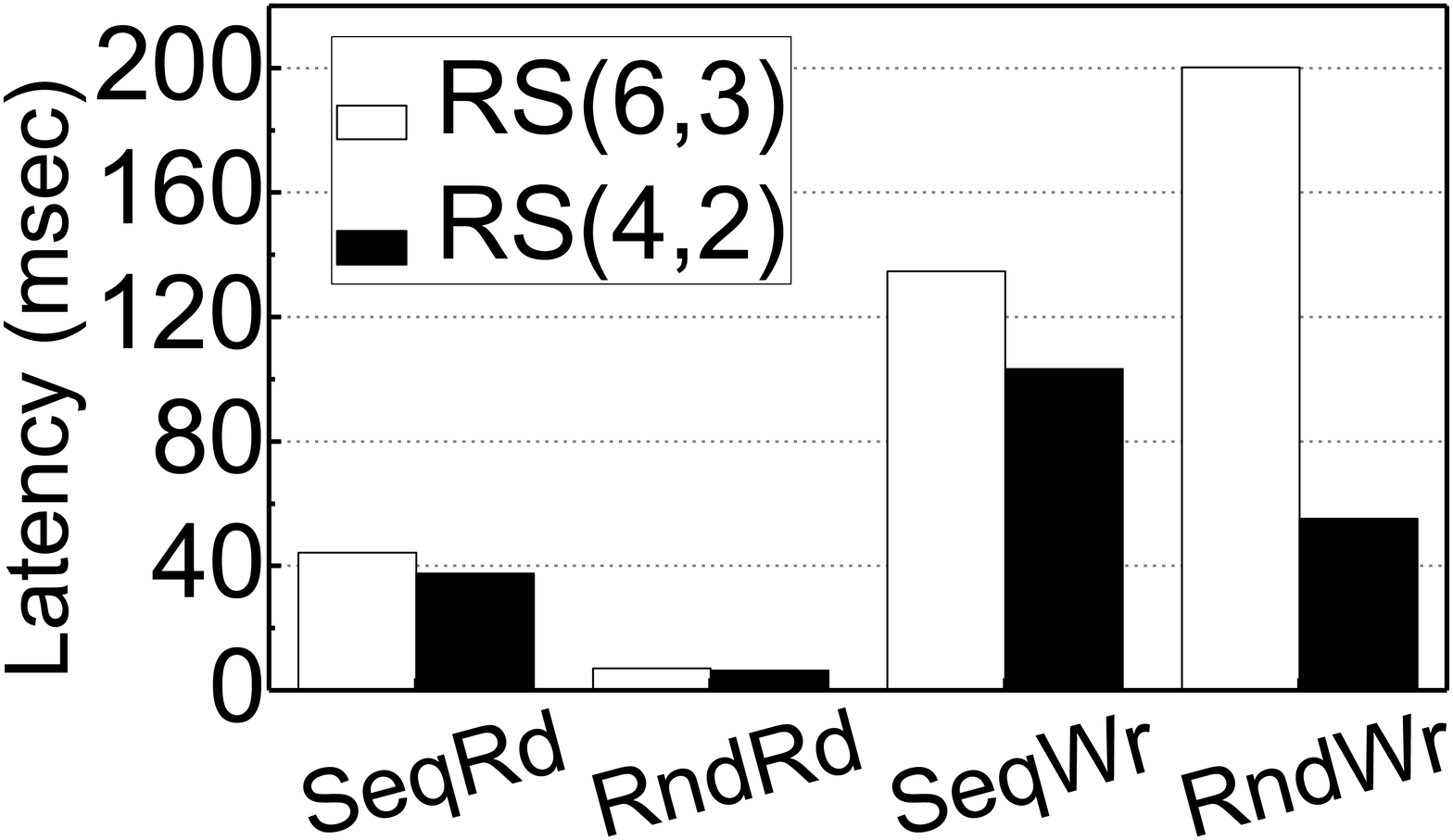}
		\caption{Latency.}
		\label{fig:42_lat}
	\end{subfigure}
	\vspace{-8pt}
	\caption{Performance of two different RS configurations.}
	\vspace{-10pt}
	\label{fig:42_perf}
\end{figure}

\begin{figure}
	\centering
	\begin{minipage}{0.50\linewidth}
	\begin{subfigure}{0.58\linewidth}
		\includegraphics[width=\linewidth]{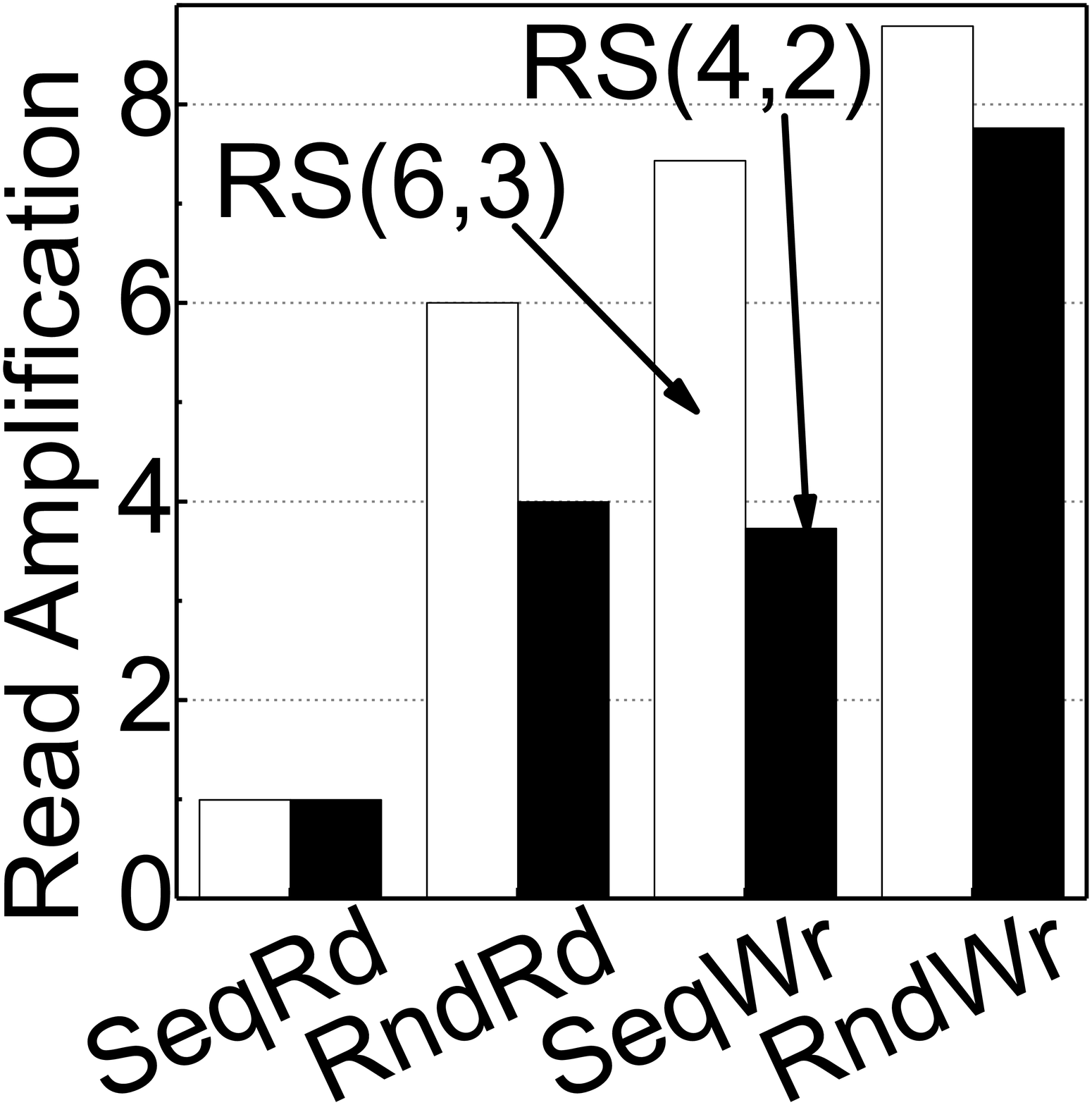}
		\caption{Read.}
		\label{fig:42_read}
	\end{subfigure}
	\begin{subfigure}{0.40\linewidth}
		\includegraphics[width=\linewidth]{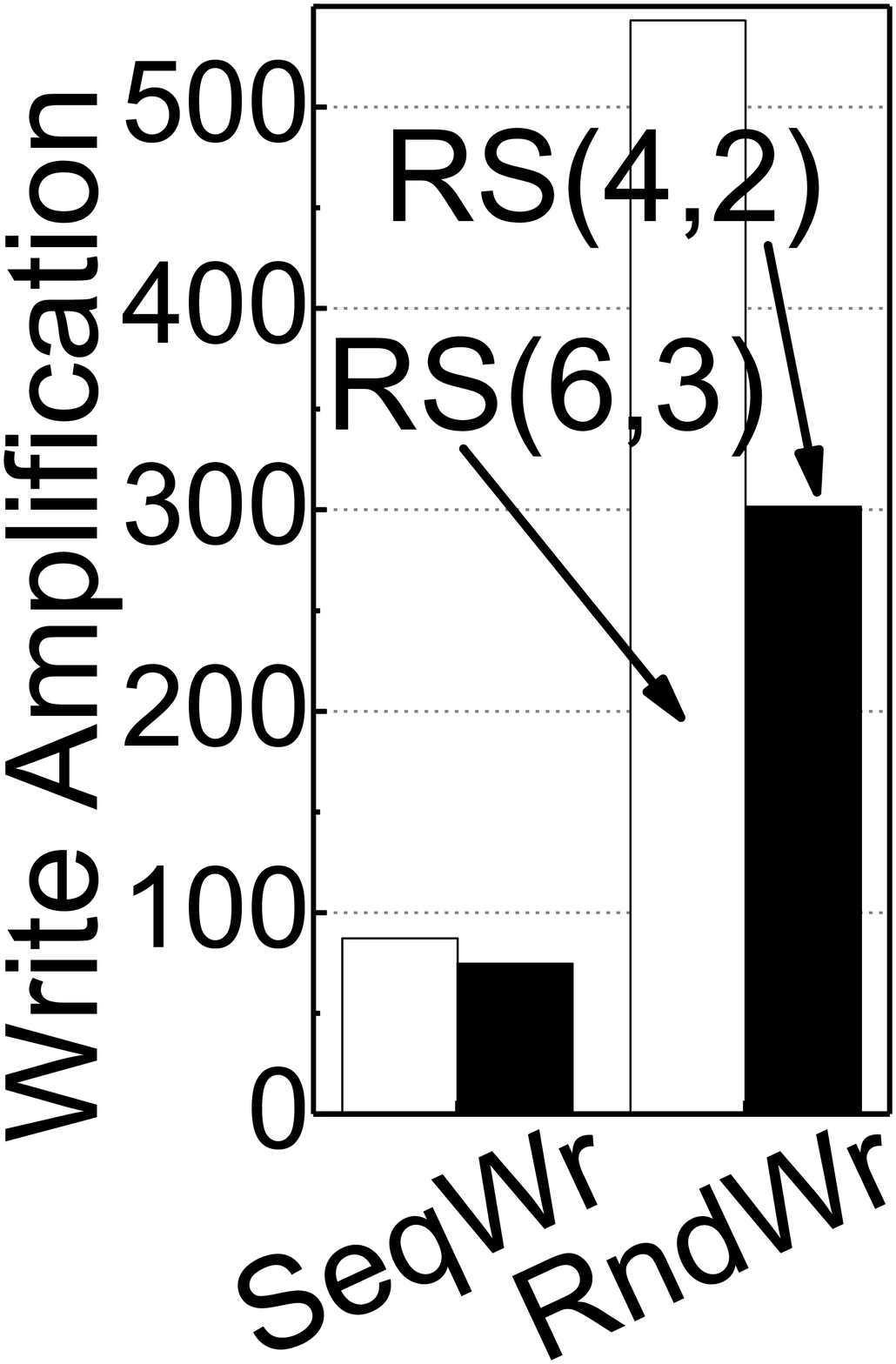}
		\caption{Write.}
		\label{fig:42_write}
	\end{subfigure}
	\vspace{-8pt}
	\caption{I/O Amplification.}
	\label{fig:42_amp}
	\end{minipage}
	\begin{minipage}{0.47\linewidth}
		\begin{subfigure}{\linewidth}
		\includegraphics[width=\linewidth]{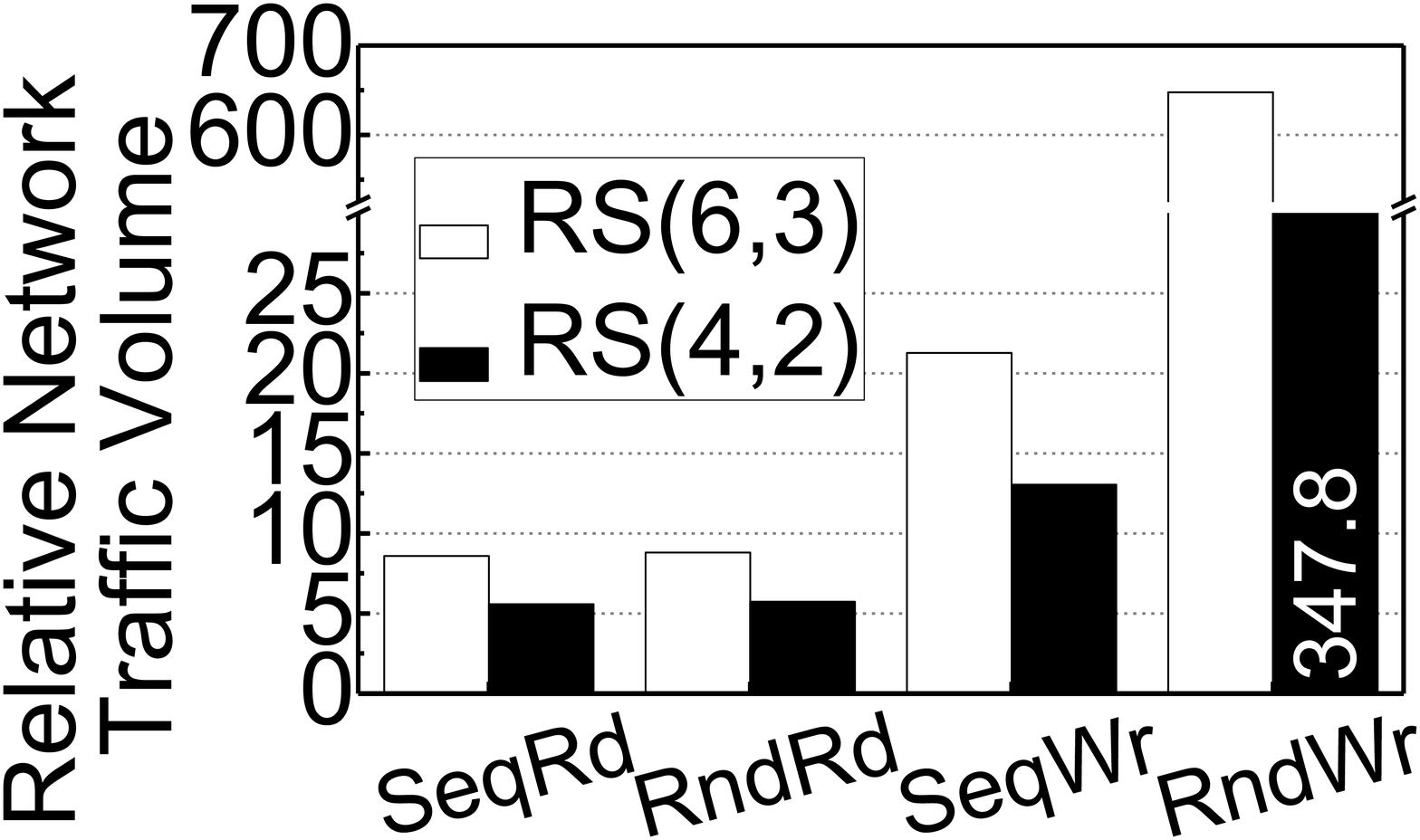}
	\end{subfigure}
	\caption{Network traffic.}
	\label{fig:42_network}
	\end{minipage}
	
	\vspace{-10pt}
\end{figure}

\noindent \textbf{RS Configuration.} Secondly, we examined the impact of $k$ and $m$, which are the main parameters of RS algorithm. We tested RS(4,2) which introduces same storage overheads (i.e., 1.5$\times$). Since RS(4,2) generates fewer chunks than RS(6,3), we can expect that much less data overheads will be incurred, and also that RS(4,2) will show better performance. Figures \ref{fig:42_perf}, \ref{fig:42_amp}, and \ref{fig:42_network} compare the performance and data overheads of RS(6,3) and RS(4,2). As we can see from the figures \ref{fig:42_thr} and \ref{fig:42_lat}, RS(4,2) shows 1.6$\times$ higher throughput and 1.6$\times$ lower latency, on average. In addition, as we expected RS(4,2) incurs 1.4$\times$ lower read amplification, and 1.5$\times$ lower write amplification than RS(6,3), on average. Moreover, 1.6$\times$ lower network traffic is occurred in RS(4,2) than RS(6,3). So we can conclude that RS with smaller $k$ and $m$ will show better performance and incur lower overheads. Note that real world clusters adopts RS(6,3) to achieve the similar or higher reliability level than 3-replication. Therefore, the overheads introduced in RS(6,3) is the minimum amount which must be considered to adopt erasure coding into the real-world systems.

\section{Real Application Usages}
\label{sec:realworkload}
\begin{table*}[]
\centering
\begin{tabular}{|c|c||c||c|c||c|c||c|c|}
\hline
                              & \textbf{Type}           & \textbf{Percentage} & \textbf{Data \%} & \textbf{Metadata \%} & \textbf{Read \%} & \textbf{Write \%} & \textbf{Sequential \%} & \textbf{Random \%} \\ \hline
\multirow{2}{*}{\textbf{DB}}  & \textbf{Table}          & 83.3\%         & 100\%         & 0\%               & 80\%          & 20\%           & 1\%                 & 99\%            \\ \cline{2-9} 
                              & \textbf{Log}            & 16.7\%          & 100\%         & 0\%               & 100\%         & 0\%            & 80\%                & 20\%            \\ \hline
\multicolumn{2}{|c||}{\textbf{VDI}}                      & 100\%          & 99            & 1\%               & 26.3\%        & 73.7\%         & 15.2\%              & 84.8\%          \\ \hline
\multirow{2}{*}{\textbf{EDA}} & \textbf{Frontend}       & 66\%           & 40\%          & 60\%              & 37.5\%        & 62.5\%         & 42.5\%              & 57.5\%          \\ \cline{2-9} 
                              & \textbf{Backend}        & 33\%           & 100\%         & 0\%               & 50\%          & 50\%           & 100\%               & 0\%             \\ \hline
\multirow{2}{*}{\textbf{VDA}} & \textbf{Data stream}    & 90\%           & 100\%         & 0\%               & 0\%           & 100\%          & 100\%               & 0\%             \\ \cline{2-9} 
                              & \textbf{Companion apps} & 10\%           & 91\%          & 9\%               & 98.9\%        & 1.1\%          & 5.5\%               & 94.5\%          \\ \hline
\end{tabular}
\caption{Workload characterization}
\vspace{-15pt}

\label{workload_char}

\end{table*}

In this section, we examine our storage system employing RS(6,3) by executing four different real application usages, and compare the results with the same system using 3-replication. 
Even though analyses of fault-tolerant mechanisms with real application executions are desirable for understanding the aforementioned key observations and insights, the system behaviors can vary according to the order of the application executions and their combinations. Thus, we use SPEC SFS (SP2) suites \cite{spec2014sfs}, which execute real applications with standardized file server usage scenarios. Specifically, in this work, the SPEC SFS captures and mimics four different types of user behaviors with a varying number of CPU processes: i) \emph{Transactional SQL database (DB)}, ii) \emph{Virtual Desktop Infrastructure (VDI)}, iii) \emph{Electronic Design Automation (EDA)} and vi) \emph{Video Data Acquisition (VDA)}.  
For these real workloads, we enable a client cache, which was disabled during FIO test. If cache hit occurs from the client cache, request can be served without asking data for the primary OSD. In this case, several overheads discussed in \ref{sec:syschar} will not be introduced. Therefore, several cache hits can reduce total RS overheads, and it needs to be considered to evaluate erasure coding in real world.

\begin{figure}
	\centering
	\begin{subfigure}{0.49\linewidth}
		\includegraphics[width=\linewidth]{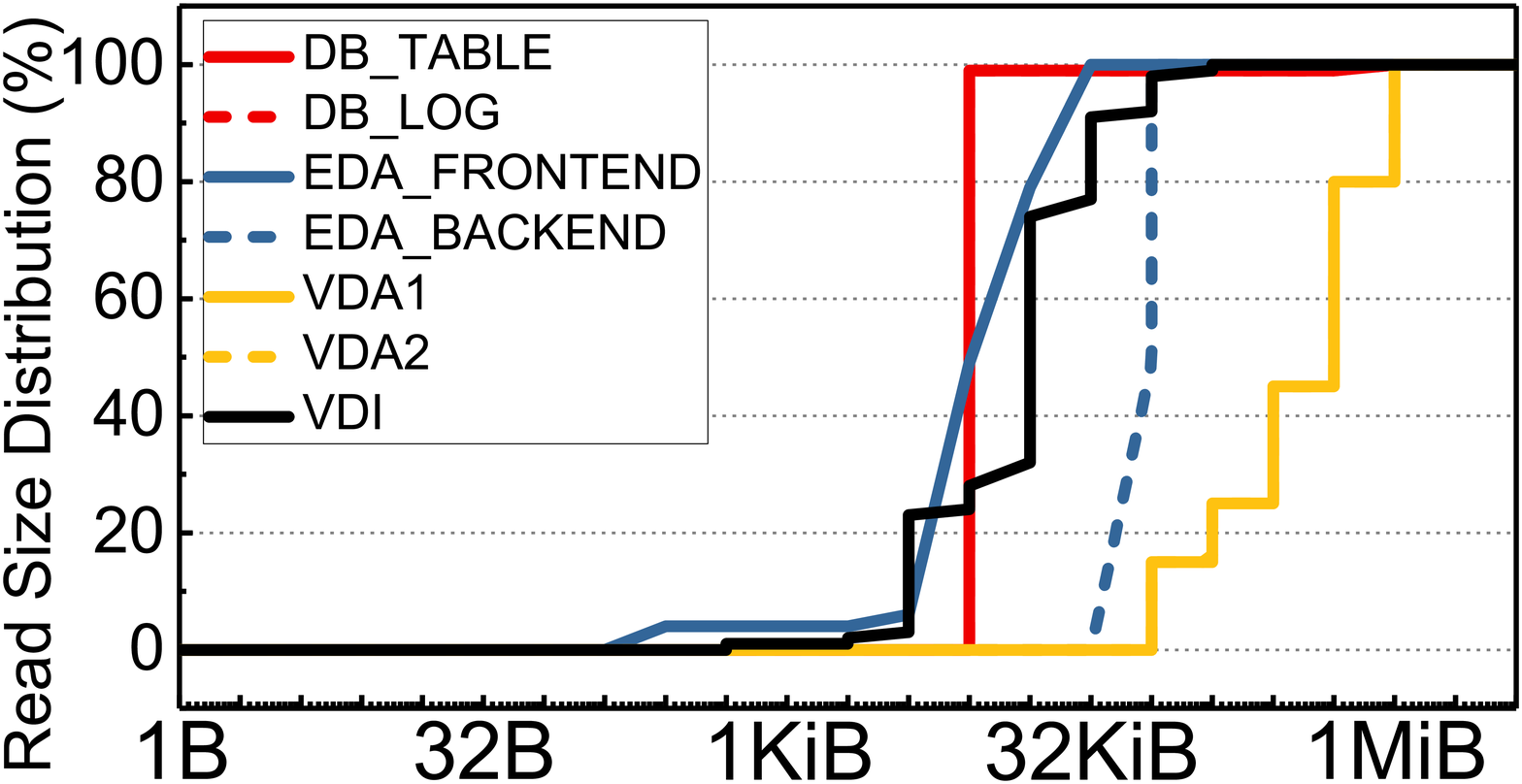}
		\caption{Read.}
		\label{fig:read_dist}
	\end{subfigure}
	\begin{subfigure}{0.49\linewidth}
		\includegraphics[width=\linewidth]{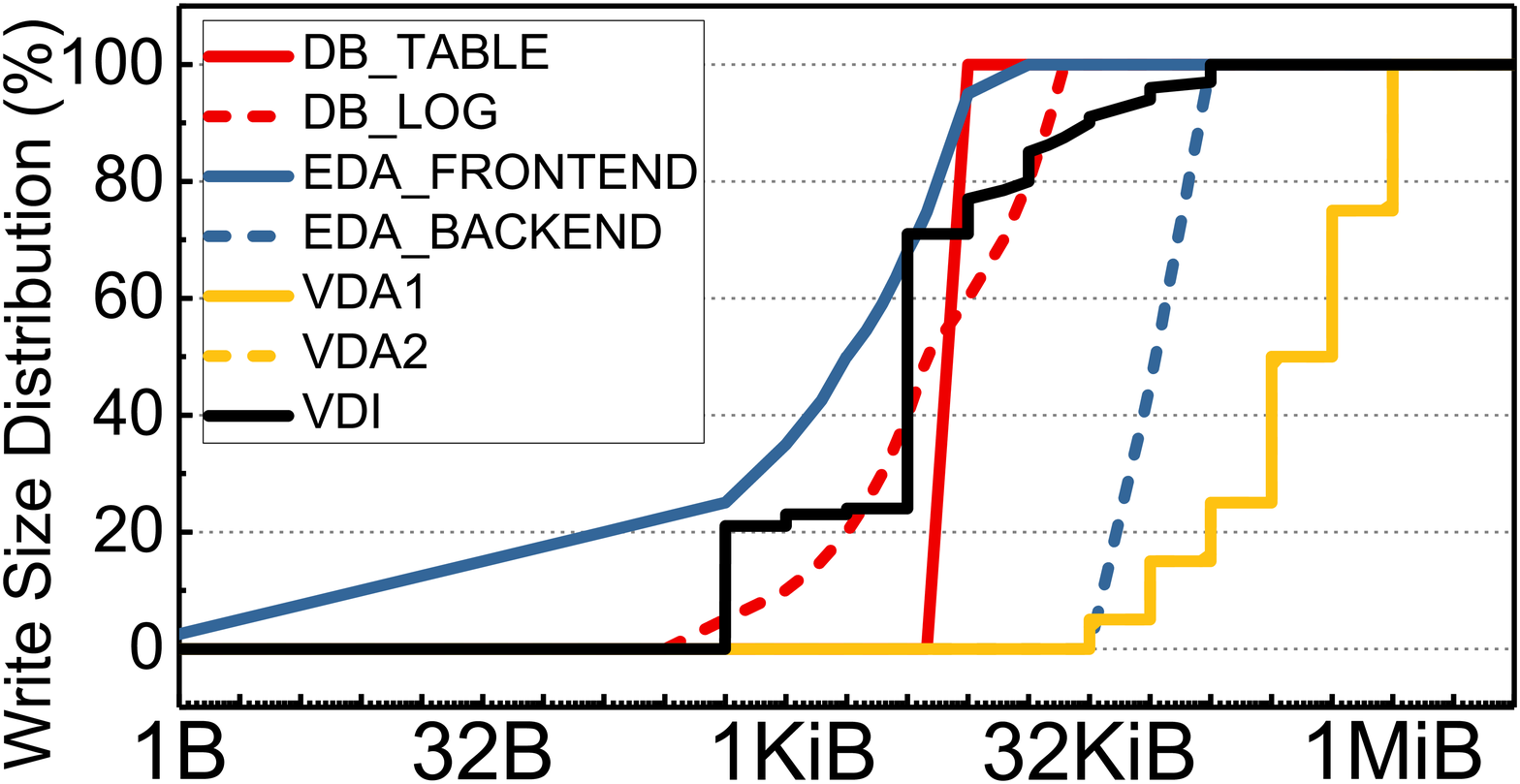}
		\caption{Write.}
		\label{fig:write_dist}
	\end{subfigure}
	\vspace{-5pt}
	\caption{Workload request size distribution.}
	\vspace{-10pt}
	\label{fig:wl_dist}
\end{figure}

\begin{figure}
	\centering
	\begin{subfigure}{0.49\linewidth}
		\includegraphics[width=\linewidth]{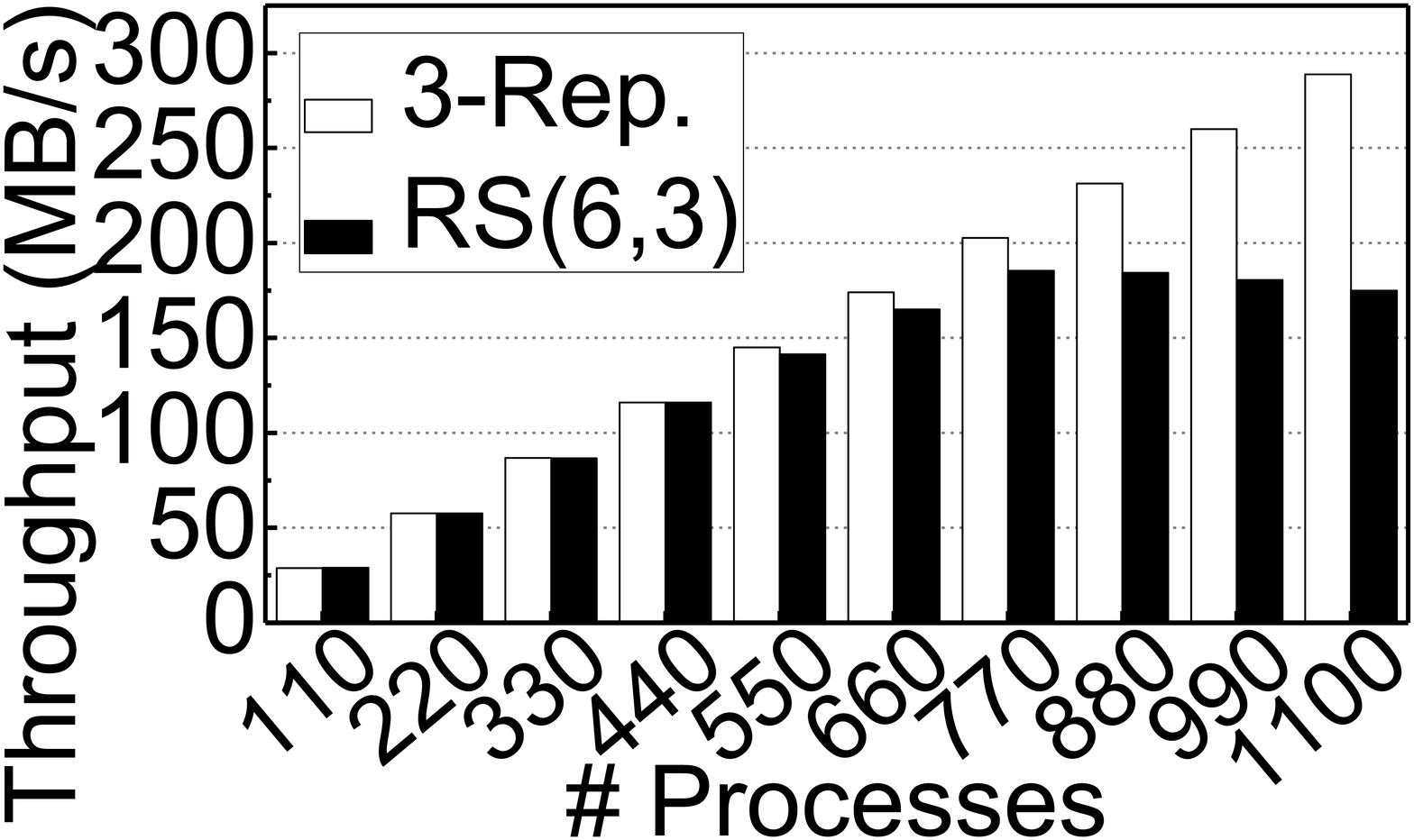}
		\caption{Throughput.}
		\label{fig:db_thr}
	\end{subfigure}
	\begin{subfigure}{0.49\linewidth}
		\includegraphics[width=\linewidth]{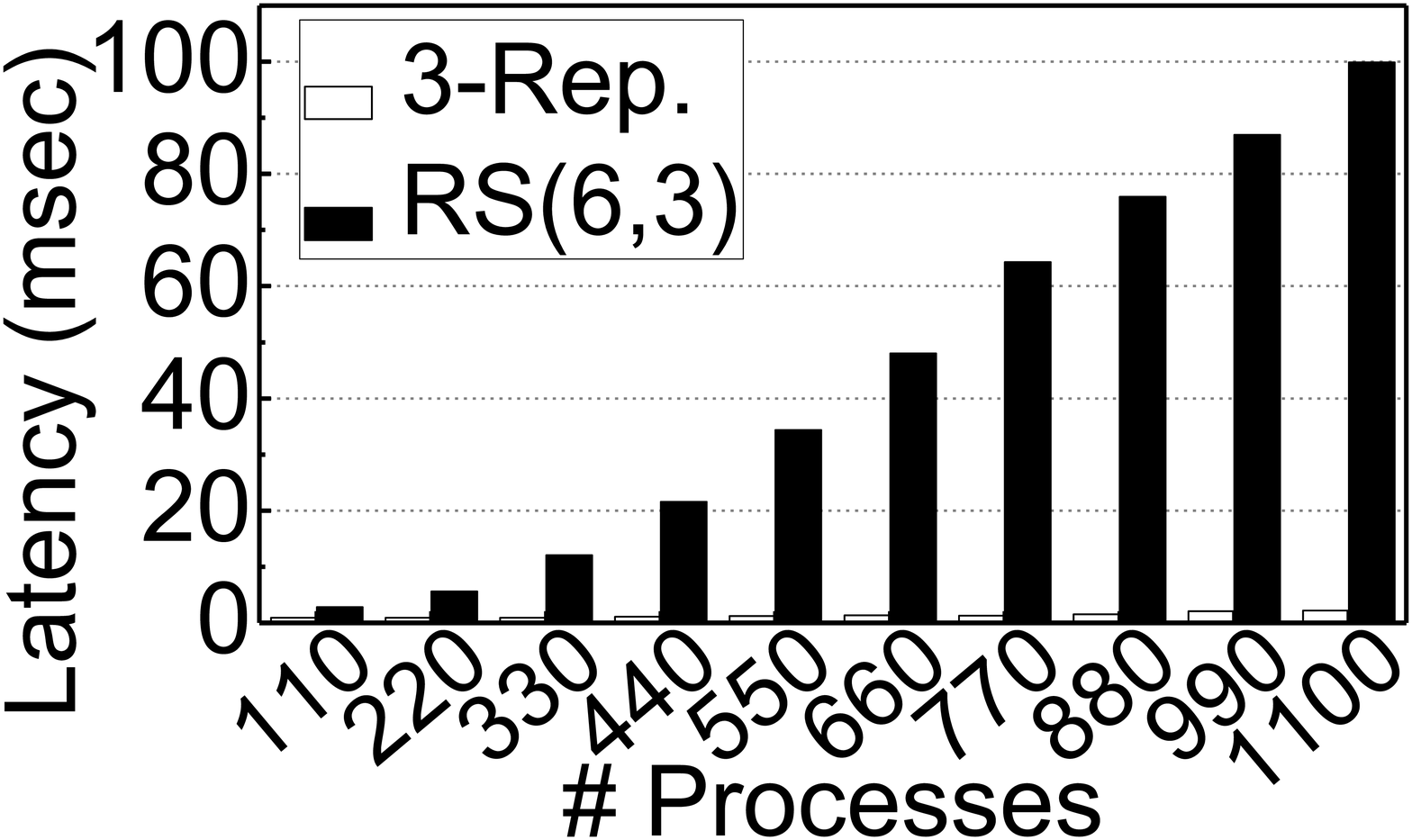}
		\caption{Latency.}
		\label{fig:db_lat}
	\end{subfigure}
	\vspace{-5pt}
	\caption{Performance of DB.}
	\vspace{-10pt}
	\label{fig:db_perf}
\end{figure}

\begin{figure}
	\centering
	\begin{subfigure}{0.49\linewidth}
		\includegraphics[width=\linewidth]{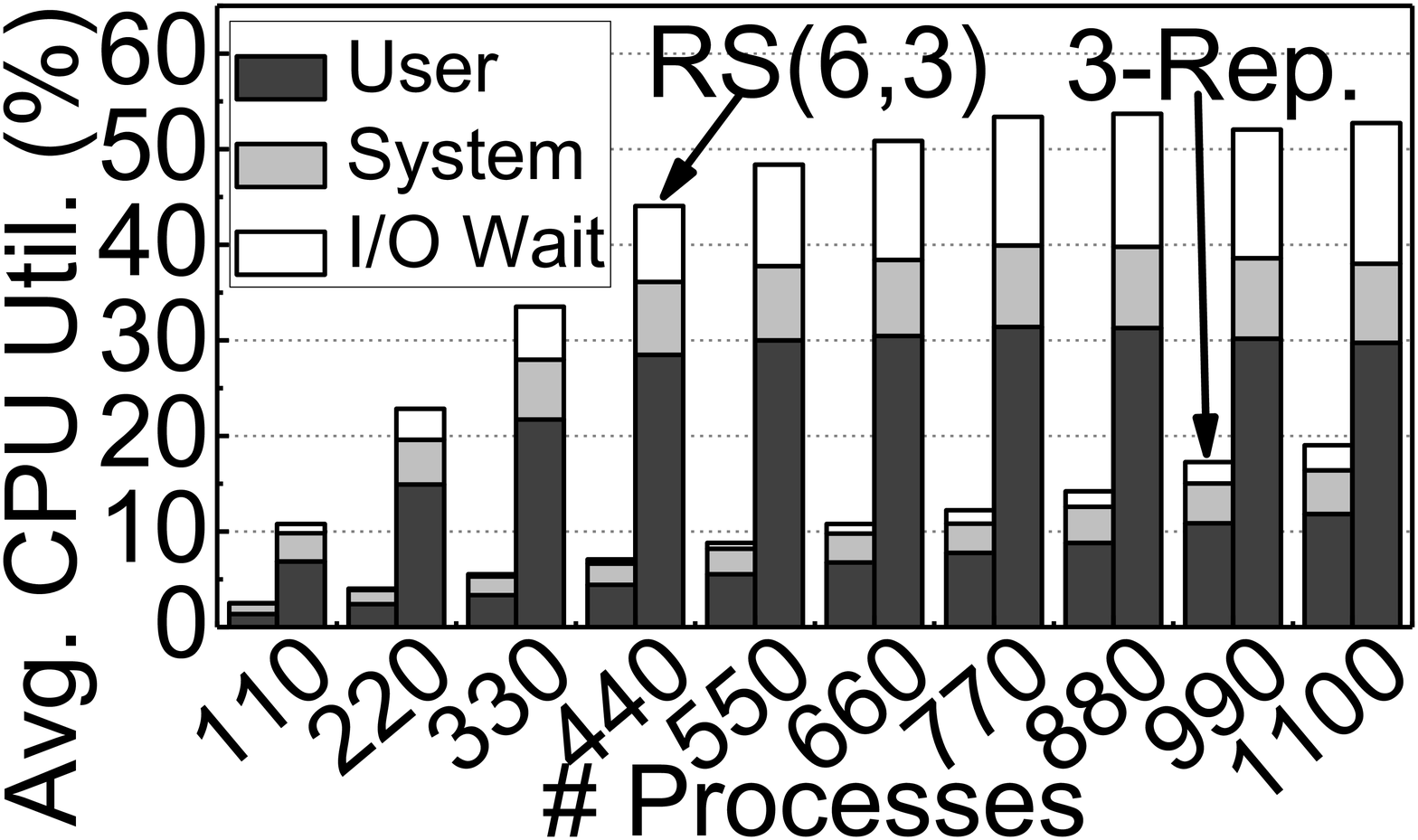}
		\caption{CPU utilization.}
		\label{fig:db_cpu}
	\end{subfigure}
	\begin{subfigure}{0.49\linewidth}
		\includegraphics[width=\linewidth]{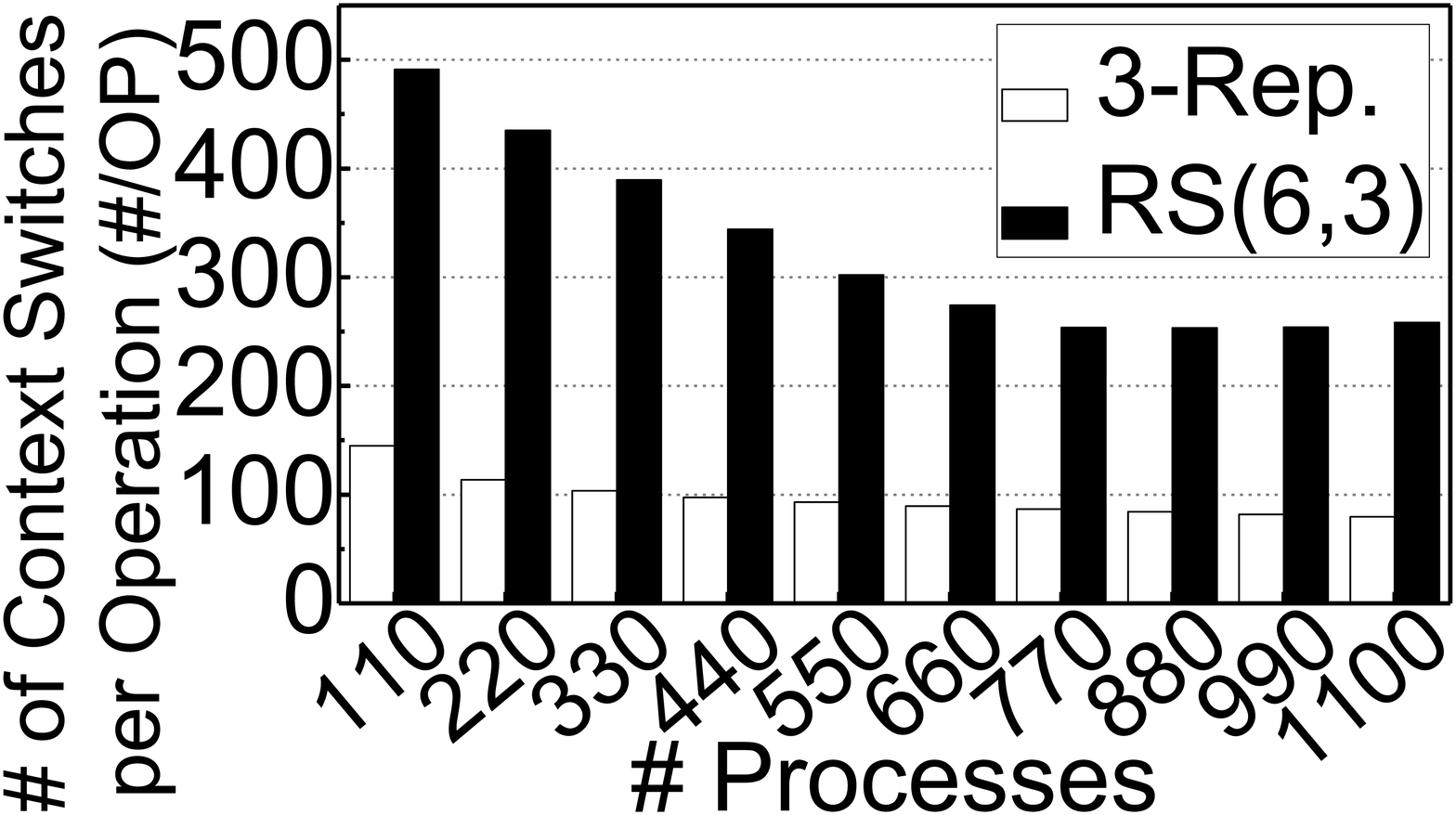}
		\caption{Context switches.}
		\label{fig:db_ctx}
	\end{subfigure}
	\vspace{-5pt}
	\caption{Computation and system overheads in DB.}
	\vspace{-10pt}
	\label{fig:db_comp}
\end{figure}

\subsection{Workload Analysis}

The type of I/O operations for each workload of SPEC SFS can be classified into ``data'' related I/O service and ``metadata'' related I/O requests; thus, we characterize the workloads based on four different viewpoints: i) the ratio of data to metadata, ii) the ratio of reads to writes, iii) the major request size, which is dominant across all request sizes, and iv) the ratio of random to sequential. The important characteristics of i), ii) and iv) are listed in Table \ref{workload_char}, whereas the distribution of request sizes is illustrated in Figure \ref{fig:wl_dist}. Since each workload also contains multiple scenario-specific processes, the table also includes the types of processes (for each workload) and their execution percentage. 
As we discussed in Section \ref{sec:overall}, VDI may degrade the system performance with RS codes more than any other workloads since write services with RS(6,3) cause significantly (encoding) overheads than reads (concatenating). In addition to the performance impact caused by the read/write ratio, the different request sizes affect the performance of RS-enabled systems due to the various I/O amplification values; generally, a large request is better from a performance angle. For example, as shown in Figure \ref{fig:wl_dist}, the dominant request size for VDA is greater than 512KB, which degrades the system performance less than other workloads. In this section, we analyze these in detail.

\subsection{Database (DB)}
\label{db}
DB consists of two types of user process, which are associated with the DB table and DB log. One process makes requests for database table, and another process writes logs to guarantee the SQL transaction. 

\noindent\textbf{Throughput and latency.} 
Figures \ref{fig:db_thr} and \ref{fig:db_lat} compare the throughput and latency, respectively, achieved by executing DB with varying numbers of process from 110 to 1100.  
Since DB has no metadata, we observe that every operation with RS(6,3) causes serious performance degradation and system-level overheads. Specifically, RS(6,3) degrades the throughput 1.2$\times$ more than 3-replication, on average, and RS(6,3) exhibits a 29.3$\times$ longer average latency compared with 3-replication. As the size of most requests is small (8KB), the overheads of RS(6,3) yield a remarkable latency increase. As shown in Figure \ref{fig:db_lat}, while the latency increase is observed from the small number of process executions (110), the throughput difference between RS(6,3) and 3-replication is not notable until after 440 processes. This is because I/O requests generated by such processes are not intensive, thereby exhibiting minor computation overheads due to the RS codes. However, the throughput of RS(6,3) degrades by having 550 process executions owing to its heavy computation loads at the user-level. 

Note that, even though the throughput difference between RS(6,3) and 3-replication with a small number of process executions may not be a major issue in SSD arrays, considering that DB applications in practice are latency critical, we believe that future systems that employ RS(6,3) should be optimized and remove the burdens of system calls, RS codes related computation, and network transfer overheads, which are notably observed for each DB transaction.

\begin{figure}
	\centering
	\begin{subfigure}{0.49\linewidth}
		\includegraphics[width=\linewidth]{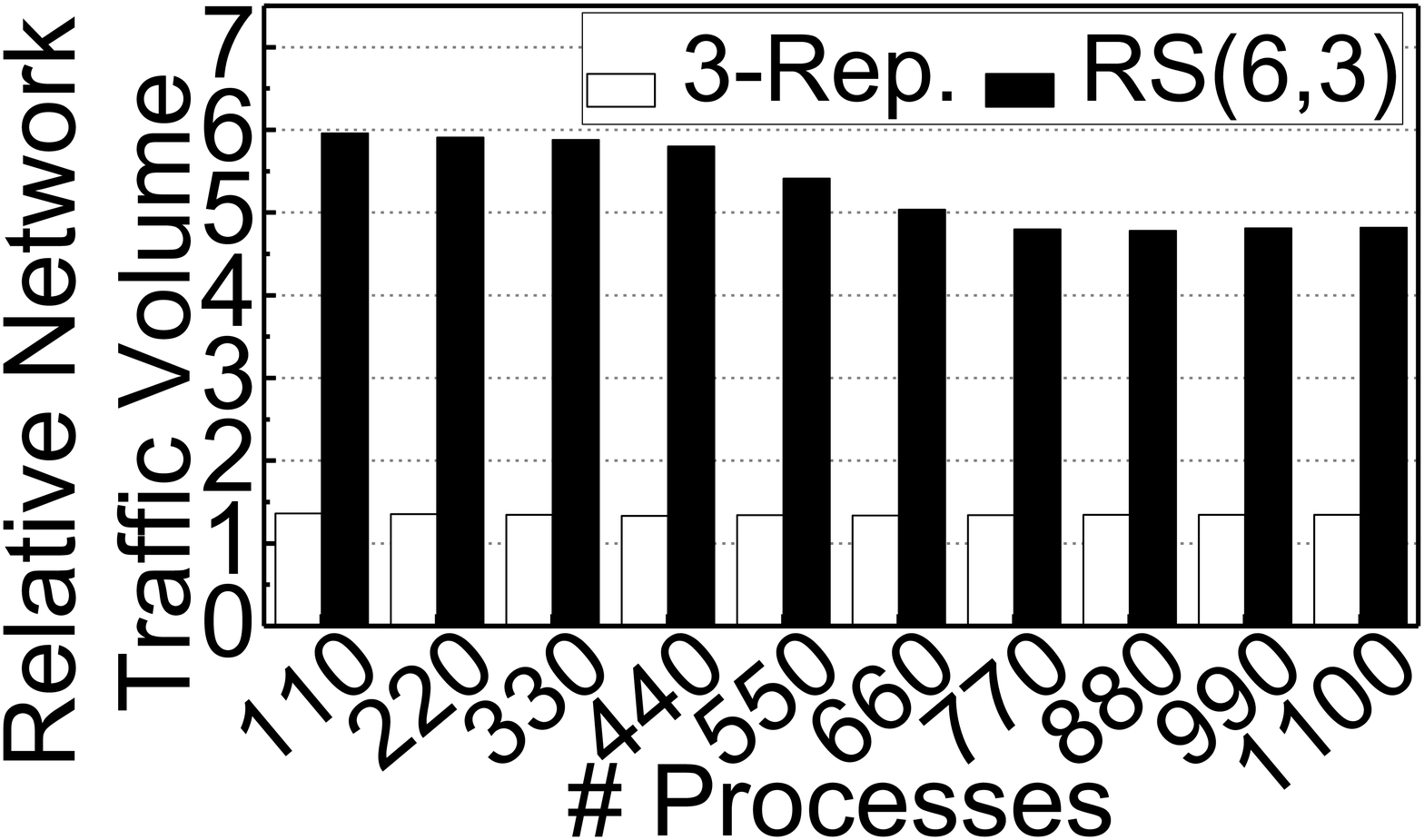}
		\caption{Network traffic.}
		\label{fig:db_network}
	\end{subfigure}
	\begin{subfigure}{0.49\linewidth}
		\includegraphics[width=\linewidth]{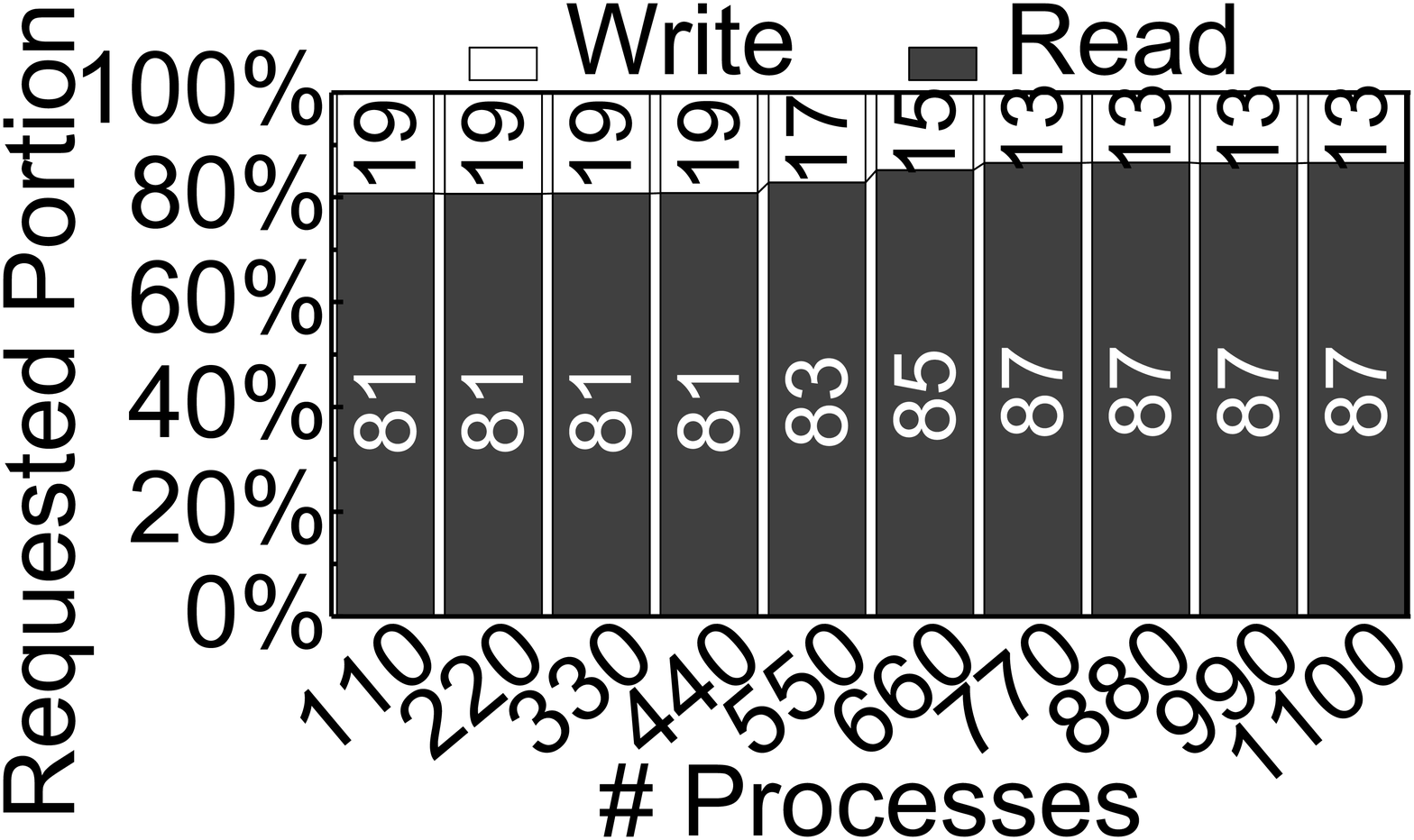}
		\caption{Read/write portion.}
		\label{fig:db_portion}
	\end{subfigure}
	\vspace{-5pt}
	\caption{Network traffic and read/write portion of DB.}
	\vspace{-10pt}
\end{figure}

\noindent\textbf{CPU utilization and context switches.} Figures \ref{fig:db_cpu} and \ref{fig:db_ctx} show the results of CPU utilization and context switch overheads, which correspond to the aforementioned performance comparison analysis. The CPU utilization analysis also decomposes the system CPU usage into user-level process cycles, system (privilege mode) process cycles, and the cycles consumed for I/O waiting (due to a delay). Unlike the throughput difference, one can see from these figures that RS(6,3) consumes 4.6$\times$ more CPU cycles and exhibits 3.3$\times$ more context switches (per operation) than 3-replication. As the number of the processes increases, the CPU utilization of RS(6,3) increases by 4.9$\times$, which also significantly impacts the server power consumption behaviors. Specifically, the CPU cycles consumed by the user level and by waiting increase to 29.8\%, and 14.7\% of the total execution, respectively. This cycle increase occurs because the requests generated by client and user-level processes must go through system calls on the storage stack distributed across different nodes and user-level functions thereby becoming the bottleneck, which is not observed for 3-replication, which simply clones the data. In contrast to the observations of CPU utilization analysis, the number of RS(6,3) context switches per operation decreases as the number of processes increases. We believe that this can be explained by the read/write ratio analysis (cf. Figure \ref{fig:db_portion}). As the number of processes increases, the total amount of reads, which impose less context switches than writes, increases, and therefore, less context switches are required on average. 

\begin{figure}
	\centering
	\begin{subfigure}{0.49\linewidth}
		\includegraphics[width=\linewidth]{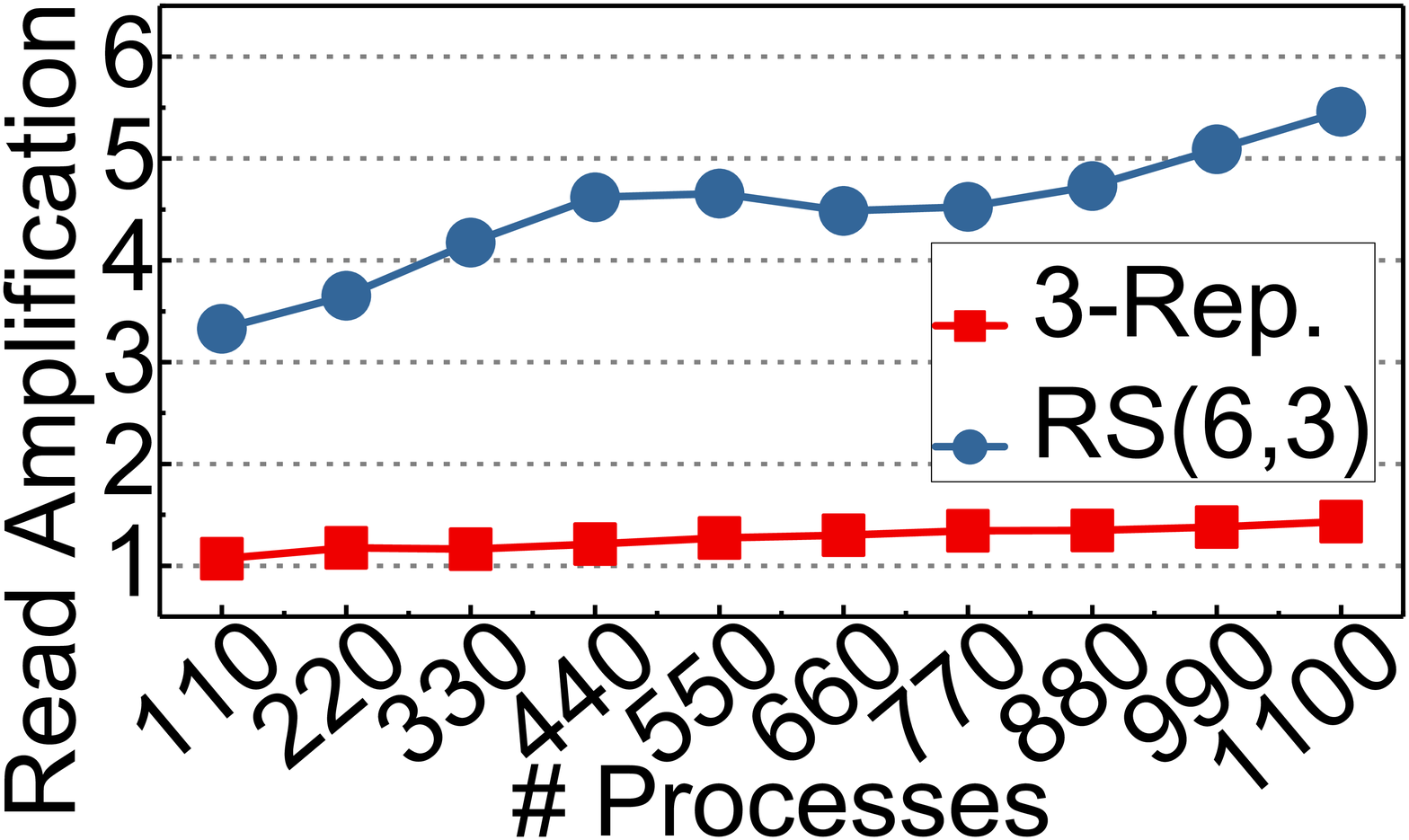}
		\caption{Read amplification.}
		\label{fig:db_read}
	\end{subfigure}
	\begin{subfigure}{0.49\linewidth}
		\includegraphics[width=\linewidth]{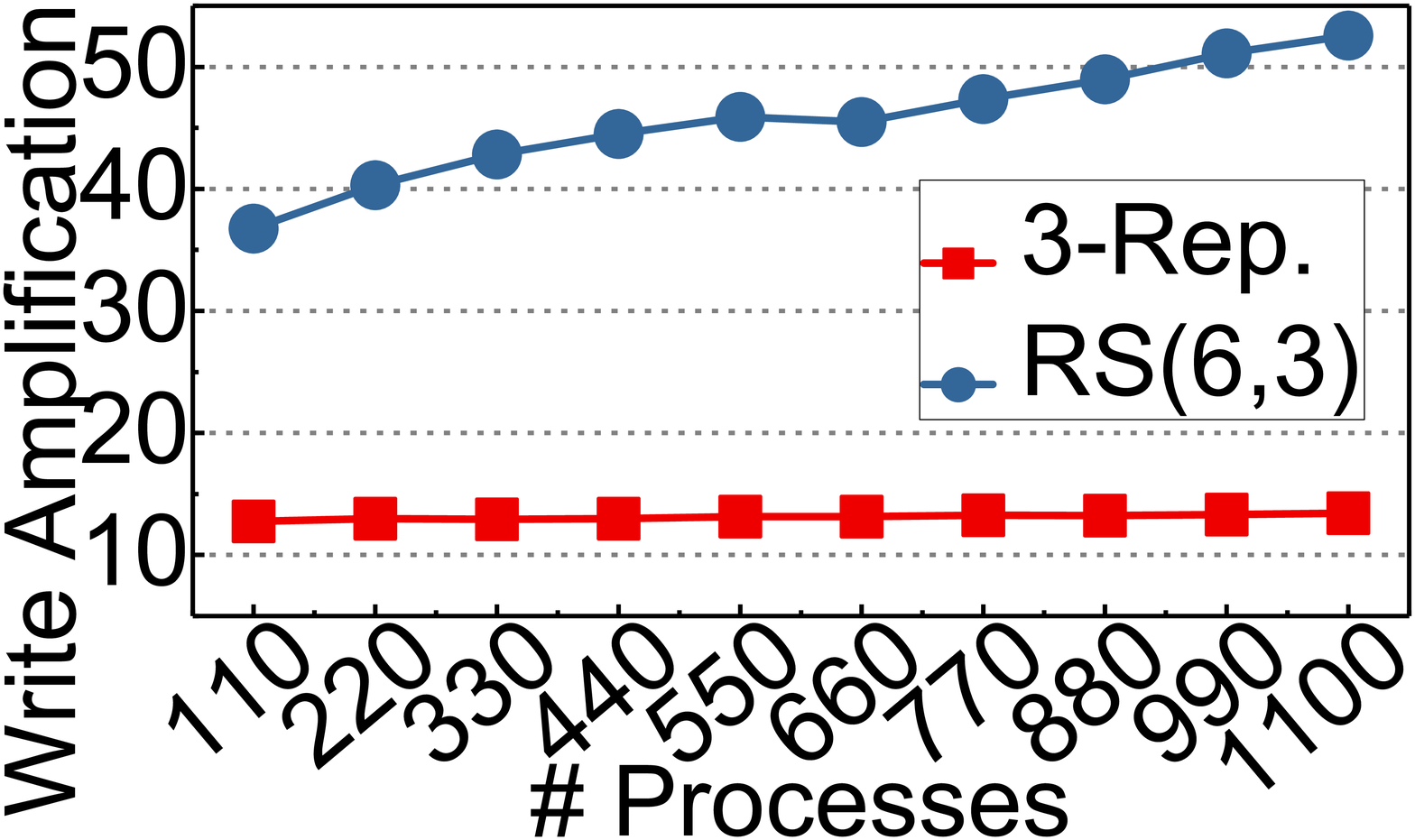}
		\caption{Write amplification.}
		\label{fig:db_write}
	\end{subfigure}
	\vspace{-5pt}
	\caption{I/O amplification in DB.}
	\vspace{-10pt}
	\label{fig:db_io}
\end{figure}

\noindent\textbf{Network traffic and I/O amplification.} Figures \ref{fig:db_network} and \ref{fig:db_io} analyze the RS(6,3) overheads of network traffic and I/O amplification, respectively. As shown in the network overhead analysis, DB with RS(6,3) results in 4$\times$ more network traffic volume (per MB) than 3-replication. These trends are similar to the overhead characteristics observed for synthetic I/O-intensive workloads in Section \ref{sec:dataoverhead}, but are not as significant as the I/O-intensive workload evaluations. This is because all the data of transactional SQL database were initialized during the pre-evaluation warm up phase, which is not included in the main evaluation phase; therefore, every write is more likely to update an object than to initialize. On the other hand, since the major request size of DB is 8KB, RS(6,3) encoding and RS-concatenation, which exhibit mostly random reads and random writes, introduce 3.5$\times$ more read amplification and 3.5$\times$ more write amplification than 3-replication, respectively.

\subsection{Virtual Desktop Infrastructure (VDI)}
\label{vdi}
VDI runs multiple full clone virtual machines to capture diverse user scenarios of virtual desktop infrastructure. As its virtual machine is built upon cloning, all the instances are executed as shared-nothing structure in VDI. Since 99\% of total operations deal with data and, only 1\% deal with metadata, one can expect severe performance degradation and significant overheads imposed by RS(6,3) compare with DB.

\begin{figure}
	\centering
	\begin{subfigure}{0.49\linewidth}
		\includegraphics[width=\linewidth]{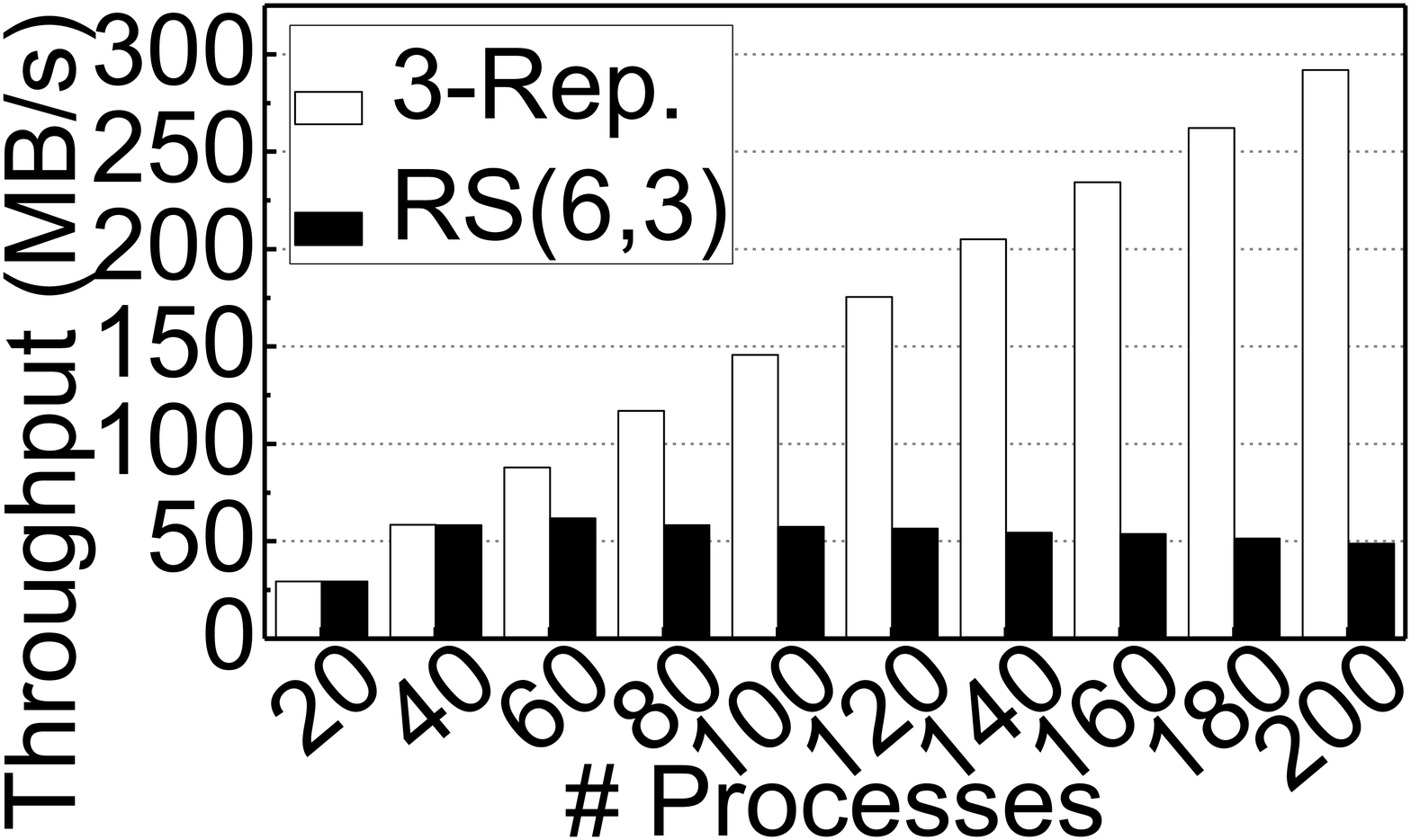}
		\caption{Throughput.}
		\label{fig:vdi_thr}
	\end{subfigure}
	\begin{subfigure}{0.49\linewidth}
		\includegraphics[width=\linewidth]{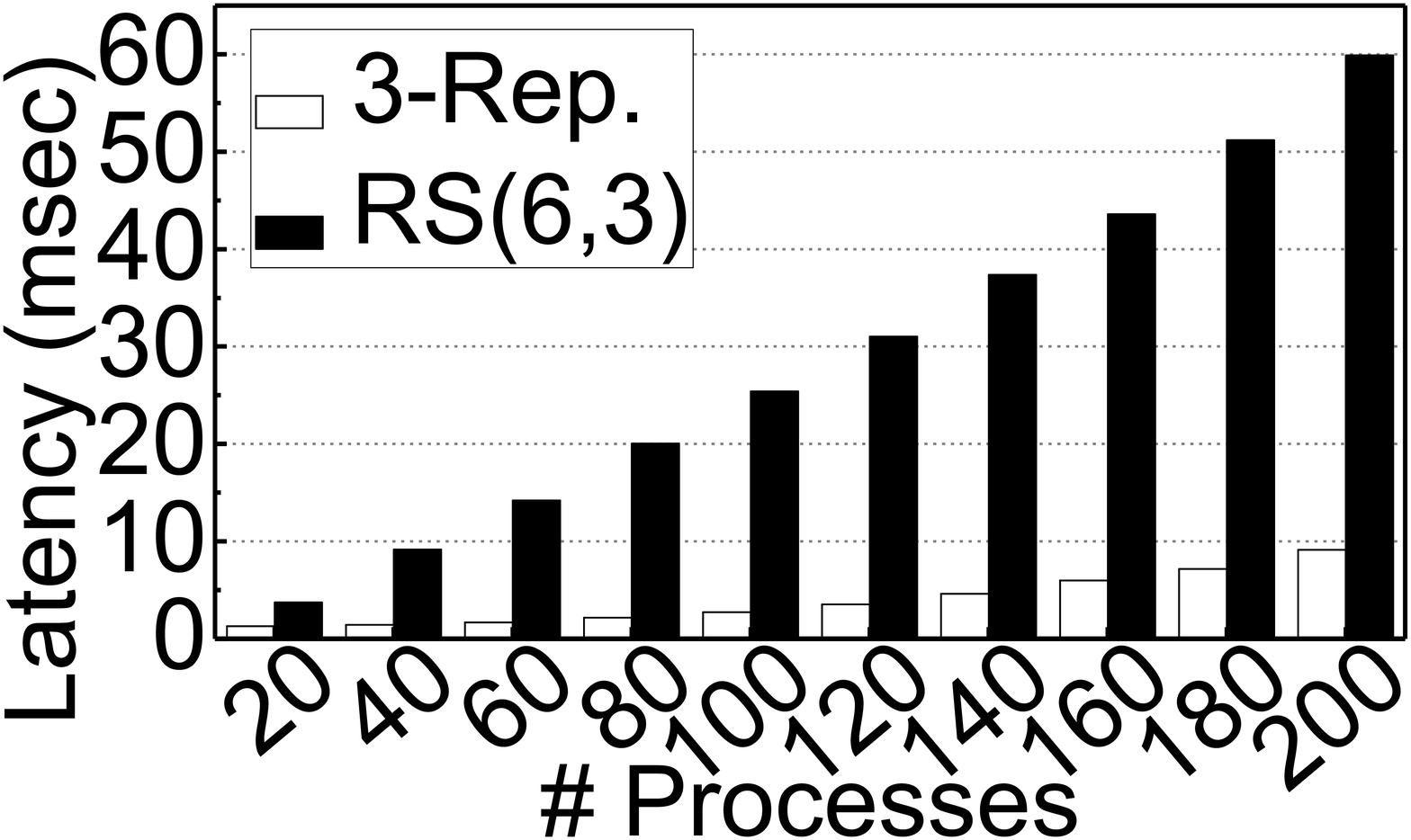}
		\caption{Latency.}
		\label{fig:vdi_lat}
	\end{subfigure}
	\vspace{-5pt}
	\caption{Performance of VDI.}
	\vspace{-10pt}
	\label{fig:vdi_perf}
\end{figure}

\noindent\textbf{Throughput and latency.} Figures \ref{fig:vdi_thr} and \ref{fig:vdi_lat} illustrate the VDI performance with a varying number of processes in terms of the throughput and latency, respectively. As expected, the RS(6,3) throughput decreases by 6$\times$ and its latency increases by 9.3$\times$ compare with 3-replication. These performance degradations are well observed in our synthesized workload evaluations (cf. Section \ref{sec:overall}). We observed that the VDI I/O operations are mostly related to writes whose major request size is 4KB, which incurs greater computation, network, and I/O application burdens. Note that 4KB random writes (discussed in Section \ref{sec:overall}) make the throughput and latency of the RS-code enabled system 18$\times$ worse than those of the system employing 3-replication. The only exception to this trend is that throughput with 20$\sim$40 processes exhibit no significant performance degradation (even though it mostly comprises writes). This is because RS(6,3), which we apply together with Jerasure and Bluestore has internal page caches whose default size is 3GB per SSD. While most reads of DB have less benefits (as the actual data is stored on the underlying SSD), writes operations of VDI can be buffered, thereby hiding most of the SSD access latency in cases of a small number of process executions. However, this buffer cache is not the ideal optimization for addressing the overheads of RS codes, for three reasons. First, such throughput benefits of buffering readily disappear as the number of processes running on the system increases. As shown in Figure \ref{fig:vdi_perf}, regardless of the throughput, the latency is high, and the page cache (of Bluestore) cannot remove the overheads, as we will explain shortly (CPU, system calls, network, etc.).   

\begin{figure}
	\centering
	\begin{subfigure}{0.49\linewidth}
		\includegraphics[width=\linewidth]{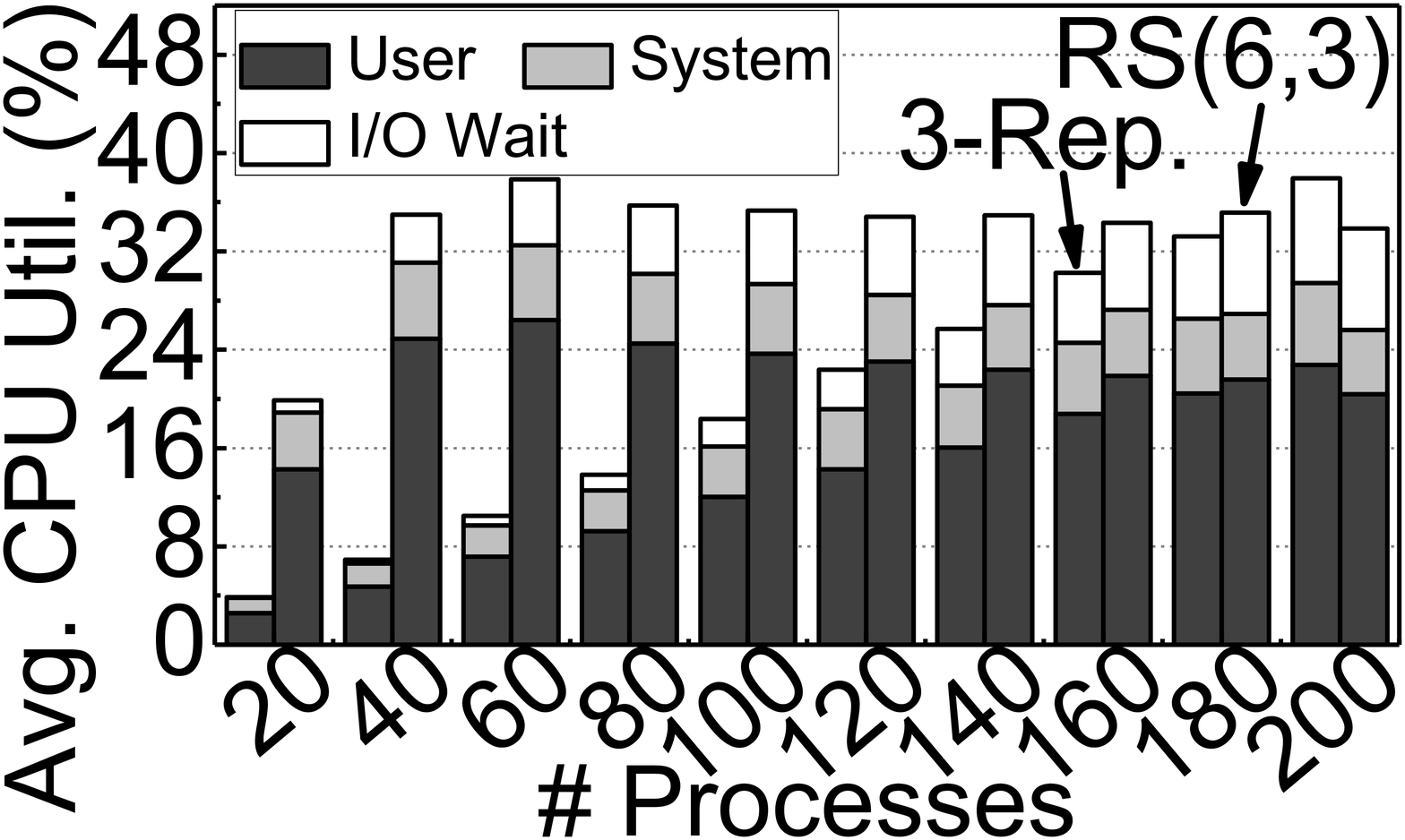}
		\caption{CPU utilization.}
		\label{fig:vdi_cpu}
	\end{subfigure}
	\begin{subfigure}{0.49\linewidth}
		\includegraphics[width=\linewidth]{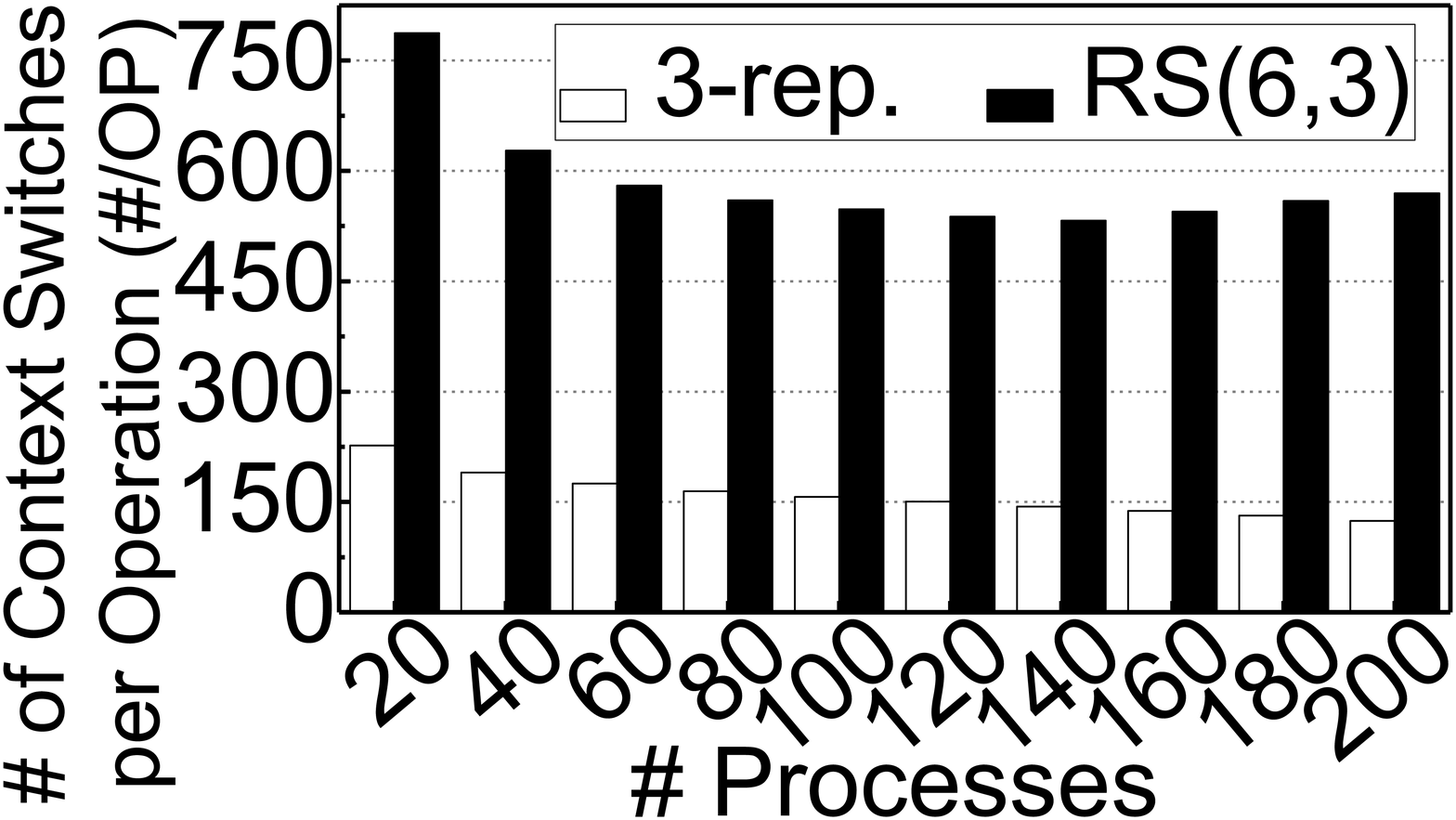}
		\caption{Context switches.}
		\label{fig:vdi_ctx}
	\end{subfigure}
	\vspace{-5pt}
	\caption{Computation and system overheads in VDI.}
	\vspace{-10pt}
	\label{fig:vdi_comp}
\end{figure}

\noindent\textbf{CPU utilization and context switches.} Figure \ref{fig:vdi_cpu} and \ref{fig:vdi_ctx} show the CPU utilization and context switching overheads in detail for the system used for the aforementioned performance analysis. Even though the VDI throughput differences between RS(6,3) and 3-replication (for 20/40 process execution) are insignificant owing to the OSD side page cache, RS(6,3) requires 5.1$\times$ more CPU cycles than 3-replication and introduces 3.4$\times$ more context switches per operation, on average. Interestingly, 3-replication consumes CPU cycles as the number of process executions increase; 3-replication exhibits similar CPU cycle overheads after 160 processes and in turn requires 11\% more cycles than RS(6,3) for executing 200 processes. We believe that this is because the replicated data are also spread across different OSDs within a node and over the other nodes, and thus, CPU requirements for copying the actual data proportionally increase. While RS(6,3) benefits from 200 process execution from the CPU utilization viewpoint, the number of context switches (per operation) is greater than that for 3-replication. Considering the slow performance of RS(6,3), the amount of CPU cycles that the entire RS(6,3) execution of 200 VDI processes requires is greater than that for 3-replication; 3-replication can finish its tasks significantly earlier than RS(6,3).

\begin{figure}
	\centering
	\begin{subfigure}{0.49\linewidth}
		\includegraphics[width=\linewidth]{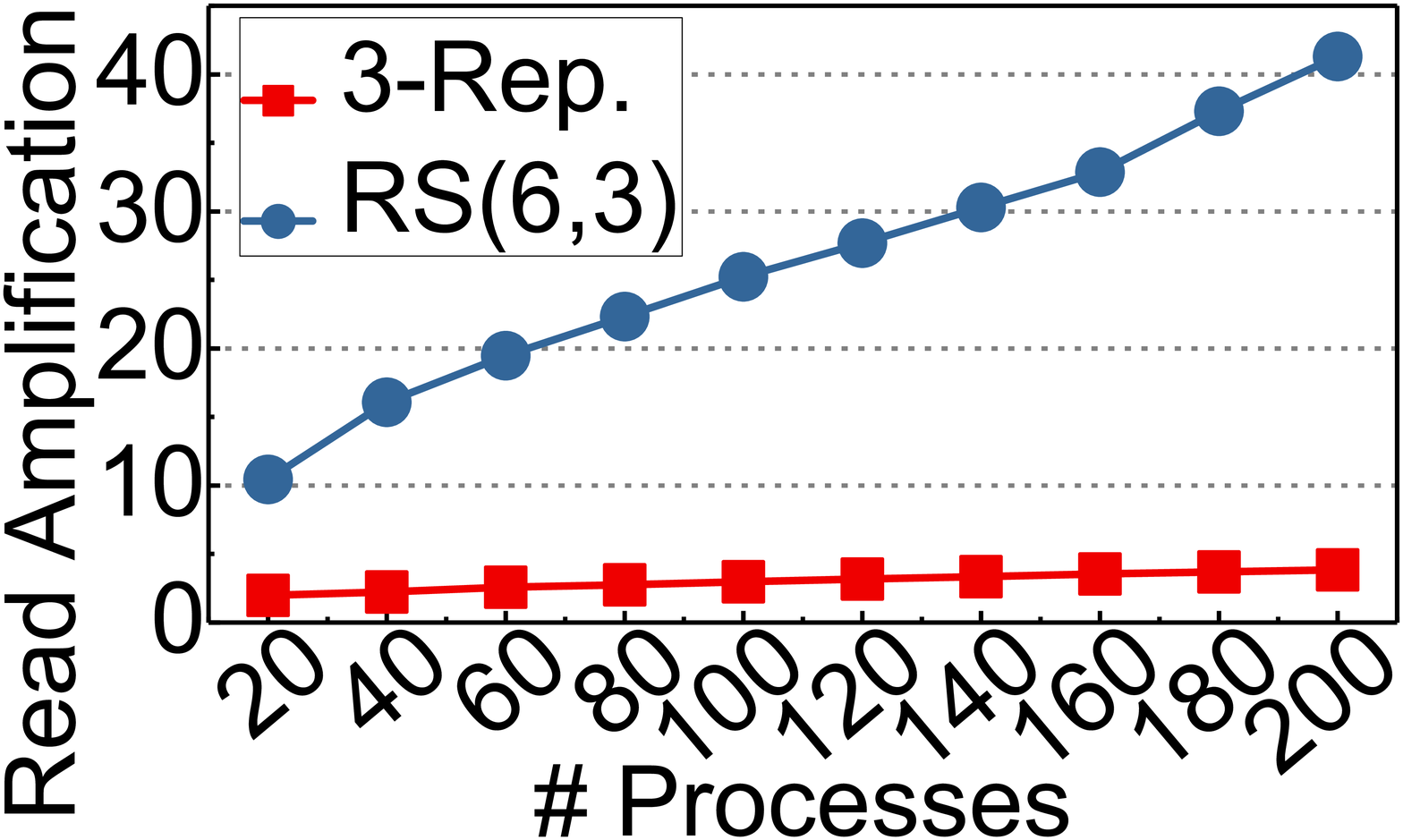}
		\caption{Read amplification.}
		\label{fig:vdi_read}
	\end{subfigure}
	\begin{subfigure}{0.49\linewidth}
		\includegraphics[width=\linewidth]{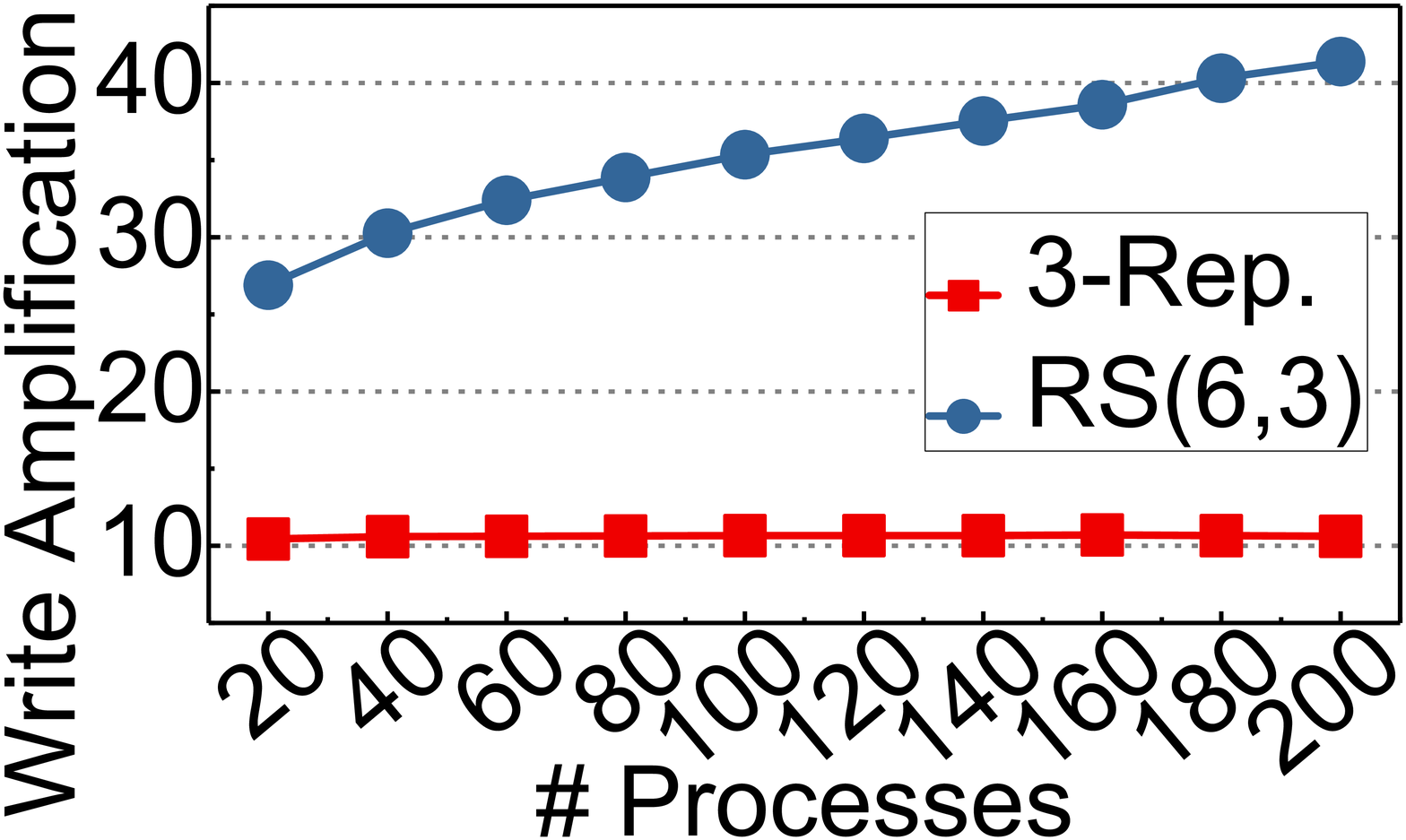}
		\caption{Write amplification.}
		\label{fig:vdi_write}
	\end{subfigure}
	\vspace{-5pt}
	\caption{I/O amplification in VDI.}
	\vspace{-10pt}
	\label{fig:vdi_io}
\end{figure}

\begin{figure}
	\centering
	\begin{subfigure}{0.49\linewidth}
		\includegraphics[width=\linewidth]{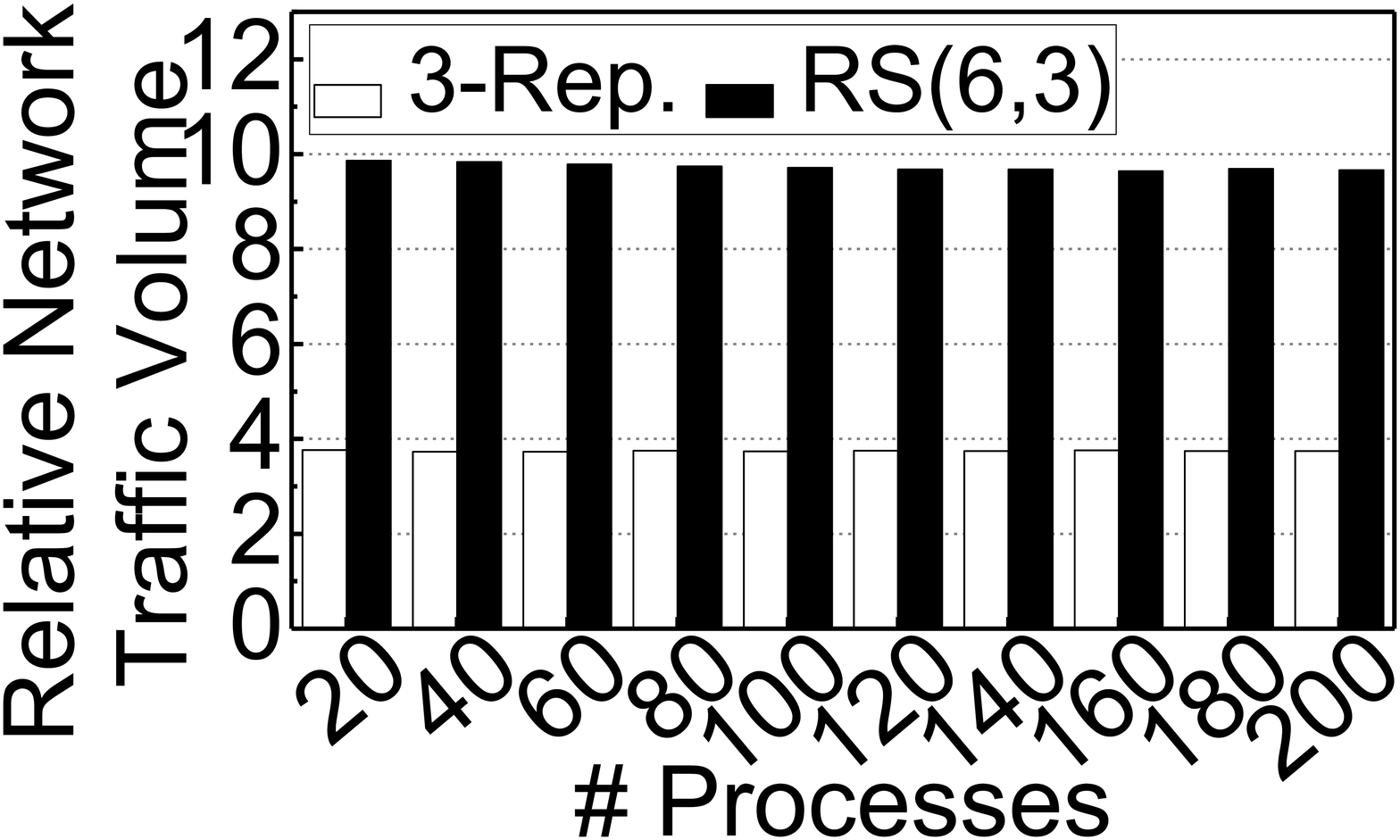}
		\caption{VDI}
		\label{fig:vdi_network}
	\end{subfigure}
	\begin{subfigure}{0.49\linewidth}
		\includegraphics[width=\linewidth]{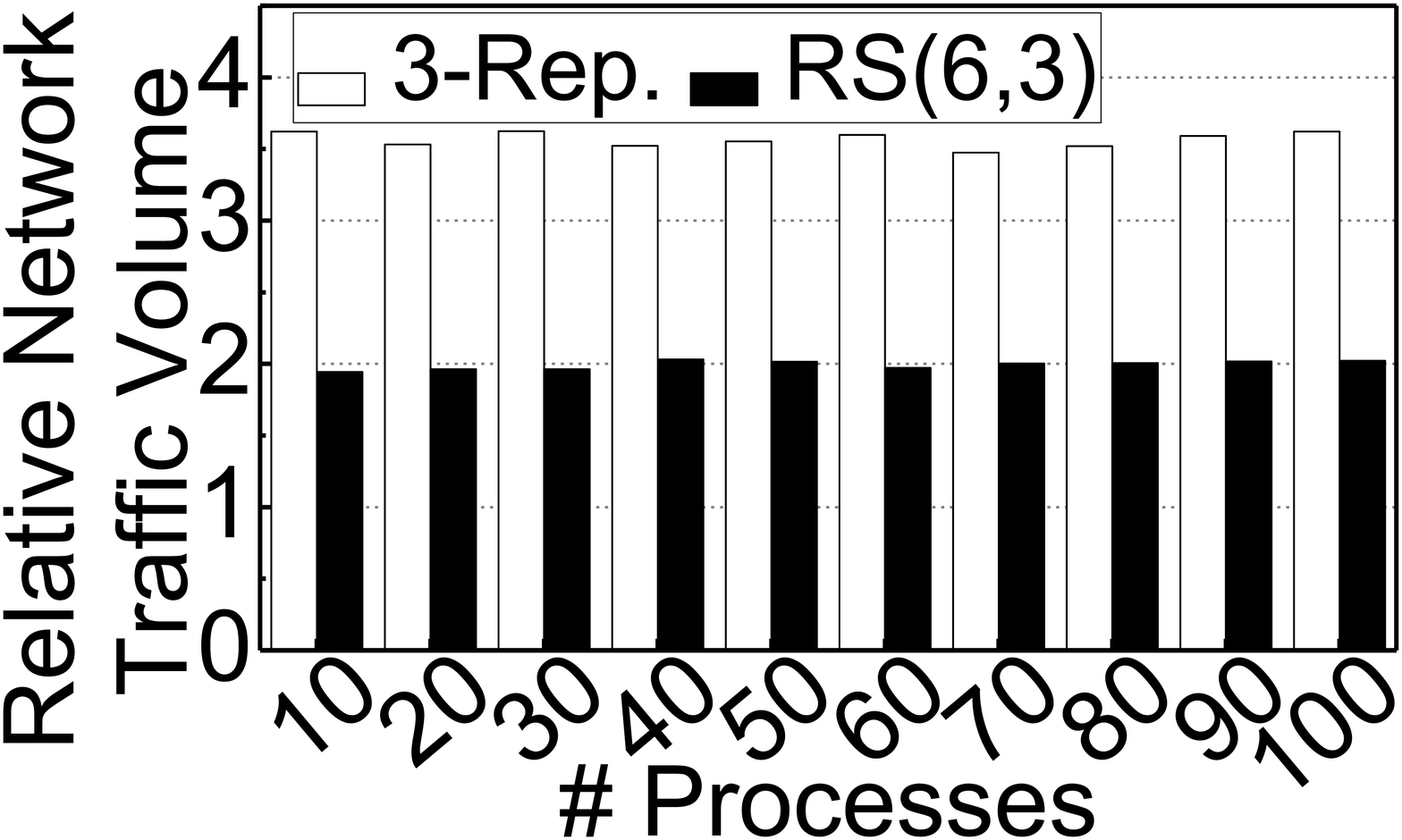}
		\caption{VDA}
		\label{fig:vda_network}
	\end{subfigure}
	\vspace{-5pt}
	\caption{Network traffic in VDI and VDA.}
	\vspace{-10pt}
\end{figure}

\noindent\textbf{Network traffic volume and I/O amplification.} 
Referring to Figure \ref{fig:vdi_io}, the read and write amplification of RS(6,3) is 8.5$\times$ and 3.3$\times$ greater, respectively, than those of 3-replication. Note that RS(6,3) read amplification is significantly greater for VDI than for DB, which is discussed in Section \ref{db}. This is because VDI is mostly composed of writes, which introduces significant additional reads to update the coding chunks, whereas the DB workload consists mostly of reads. For the same reason in RS(6,3), more network traffic is incurred under VDI than under DB, as shown in Figure \ref{fig:vdi_network}. Not only do additional reads increase the number of requests to be served in storage but also mixed reads and writes incur significant degradation in SSDs. In addition, a notable increment is observed on the read amplification of RS(6,3). This is because the increment in the total requested write volume incurs an increment in the read volume. (as the read amplification is calculated by dividing the read volume from storage by the total requested read volume.)

\subsection{Electronic Design Automation (EDA)}
EDA mimics the user scenarios that execute a mixture of electronic design automation applications. EDA execution consists of frontend processes and backend processes, which are mainly related to processing data and generating the output files, respectively. In a nutshell, EDA can be classified as data-intensive applications whose read and write ratios are well balanced, and we observe that the major request size of EDA is 64KB (cf. Table \ref{workload_char}).

\begin{figure}
	\centering
	\begin{subfigure}{0.49\linewidth}
		\includegraphics[width=\linewidth]{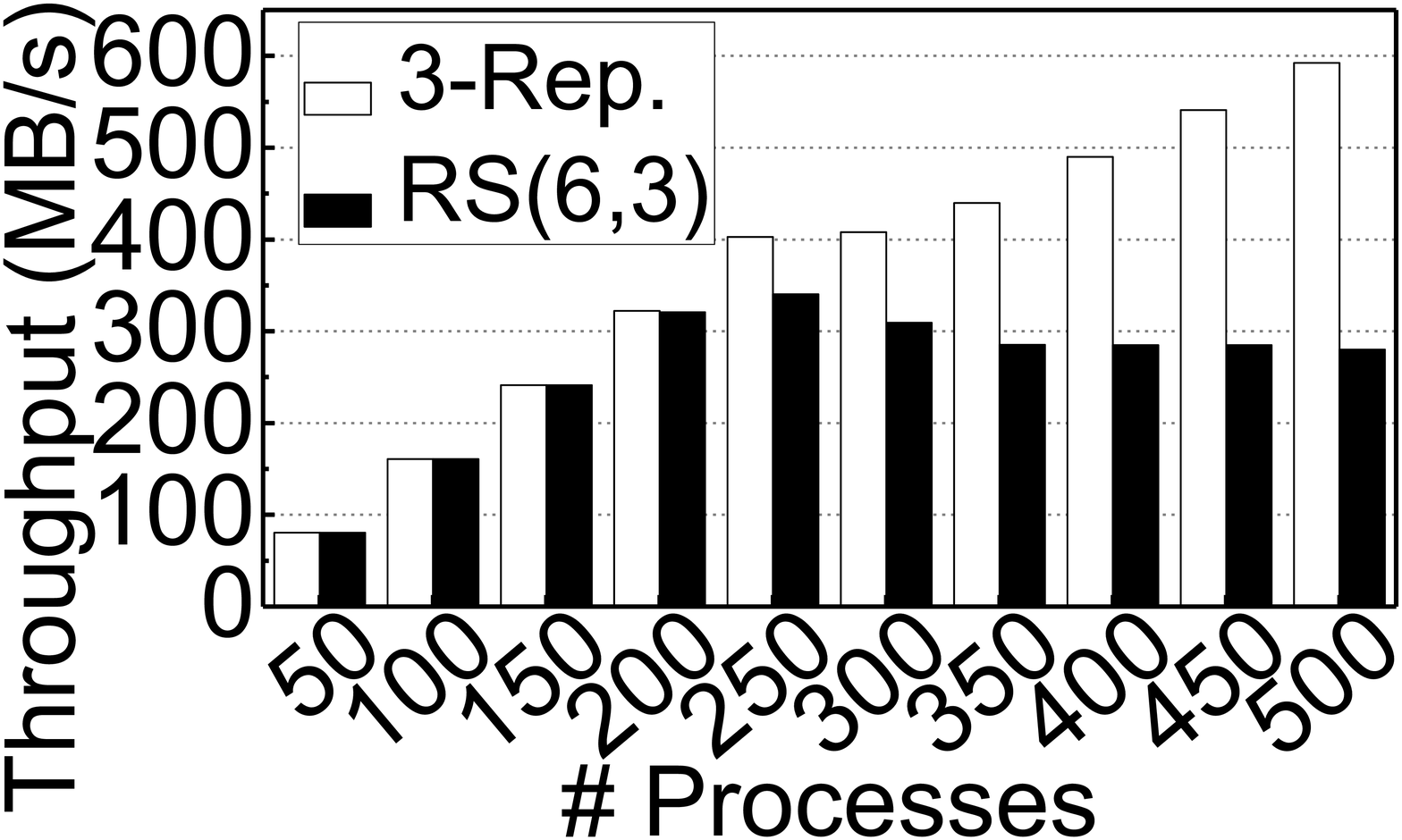}
		\caption{Throughput.}
		\label{fig:eda_thr}
	\end{subfigure}
	\begin{subfigure}{0.49\linewidth}
		\includegraphics[width=\linewidth]{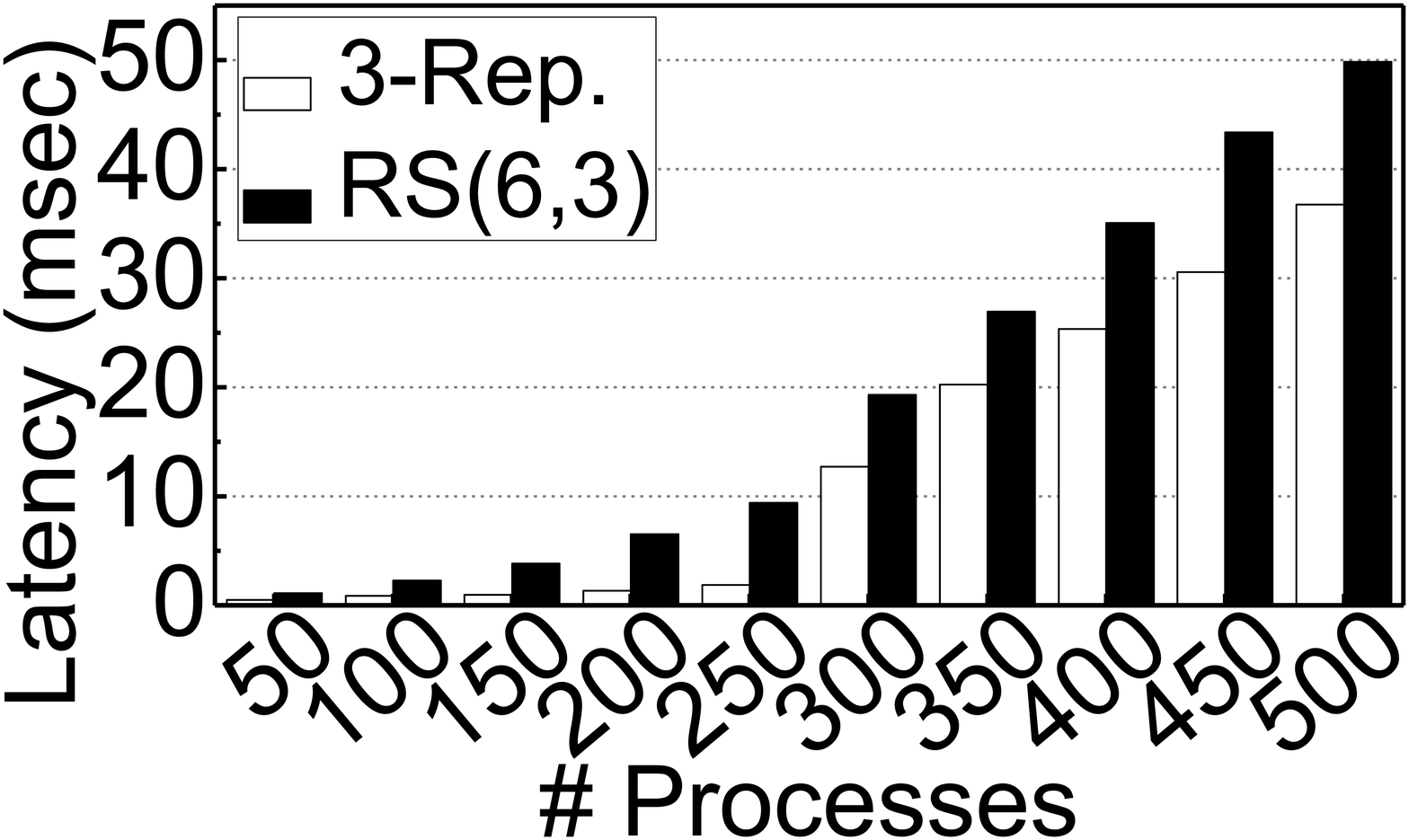}
		\caption{Latency.}
		\label{fig:eda_lat}
	\end{subfigure}
	\vspace{-5pt}
	\caption{Performance of EDA.}
	\vspace{-10pt}
	\label{fig:eda_perf}
\end{figure}

\begin{figure}
	\centering
	\begin{subfigure}{0.49\linewidth}
		\includegraphics[width=\linewidth]{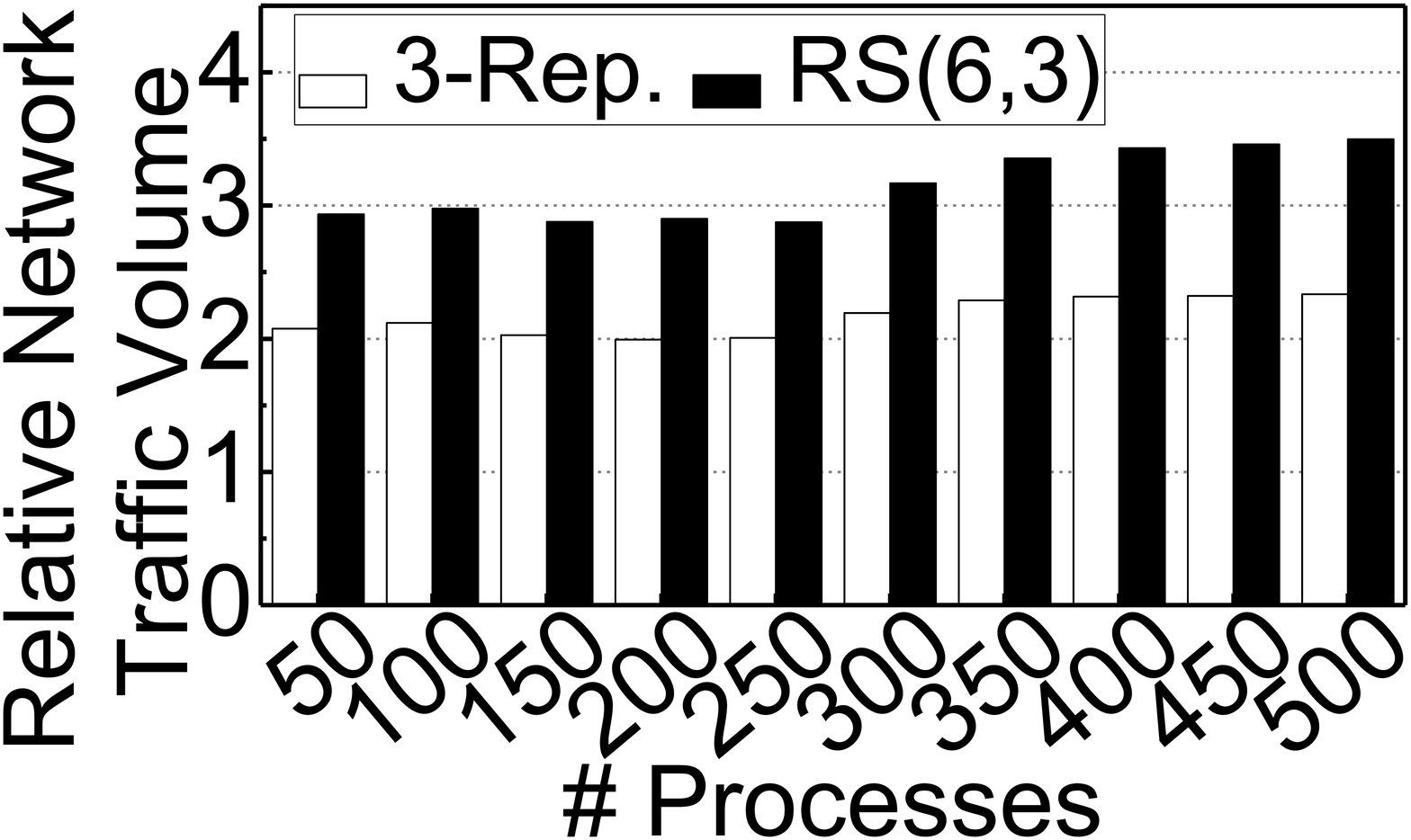}
		\caption{Network traffic.}
		\label{fig:eda_network}
	\end{subfigure}
	\begin{subfigure}{0.49\linewidth}
		\includegraphics[width=\linewidth]{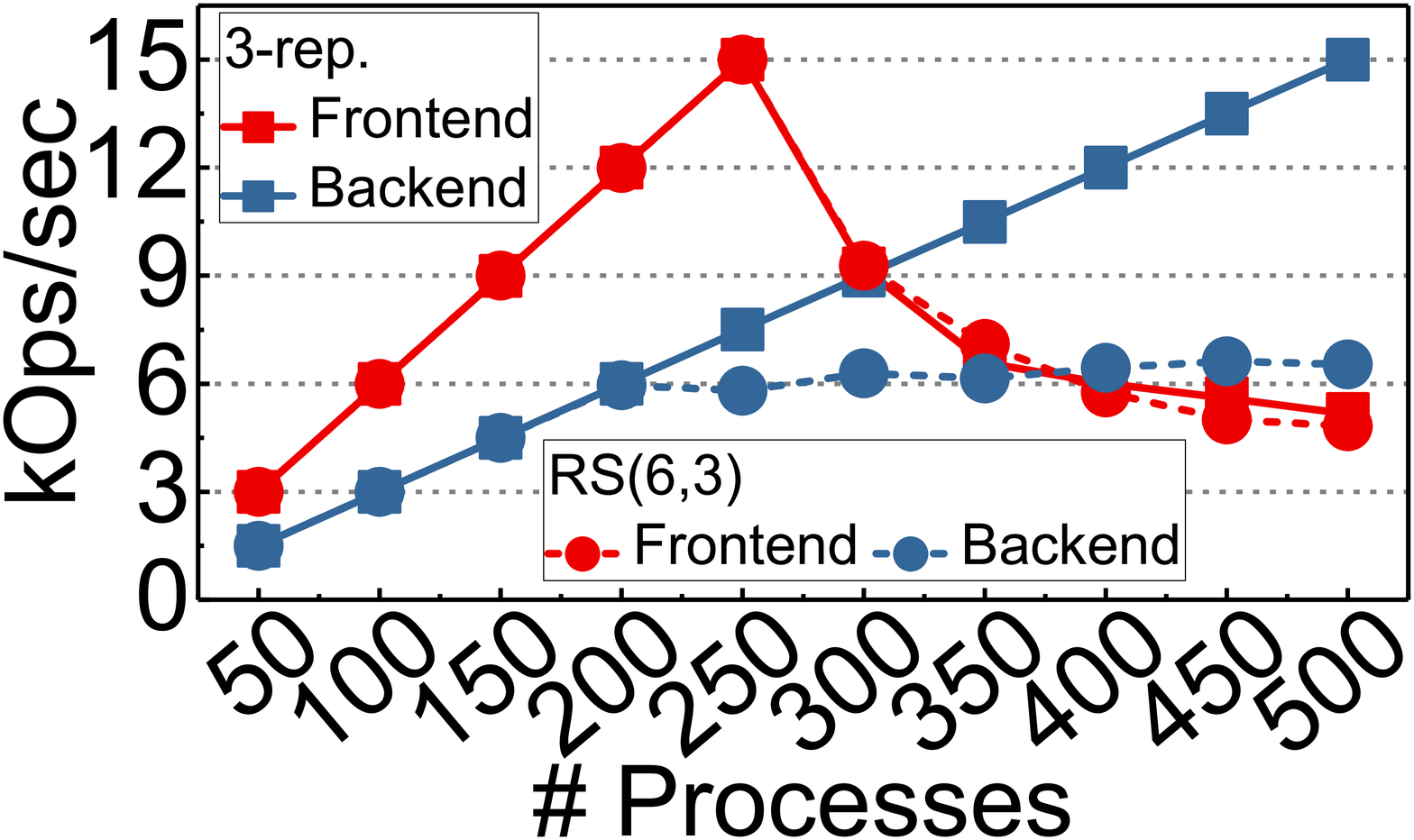}
		\caption{Performance in detail.}
		\label{fig:eda_portion}
	\end{subfigure}
	\vspace{-5pt}
	\caption{Network traffic and performance portion in EDA.}
	\vspace{-10pt}
\end{figure}

\noindent\textbf{Throughput and latency.} As shown in Figures \ref{fig:eda_thr} and \ref{fig:eda_lat}, at most, 2.1$\times$ throughput degradation and 5$\times$ latency degradation, respectively, are observed in RS(6,3) compared with 3-replication. Since the average request size of the EDA workload is larger than that for the previous two workloads (i.e., DB, VDI), the degradation impact is smaller than those workloads, as observed in Section \ref{sec:overall}. 
As the number of processes increases to 250, both throughput and latency of the system deploying erasure coding increase. However, when the number of processes exceeds 250, the throughput of RS(6,3) decreases, the throughput of 3-replication decreases and the latency of both RS(6,3) and 3-replication radically increases. In particular, the number of operations, which are requested by frontend processes, served per unit time radically decreases, in both 3-replication and RS(6,3), as shown in Figure \ref{fig:eda_portion}. Frontend processes are mostly composed of metadata related I/O requests, and therefore, we believe that the MDS of Ceph becomes a bottleneck. Because MDS shares SSDs with OSDs, if the amount of metadata to be processed exceeds the cache size of the MDS as the number of processes increases, metadata related requests are delayed by the significant I/Os incurred by data related requests. Therefore, the throughput of metadata related requests drastically decreases and the latency drastically increases irrespective of the type of data pool. However, far more requests from backend processes, which only consist of data-related requests, are served in the same amount of time in 3-replication than in RS(6,3), as shown in Figure \ref{fig:eda_portion}.


\begin{figure}
	\centering
	\begin{subfigure}{0.49\linewidth}
		\includegraphics[width=\linewidth]{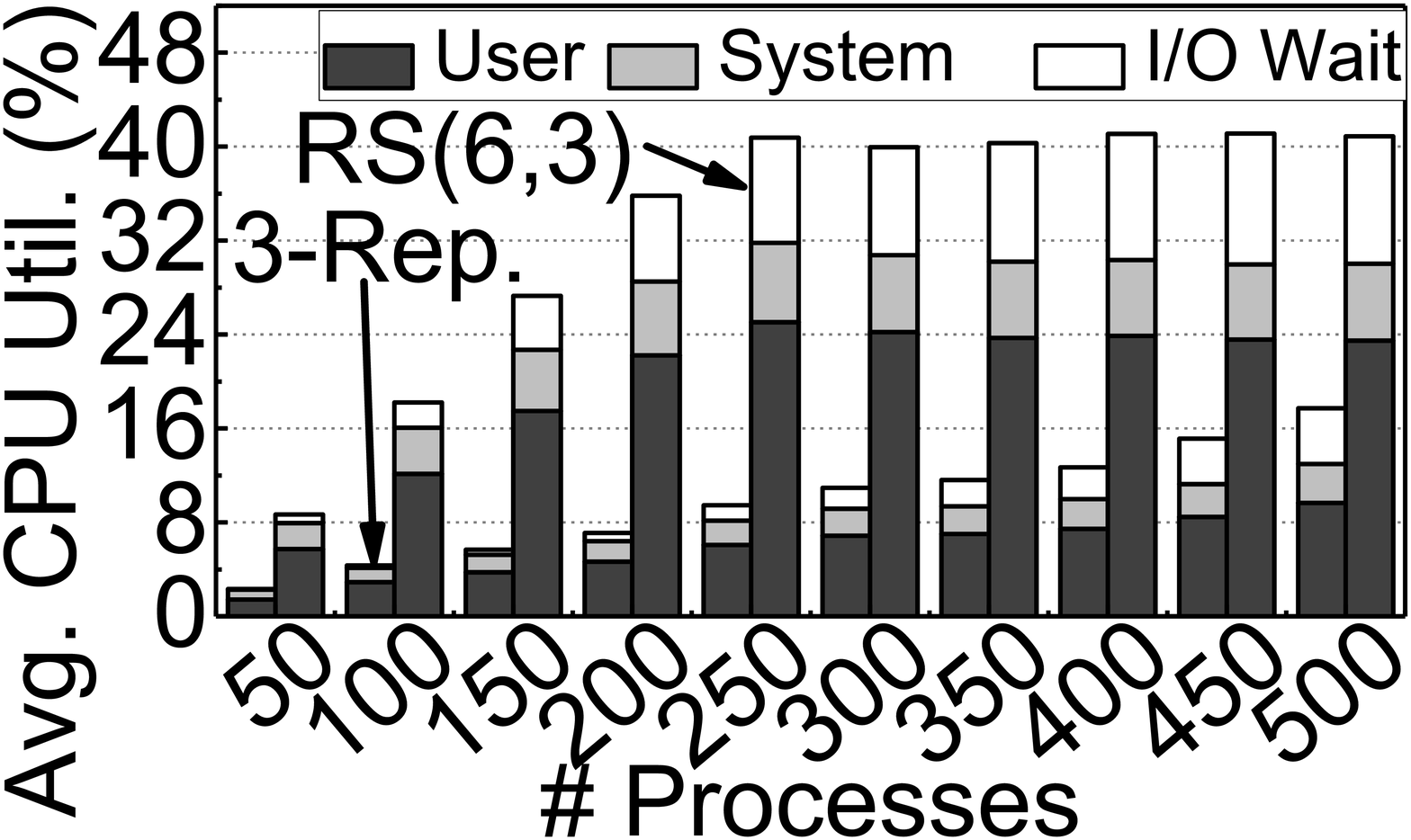}
		\caption{CPU utilization.}
		\label{fig:eda_cpu}
	\end{subfigure}
	\begin{subfigure}{0.49\linewidth}
		\includegraphics[width=\linewidth]{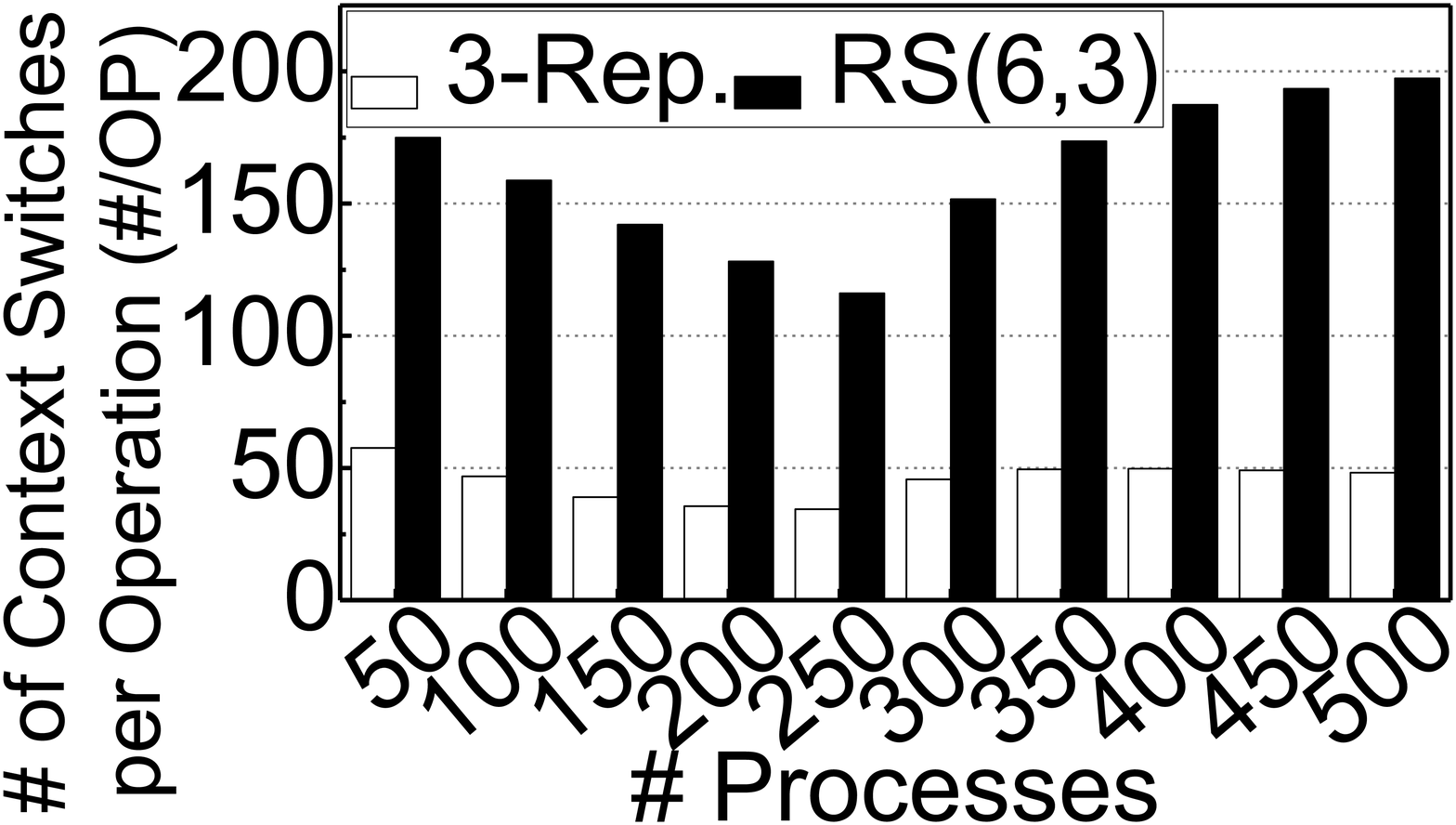}
		\caption{Context switches.}
		\label{fig:eda_ctx}
	\end{subfigure}
	\vspace{-5pt}
	\caption{Computation and system overheads in EDA.}
	\vspace{-10pt}
	\label{fig:eda_comp}
\end{figure}

\noindent\textbf{CPU utilization and context switches.} Figure \ref{fig:eda_cpu} indicates that RS(6,3) utilizes 3.7$\times$ more CPU cycles than 3-replication, on average. For the same reasons discussed in subsections \ref{db} and \ref{vdi}, the CPU utilization of RS(6,3) increases and becomes saturated as the number of processes increases. The CPU utilization of 3-replication is significantly smaller under VDI than under EDA since this workload includes far more metadata-related requests which requires less CPU cycles than data related requests. As shown in Figure \ref{fig:eda_ctx}, RS(6,3) requires 3.6$\times$ more context switches per operation than 3-replication, on average, but this is smaller than the values reported in Section \ref{db}, for the read-intensive DB workload. This is because far more metadata-related requests that require less context switches per request are included in this workload than in the DB workload. The trend of relative number of context switches is not consistent as the number of processes increases. As long as the number of processes is smaller than 250, an increase in the number of processes results in the increase of the number of metadata related requests, therefore the relative number of context switches decreases. When the number of processes exceeds 300, the requests from the frontend, which contain numerous metadata operations decrease, and the relative number of context switches increases.

\begin{figure}
	\centering
	\begin{subfigure}{0.49\linewidth}
		\includegraphics[width=\linewidth]{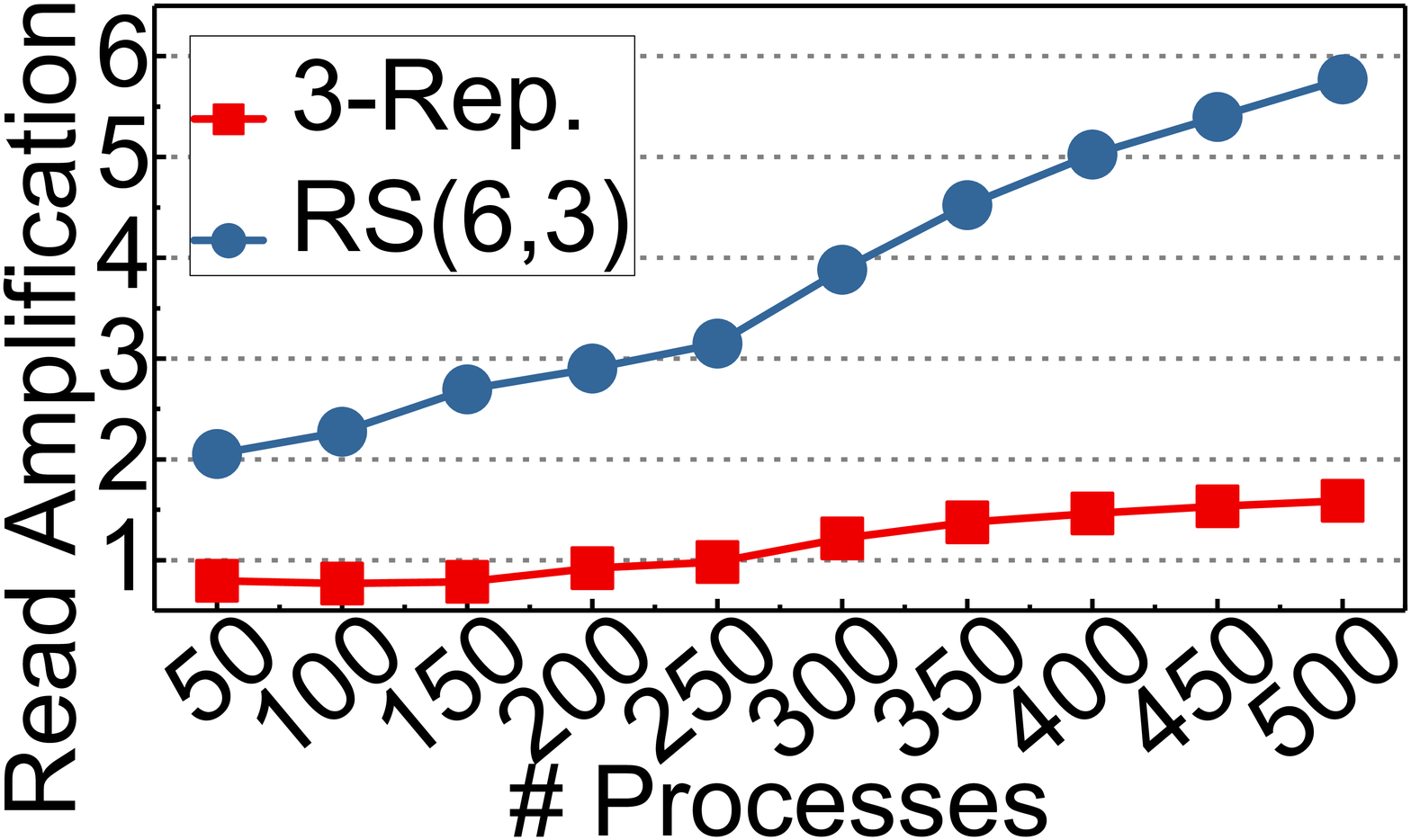}
		\caption{Read amplification.}
		\label{fig:eda_read}
	\end{subfigure}
	\begin{subfigure}{0.49\linewidth}
		\includegraphics[width=\linewidth]{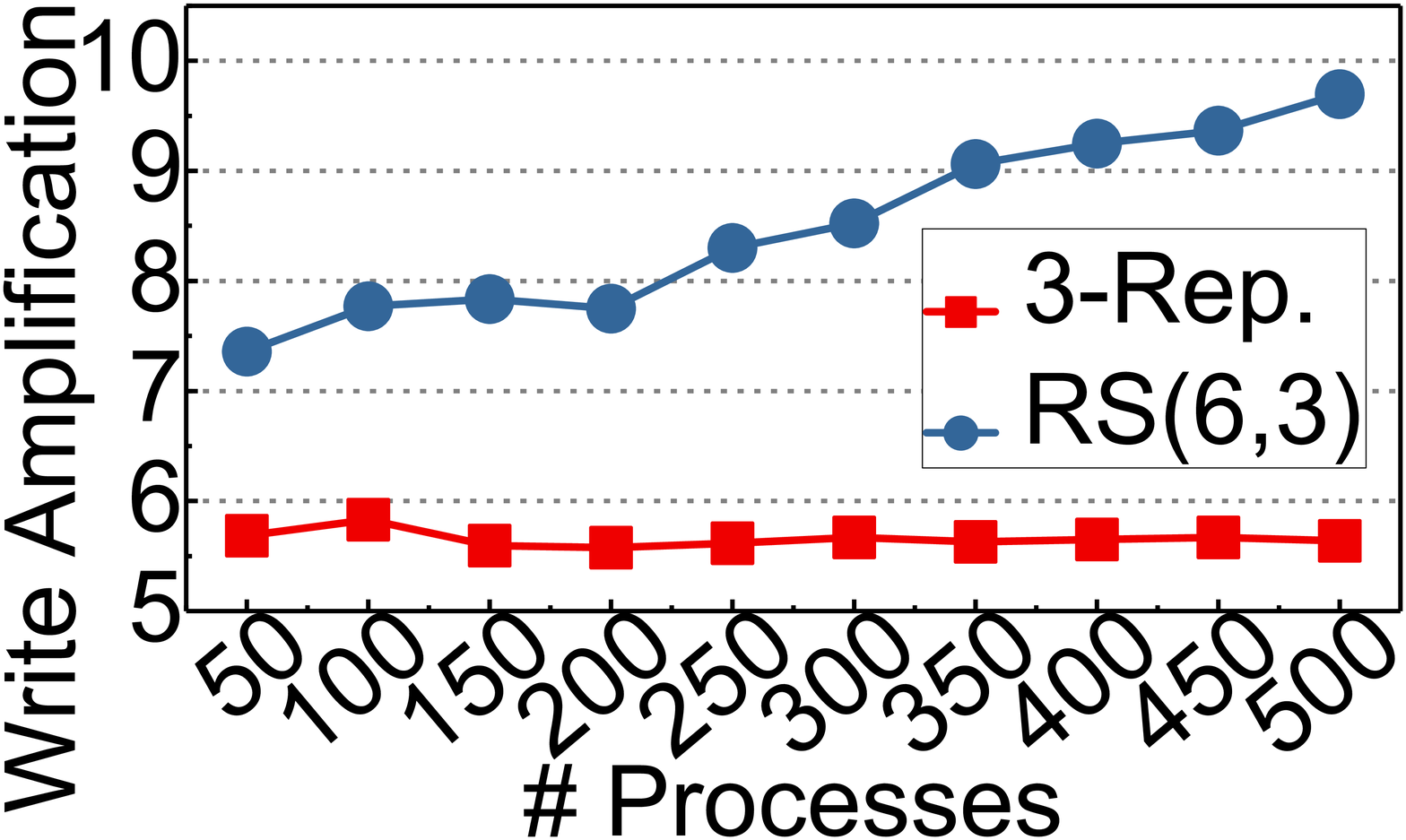}
		\caption{Write amplification.}
		\label{fig:eda_write}
	\end{subfigure}
	\vspace{-5pt}
	\caption{I/O amplification in EDA.}
	\vspace{-10pt}
	\label{fig:eda_io}
\end{figure}

\noindent\textbf{Network traffic volume and I/O amplification.} As shown in Figure \ref{fig:eda_network}, RS(6,3) generates 3.1$\times$ as much network traffic volumes as the requested volumes which is 1.5$\times$ more than replication, on average. Referring to figures \ref{fig:eda_io} and \ref{fig:eda_network}, the magnitude of I/O amplification and the relative network traffic volume of RS(6,3) are far smaller than those of the previous workloads. This is because the major request size is bigger than that of the previous workloads, occupying a larger portion of the stripe; therefore, the rest of the stripe, which needs to be additionally read/written, becomes smaller.


\subsection{Video Data Acquisition (VDA)}
VDA executes two different processes; one for storing the data stream acquired from a volatile storage, and another for running companion/user applications. Even though VDA has more data than metadata, we expect that the performance of RS(6,3) is not severely degraded compared with 3-replication. We believe that this is because, to deliver a high-resolution video service, each stream of VDA is only required to serve video with a 36 Mbit/s data transfer rate. Moreover, as shown in Table \ref{workload_char}, more than 50\% of the I/O services comprise 512KB requests, which can reduce the throughput overheads incurred by RS codes while increasing the latency of 3-replication to clone data for each operation.

\begin{figure}
	\centering
	\begin{subfigure}{0.49\linewidth}
		\includegraphics[width=\linewidth]{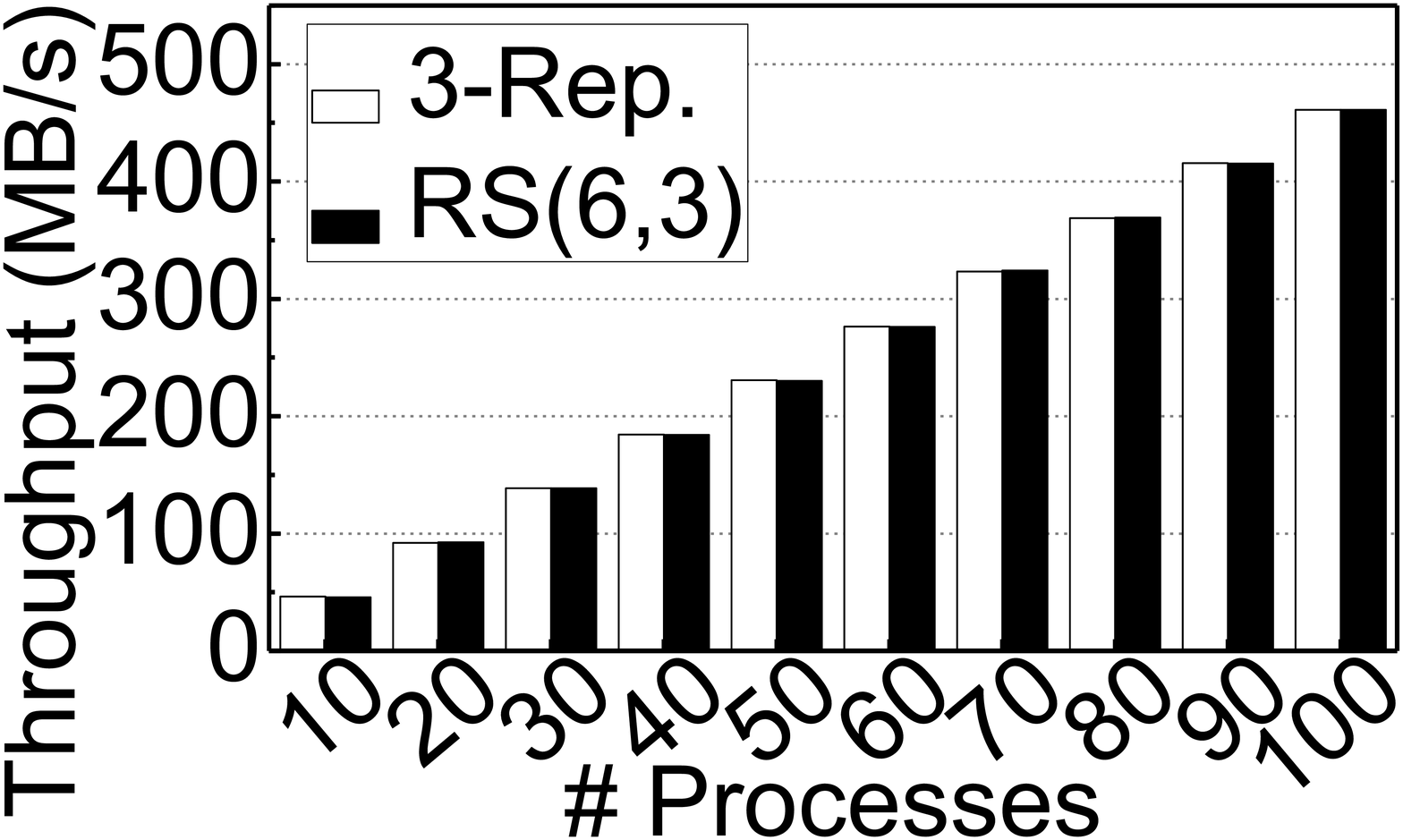}
		\caption{Throughput.}
		\label{fig:vda_thr}
	\end{subfigure}
	\begin{subfigure}{0.49\linewidth}
		\includegraphics[width=\linewidth]{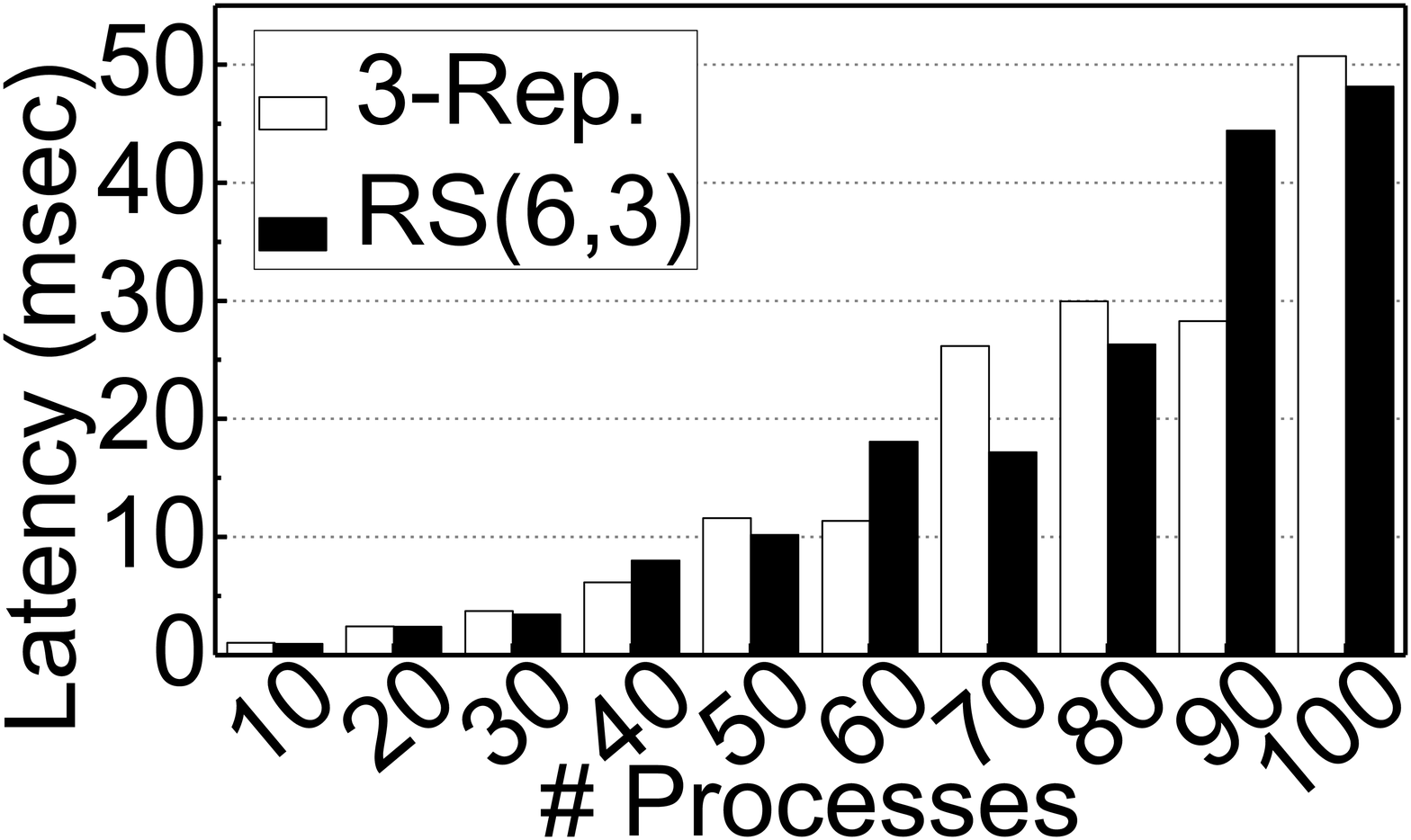}
		\caption{Latency.}
		\label{fig:vda_lat}
	\end{subfigure}
	\vspace{-5pt}
	\caption{Performance of VDA.}
	\vspace{-10pt}
	\label{fig:vda_perf}
\end{figure}

\noindent\textbf{Throughput and latency.} As shown in Figure \ref{fig:vda_perf}, the performance difference between RS(6,3) and 3-replication is insignificant compared with DB, VDI, and EDA. The throughput and latency of RS(6,3) are hardly worse than those of 3-replication. We propose two reasons for this performance characteristic. First, the major request size is sufficiently large to achieve the benefits of the high redundancy reduction of RS codes and minimize the overheads incurred by RS-concatenation. As discussed in Section \ref{sec:overall}, a large request size (equal to or greater than 512KB) makes RS codes competitive to 3-replication, except for sequential writes. Even though 50\% of VDA exhibits sequential writes, RS codes exhibit no performance degradation, compared with 3-replication, as each stream only performs I/O services with 4.5MB/s service-level agreement for high-resolution video service.

\begin{figure}
	\centering
	\begin{subfigure}{0.49\linewidth}
		\includegraphics[width=\linewidth]{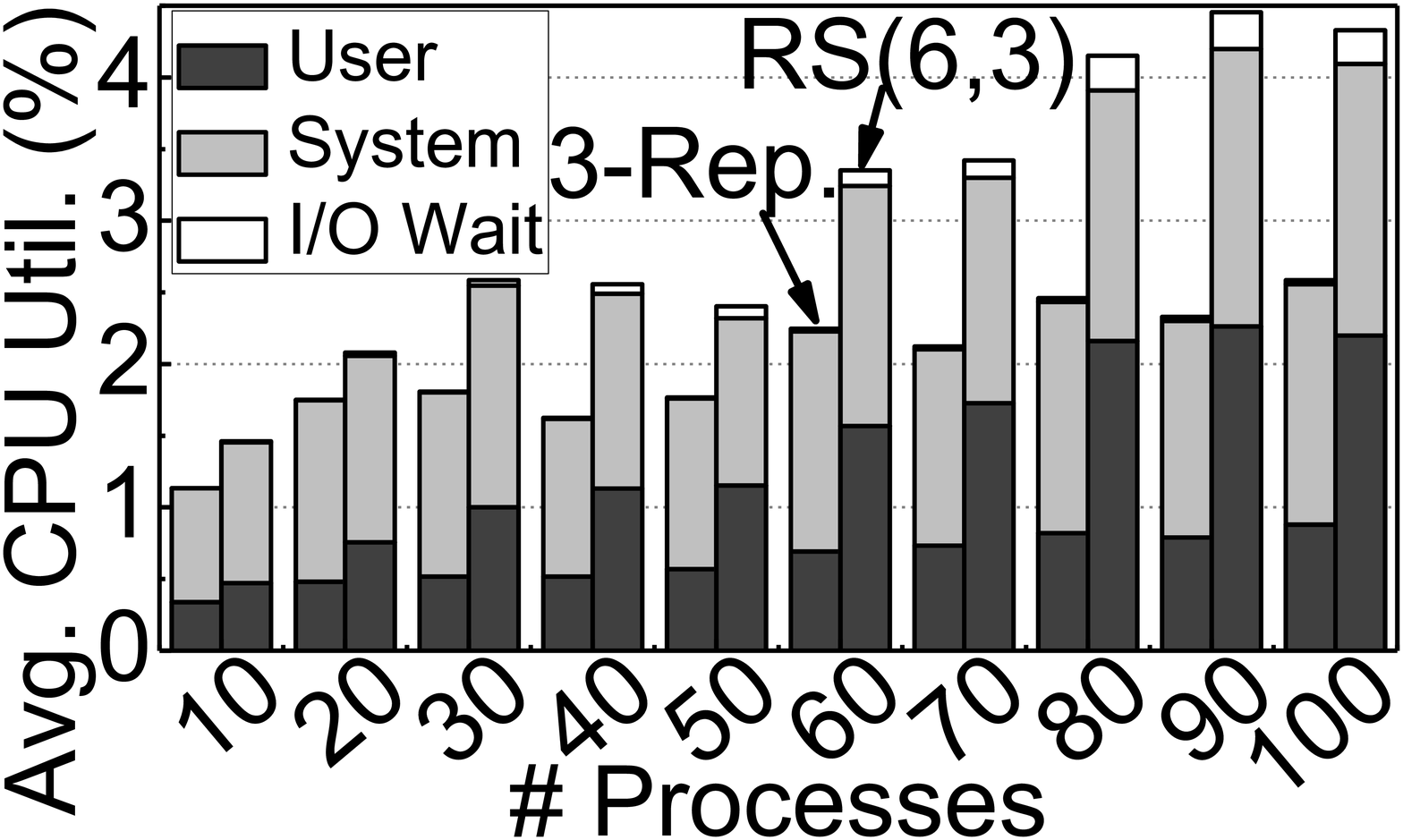}
		\caption{CPU utilization.}
		\label{fig:vda_cpu}
	\end{subfigure}
	\begin{subfigure}{0.49\linewidth}
		\includegraphics[width=\linewidth]{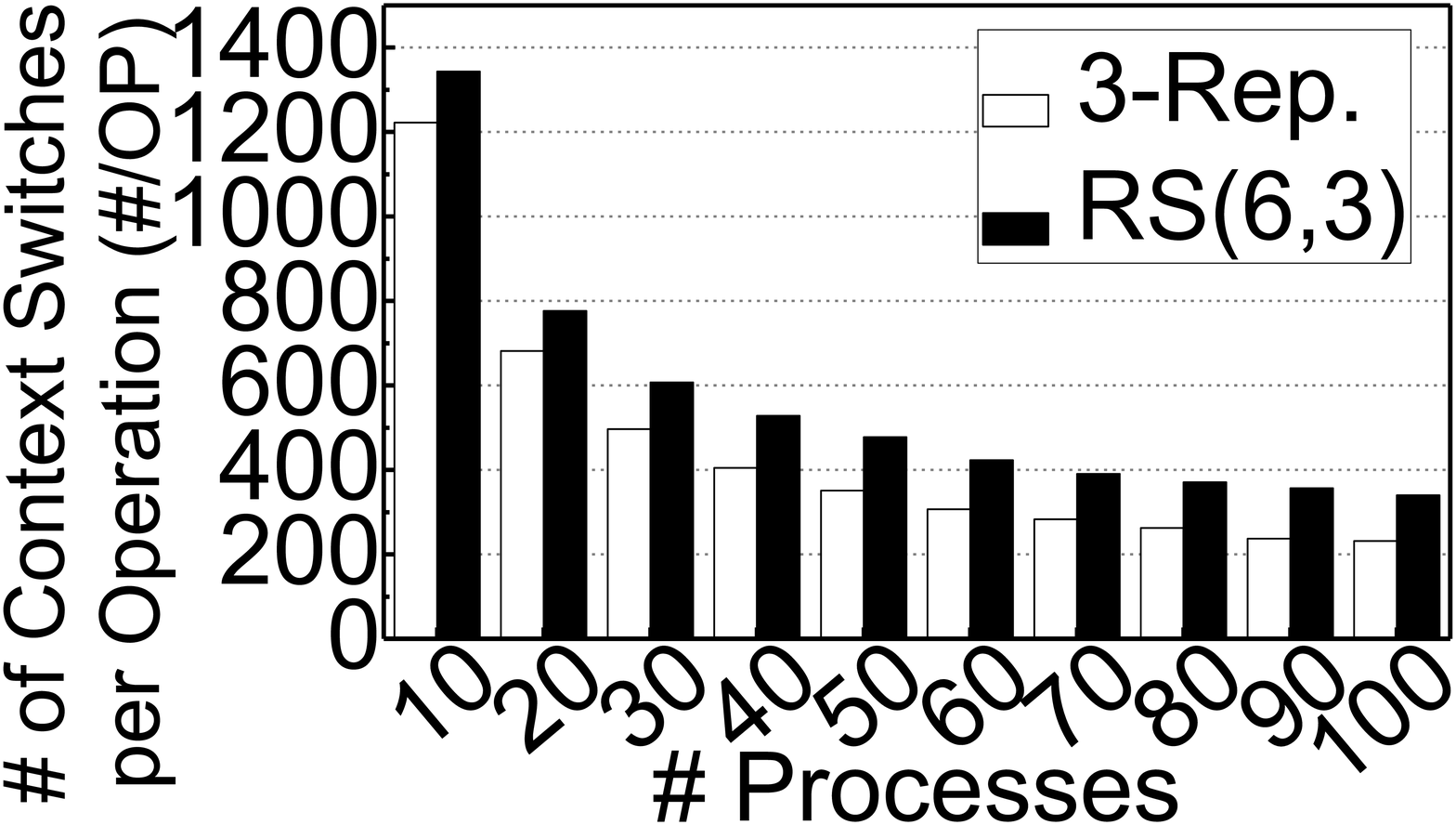}
		\caption{Context switches.}
		\label{fig:vda_ctx}
	\end{subfigure}
	\vspace{-5pt}
	\caption{Computation and system overheads in VDA.}
	\vspace{-10pt}
	\label{fig:vda_comp}
\end{figure}

\noindent\textbf{CPU utilization and context switches.} Figure \ref{fig:vda_comp} illustrates the CPU utilization and context switch overheads of VDI. We observe from this figure that the CPU cycles consumed by RS(6,3) and 3-replication are less than 4\% of the total execution cycles, owing to the non-intensive I/O services of the video stream. This can also be observed in the system-level CPU cycles required by kernel, which account for 67.7\% and 51.6\% of the total CPU cycles on average, respectively, in contrast to other applications. However, the CPU utilization of RS(6,3) is 35.2\% higher than that of 3-replication, on average for encoding the stripe and RS-concatenation. RS codes also need to manage more information for data chunks at the user-level compared with 3-replication, and thus switching the contexts 1.3$\times$ more than 3-replication. 

\begin{figure}
	\centering
	\begin{subfigure}{0.49\linewidth}
		\includegraphics[width=\linewidth]{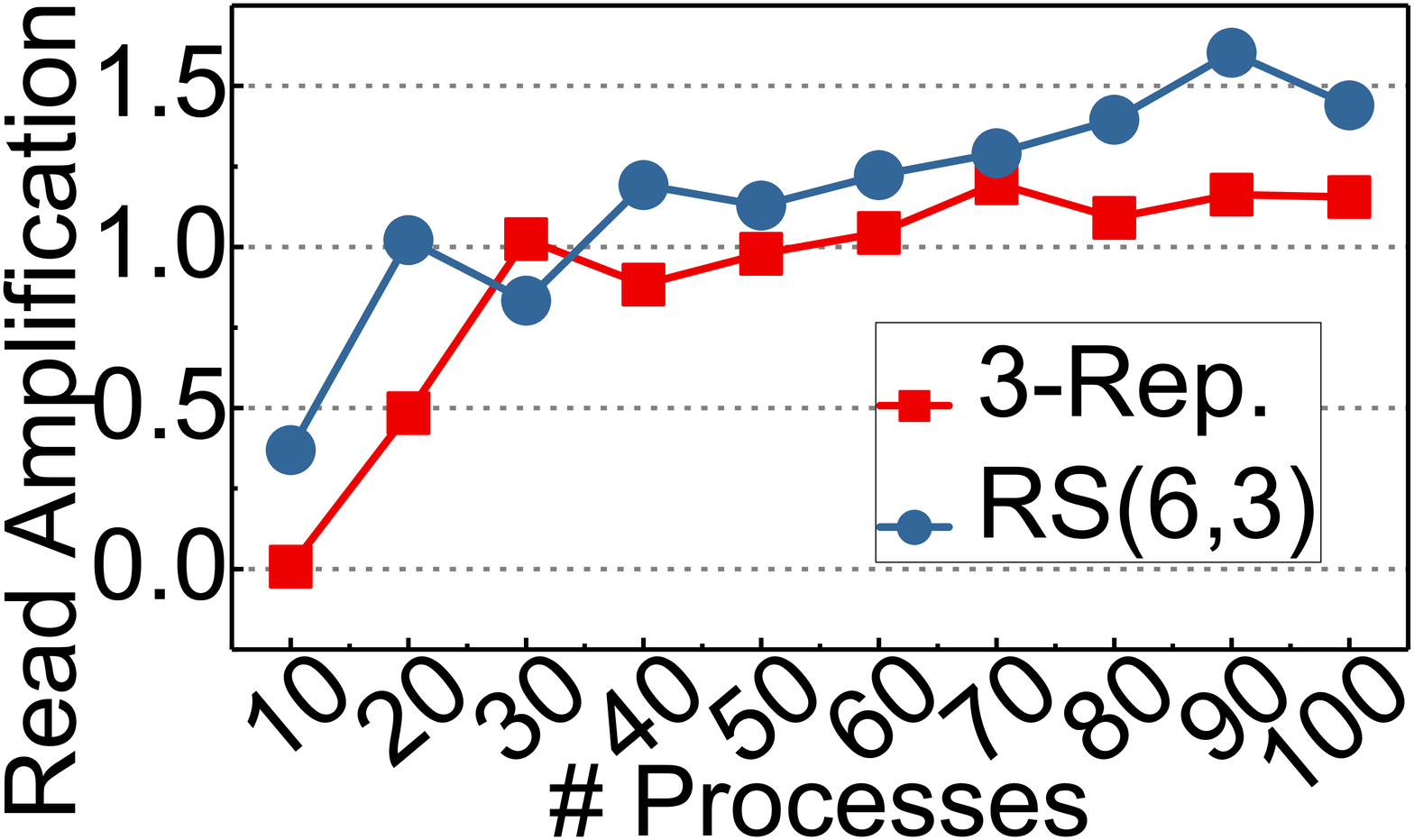}
		\caption{Read amplification.}
		\label{fig:vda_read}
	\end{subfigure}
	\begin{subfigure}{0.49\linewidth}
		\includegraphics[width=\linewidth]{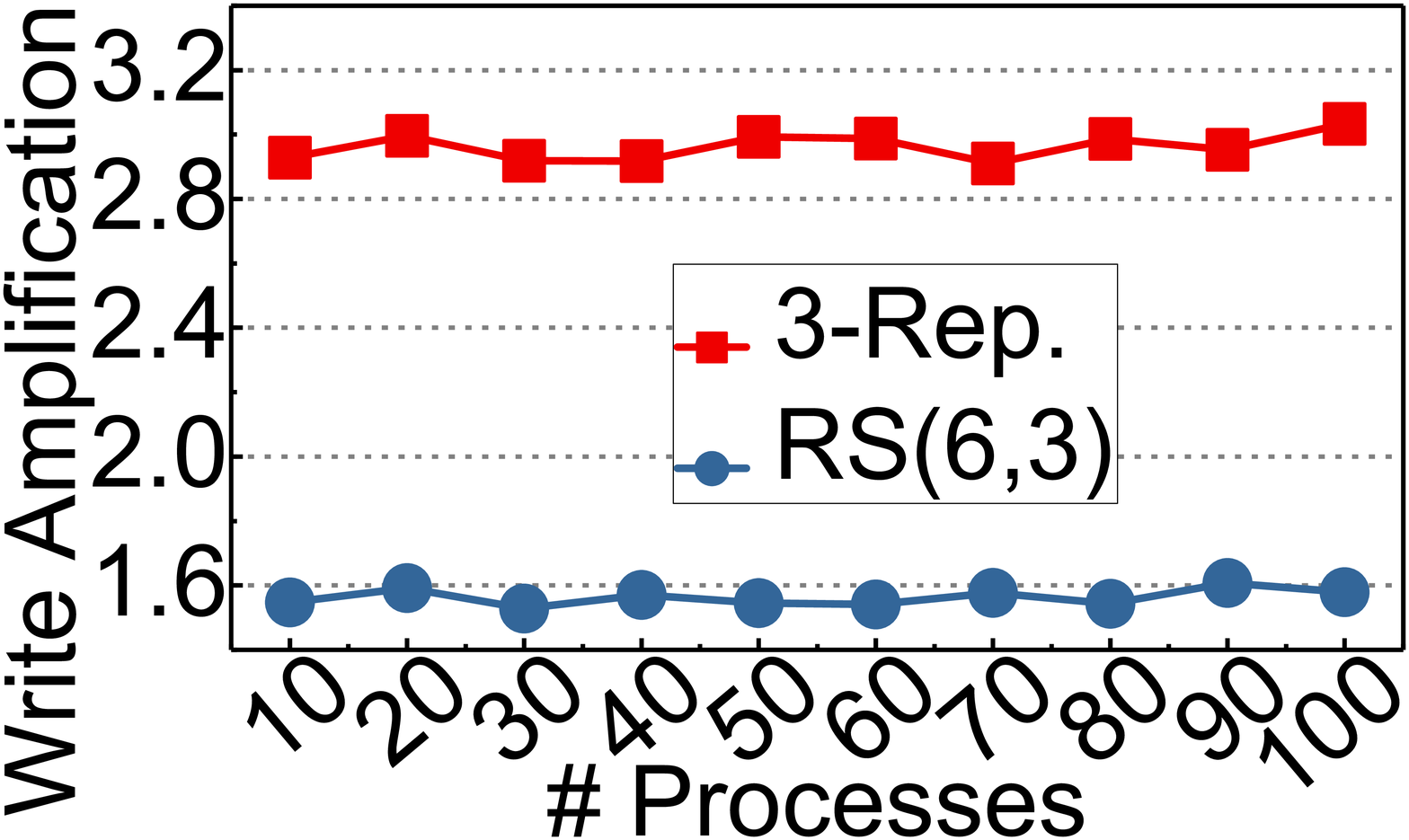}
		\caption{Write amplification.}
		\label{fig:vda_write}
	\end{subfigure}
	\vspace{-5pt}
	\caption{I/O amplification in VDA.}
	\vspace{-10pt}
	\label{fig:vda_io}
\end{figure}

\noindent\textbf{Network traffic volume and I/O amplification.} 
The network traffic and I/O amplification analysis results for VDA also exhibit a different trend, from the other application analyses. 
As shown in Figures \ref{fig:vda_network} and \ref{fig:vda_io}, 3-replication exhibits relative network traffic and write amplification worse than RS codes by 1.8$\times$ and 1.9$\times$, respectively. Even though VDA needs I/O services less than other applications, the small number of I/O requests still require redundancy. Since the size of most I/O requests of VDA is equal to or greater than 512KB, the redundancy reduction rates of RS(6,3) brings more benefits compared with 3-replication. In particular, writes are only generated by data stream processes, and their request pattern allows 100\% 512KB sequential access. This in turn reduces the overheads of the RS codes, and their small number of I/O requests are mostly buffered in the page cache of Bluestore thereby eliminating the network traffic and context switching overheads as well. 
In addition, as most of the requests generated by the companion processes are reads ($\geq$ 98\%), the overheads incurred by the online encoding of RS codes are not present.

\begin{figure}
	\centering
	\includegraphics[width=\linewidth]{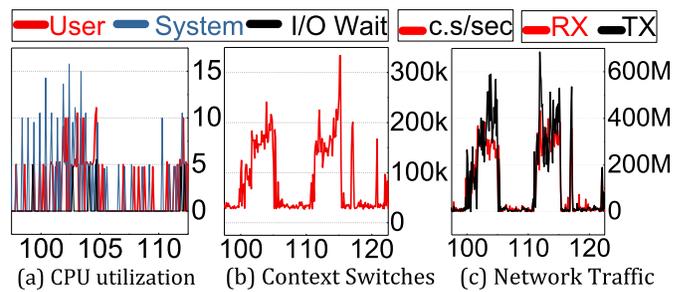}
	\vspace{-5pt}
	\caption{Times series analysis of the VDA workload from three viewpoints.}
	\vspace{-10pt}
	\label{fig:VDA_tsa}
\end{figure}

\noindent \textbf{Dynamics.}
To confirm the results for the VDA system-level characteristics, we perform a time series analysis of the CPU utilization, context switches and network traffic for RS(6,3), as illustrated in Figures \ref{fig:VDA_tsa}a, \ref{fig:VDA_tsa}b, and \ref{fig:VDA_tsa}c, respectively. As shown in the figures, the CPU cycles consumed by the RS codes are not over 25\% of total CPU execution, and the context switches and I/O traffic are 4\% and 80\% lower than those of DB, on average, respectively. We observed that, for the entire application execution process, the curve patterns of Figure \ref{fig:VDA_tsa} are repeated, making this time series analysis clear evidence of the small performance degradation of RS, compared with other applications. 

\section{System Implication and Future Work}
\label{sec:optimization}
In theory, erasure codes can guarantee a same level of system reliability while removing the heavy data redundancy imposed by 3-replication. However, in practice, we revealed that SSD arrays with the erasure codes significantly deteriorate their system performance. This performance degradation is mainly caused by frequent context switches, which deal with computing and I/O boundaries for RS en/decoding. In addition, erasure codes introduce excessive network traffic and increase the amount of writes, which can shorten the lifetime of the underlying SSDs. We believe that these performance degradations and system-level overheads brought by the erasure codes should be addressed for SSD arrays in distributed systems. In this section, we discuss system implications and improvement points to appropriately adopt erasure codes in SSD arrays.

\noindent \textbf{Performance degradation caused by computing.} As we discussed in Sections \ref{sec:overall} and \ref{sec:syschar}, the computing power itself for erasure codes is not on the critical path, but the location of en/decoding RS introduces serious system-level challenges. Specifically, the en/decoding processes for RS are mainly performed in userland, and therefore, all computations of erasure codes introduce kernel and user mode switching, including redundant memory copies from virtual address spaces to physical spaces. We believe that there are three possible approaches to address this challenge: i) \emph{user-mode storage stack}, ii) \emph{kernel-level en/decoding} and iii) \emph{en/decoding offloading}. Employing user-mode storage stack can remove the overheads of context switching and data copies between user and kernel buffers. For example, storage performance development kit (SPDK) moves all necessary storage software components and drivers from the kernel to userland. This in turn allows en/decoding processes to bypass the underlying kernel drivers thereby removing the overheads imposed by the context switches. Even though the adoption of SPDK is an efficient mechanism to enable erasure codes on SSD arrays, it in practice requires significant engineering efforts and introduces many technical challenges to redesign the current storage stack of distributed storage systems. A different approach to address the performance degradation caused by erasure codes is to migrate all RS en/decoding processes from userland to kernel. This kernel-level en/decoding method can directly perform I/O services related to erasure codes without a data copy or context switch. While those two user or kernel-level approaches consider eliminating data copies and context switches, our on-going project is to offload all the computation of erasure codes into a FPGA-based hardware accelerator. This hardware acceleration approach not only can remove aforementioned system-level overheads but also efficiently perform matrix calculations of RS (cf. Section \ref{sec:computation}) and perform I/O services, which can free the current storage stack from the reliability management.

\noindent \textbf{Network traffic and write amplification.} In contrast to the architectural challenges that introduce context switches and data copies, the inefficiency of network traffic management and write amplification problems, which are characterized by Section \ref{sec:dataoverhead}, can be addressed by tailoring the system parameters and software module design of the current distributed systems. Specifically, the size of an I/O object that Ceph reads and writes is 4MB. Even though this default object size is suggested by considering system-level parallelism and bandwidth of the underlying storage systems, the object is greater than a stripe width (cf. Section \ref{sec:distribution}) that the current en/decoding uses (24KB). Thus, this size difference between the object and stripe width introduces buffer cache corruptions and unnecessary I/O requests, which in turn imposes heavy traffic and write amplification at the storage node side. Motivated by this, one can change the system parameters (the object size or stripe width) based on the system's performance demand. However, simply reducing the object size can unfortunately increase the amount of metadata, which in turn introduces different types of challenges such as heavy memory traffic and frequent metadata accesses. Thus, we believe that the distributed system designer carefully studies the performance trade-off by considering the difference between the object and stripe sizes. The alternative option to address this challenge is to modify the current software module of the distributed system. For example, even though the object size is greater than the stripe size, the erasure code related I/O requests can be delivered to the underlying software modules with a hint. Based on the hint, the modules do not read or write data based on the object (if the I/O service is related to RS en/decoding operations). It would be also possible to modify the software stack modules in dynamically adjusting the stripe (or object) size being aware of RS en/decoding processes. 

\section{Related Work and Discussion}
\label{sec:relatedwork}
Replication is the most common but effective and practical method for maintaining storage reliability and resilience \cite{tanenbaum2007distributed}. Since the storage overheads imposed by replicas make distributed storage systems non-scalable and the network overly crowded, many studies focus on erasure coding for distributed file systems \cite{rashmi2013solution, rashmi2016ec, mitra2016partial, silberstein2014lazy}. However, erasure codes suffer from heavy overheads in cases where they must reconstruct data on a node or in cases of device failure. This is a well known problem, called repair bandwidth, which not only significantly degrades the system performance \cite{rashmi2013solution} but also severely amplifies the data volume and network traffic. To address these challenges, new techniques on erasure codes are proposed \cite{rashmi2015having, rashmi2016ec, mitra2016partial}. EC-Cache \cite{rashmi2016ec} is designed to maximize the load balancing and reduce the latency for erasure codes. In \cite{silberstein2014lazy} the repair bandwidth issue was addressed via a lazy recovery mechanism. While all these studies analyzed the overheads on data reconstruction and optimized the decoding procedures of erasure coding applied to disk-based storage clusters, our work highlights many observations with in-depth study for encoding mechanisms on distributed storage systems and parallel file system implementation.

On the other hand, \cite{mohan2015benchmarking} reports that the performance of erasure coding on the Hadoop distributed file system (HDFS) does not degrade, compared with that of replications.
This study was performed on a HDD-based cluster, and HDFS applies I/O striping to both erasure coding and replication. In contrast, the Ceph file system used in the present work is optimized for a SSD-based cluster, which directly services I/Os from a target OSD. However, erasure coding has software interventions due to its en/decoding related computations and raises synchronization/barrier issues, which makes it employ I/O striping rather than the direct I/O services. This introduces further performance differences between erasure coding and replication on SSDs, which were not observed in previous studies (on HDDs). 

\noindent \textbf{Limits of this study.} We believe that applying erasure coding to SSD arrays is in an early stage for parallel file systems, and some observations of this work may depend on various implementations of them. However, several fundamental behaviors that erasure coding requires, such as matrix multiplication with new and old data, introduce challenges similar to those reported herein. For example, erasure coding implemented at the user-level needs to read data and generates new parity bits, which introduces not only extra I/O activities such as network service and storage I/O requests but also system overheads such as kernel mode switching and memory copies. In addition, if the target file system changes its configuration such as a stripe width, the values of our observations will also change, but their trend will not vary greatly. For example, in cases where the target system increases the stripe width, the latencies for both encoding and decoding increase almost linearly, which has been observed in other studies \cite{ hu2012nccloud, papailiopoulos2012simple}.

\noindent \textbf{Reason why read operation is managed in a stripe.} Ceph handles read requests in a stripe. We believe that this is because of the architecture of Ceph. Every request is served through primary OSD by pulling data from other OSDs within the same PG. However, in some cases OSDs can be failed or be busy due to lots of pending requests. OSDs periodically checks each other's heartbeats to check whether failure occurs or not. However, during this interval, failure cannot be detected. When OSD with requested data chunk suffers from undetected failure, primary OSD has to wait until the failure is detected. In this case, the read request will be significantly delayed. So we concluded that primary OSD constructs stripe for every read request to prevent from these problems.

\section{Conclusion}
\label{sec:conclusion}
We studied the overheads imposed by erasure coding on a distributed SSD array system. In contrast to the common expectation for erasure codes, we observed that they exhibit heavy network traffic (which is invisible to users) up to 142$\times$ greater than a popular replication method (triple replication), and increases the amount of data that the underlying SSD needs to manage. Our results reveal that the erasure coding mechanisms on the distributed SSD array systems introduce 10.8$\times$ more context switches per operation and require 6.6$\times$ more CPU cycles at most, than the replication due to the user-level implementation and storage cluster management. 
Disregarding to the several overheads, distributed SSD array system should employ erasure codes due to its cost. Therefore, the overheads must be reduced, and it can be done by several approaches such as employing user-mode storage stack, kernel-level en/decoding, en/decoding offloading, and system-level optimization.


%



  \section*{Acknowledgments}
	This is a full version of a conference paper \cite{koh2017iiswc}. In this work, we completely revised all the previous evaluation results from scratch by replacing a block interface Ceph with filesystem-enabled Ceph, and evaluated our all-flash array clusters with an on-line erasure coding mechanism by executing diverse real application scenario \cite{spec2014sfs}. 
This research is mainly supported by  NRF 2016R1C1B2015312.
This work is also supported in part by Yonsei Future Research Grant (2017-22-0105), IITP-2017-2017-0-01015, NRF-2015M3C4A7065645, DOE DE-AC02-05CH 11231, and MemRay grant (2015-11-1731). Dr. Kim is supported in part by NSF 1640196 and SRC/NRC NERC 2016-NE-2697-A. Myoungsoo Jung is the corresponding author.

\ifCLASSOPTIONcaptionsoff
  \newpage
\fi



%



\bibliographystyle{IEEEtran}
\bibliography{reference}

\end{document}